\documentclass[a4paper]{jpconf} 
\usepackage{graphicx}
\usepackage{amsmath}
\usepackage{algorithm}
\usepackage{algorithmicx}
\usepackage{algpseudocode}
\usepackage{multirow}
\usepackage{url}

\begin{document}

\title{Finding high-order Hadamard matrices by using quantum computers}

\author{Andriyan Bayu Suksmono*}
  \address{School of Electrical Engineering and Informatics, \\Institut Teknologi Bandung, Indonesia}
\ead{suksmono@stei.itb.ac.id, absuksmono@gmail.com}

\author{Yuichiro Minato}
 \address{Blueqat Inc., Tokyo, Japan}

\begin{abstract}
Solving hard problems is one of the most important issues in computing to be addressed by a quantum computer. Previously, we have shown that the H-SEARCH; which is the problem of finding a Hadamard matrix (H-matrix) among all possible binary matrices of corresponding order, is a hard problem that can be solved by a quantum computer. However, due to the limitation on the number of qubits and connections in present day quantum processors, only low orders H-SEARCH are implementable. In this paper, we show that by adopting classical construction/search techniques of the H-matrix, we can develop new quantum computing methods to find higher order H-matrices. Especially, the Turyn-based quantum computing method can be further developed to find an arbitrarily high order H-matrix by balancing the classical and quantum resources. This  method is potentially capable to find some unknown H-matrices of practical and scientific interests, where a classical computer alone cannot do because of the exponential grow of the complexity. We present some results of finding H-matrix of order more than one hundred and a prototypical experiment to find even higher order matrix by using the classical-quantum resource balancing method. Although heuristic optimizations generally only achieve approximate solutions, whereas the exact one should be determined by exhaustive listing; which is difficult to perform, in the H-SEARCH we can assure such exactness in polynomial time by checking the orthogonality of the solution. Since quantum advantage over the classical computing should have been measured by comparing the performance in solving a problem up to a definitive solution, the proposed method may lead to an alternate route for demonstrating practical quantum supremacy in the near future.
\end{abstract}

\section{Introduction}

\subsection{Background}
A Hadamard matrix (H-matrix) is a binary orthogonal matrix with $\{-1, +1\}$ elements whose any distinct pair of its columns (or rows) are orthogonal to each other. Such a matrix only exists when it is square and the length of its column (row) is a multiple of four; i.e., for an $M \times M$ dimension H-matrix, then $M=4k$ for a positive integer $k$. The reversed statement that for any positive integer $k$ there is a H-matrix is also believed to be true, although a mathematical proof (nor disproof) not yet exists. This is a long standing problem of the \emph{Hadamard Matrix Conjecture}.

The H-matrix has been a subject of scientific and practical interests. First discovered and described by Sylverster in 1867 \cite{sylvester1867}, it is further studied by Hadamard  concerning its relationship with the determinant problem \cite{hadamard_1893}. The orthogonal property and binaryness of its elements make it widely used in information processing and digital communications. The CDMA (Code Division Multiple Access) system employs Hadamard-Walsh code to reduce interference among their users, so that the capacity of the communication system 
is not badly deteriorated by the increasing number of its users. The H-matrix was also used by Mariner 9 space-craft as its ECC (Error Correcting Code) for sending images of Mars to a receiving station located on Earth, thanks to its capability for long error correction.

Some particular kinds of H-matrices can be found (constructed) easily, while others need huge computational resource to do. An H-matrix of size $M\times M$ is also called an $M$-order H-matrix. When $M$ follows a particular pattern of $M=2^n$, where $n$ is a positive integer, the matrix can be easily constructed by the Sylvester's method of tensor product. Hadamard \cite{hadamard_1893} constructed the H-matrices of order $12$ and $20$, whose orders do not follow the $2^n$ pattern. It indicates that other orders than prescribed by the Sylvester's method do exist. Paley showed the construction of H-matrix of order $M=4k$ where $k \equiv 1 \mod 4 $ and $k \equiv 3 \mod 4 $, which are known as the Paley Type I and Type II H-matrices, respectively \cite{paley_1933}. In the formulation, he  employed the method of quadratic residues in a Galois field $GF(q)$, where $q$ is a power of an odd prime number. Various kinds of construction methods have also been proposed, among others which are related to the method described in this paper, are the Williamson \cite{williamson_1944}, Baumert-Hall \cite{baumert_hall_1965}, and Turyn \cite{turyn_1974}. These three later methods involve searching of particular binary sequences as an essential stage; therefore, we will also refer them to as (classical) H-matrix searching methods. 

Although at a glance it looks simple, finding a H-matrix is actually a challenging task. Low orders H-matrices can be calculated by hand, such as the $12$ and $20$ cases found by Hadamard \cite{hadamard_1893}. To find a H-matrix of order $92$, in 1961 three JPL (Jet Propulsion Laboratory) researchers employed a state of the art computer at that time, i.e. the  IBM/7090 Mainframe \cite{baumert_golomb_hall_1962}. For the order under $1,000$, the most recent unknown H-matrix successfully found is the one with order $428$, which was discovered in 2005 by using computer search of a particular binary sequence \cite{kharaghani_rezaie_2004}. The method described in the paper is of particular interest because the next unknown H-matrices, such as the one with order $668$, can possibly can be found by using the same method. The main reason they have not been found at this time is because of the huge computational resource needed to find such a matrix, which grows exponentially by the order of the matrix. 

Finding a H-matrix of order $M$ among all of about $2^{M^2}$ binary matrices, which we refer to as \emph{H-SEARCH}, is a hard problem. We have proposed to find such a matrix by using a quantum computer due to its capability in solving hard problems \cite{suksmono_minato_2019}. Theoretically, a quantum computer will need around $M^2$ qubits in superposition to solve such a problem. However, in the existing quantum annealing processor, we need around $M^3$ due to extra ancillary qubits required to translate $k$-body terms into $2$-body Ising Hamiltonian model. In this paper, we show that by adopting the classical searching methods, we can reduce the required computing resource, which for a quantum annealing processor implementing the Ising model, will become around $M^2$. We describe how to formulate the corresponding Hamiltonian related to the classical methods and shows some results of order up to more than one hundred. We also describe how to further develop this technique to find even higher order matrices, by managing the classical and quantum computing resources.

Usually, solving an optimization problem by annealing or heuristic methods yields only an approximate solution,  i.e., we can not sure that it is actually the optimal point, unless all of possible solutions are enumerated. However, enumeration of all possible solutions of a hard problem is an extremely laborious task. In contrast, the correctness of a solution in H-SEARCH can be verified easily in polynomial time; i.e., by evaluating the orthogonality of the found matrix (solution). If we consider the solution as a certificate, H-SEARCH behaves like an NP-complete problem because finding the solution is hard, but checking its correctness is easy. In this particular point of view, H-SEARCH is an interesting hard problem worth to consider in addressing practical quantum supremacy.

\subsection{A Brief on Quantum Computing}

Quantum computers are expected to have computational capability beyond the classical ones;
a feature which is well known as \emph{quantum supremacy} \cite{harrow_2017}. An important progress regarding this issue is the achievement of the Google team in 2019. It was claimed that a Sycamore quantum processor needs only about $200$ seconds to do a particular computational task; which is sampling random quantum circuits in this case, where a classical supercomputer would take about $10,000$ years \cite{arute_2019} to perform. In the next step, a capability of solving a closer to real-life problems, where a classical computer cannot do, rather than generating random numbers is desired. A creative thinking of building an algorithm that can demonstrate such practical supremacy is needed.

\begin{figure*}[t]
    \centering
    \includegraphics[width=1.0\columnwidth]{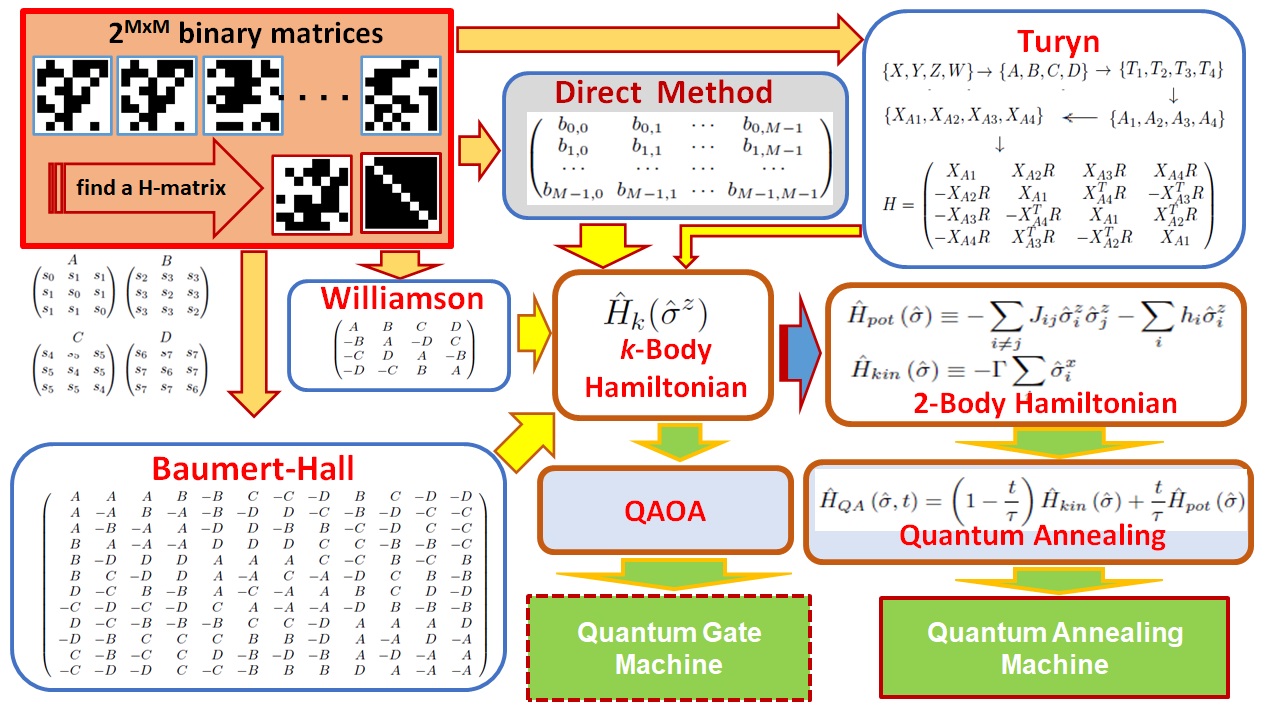}
    \protect\caption{\label{FIG_Method} Quantum computing methods for solving the problem of finding high-orders H-matrix developed from classical methods. In the previous \emph{direct method} \cite{suksmono_minato_2019}, we represent an $M$-order H-matrix to be found by an $M\times M$ binary variables, which becomes $O(M^2)$ logical qubits and $O(M^3)$ physical qubits including the ancillas, to be implemented on a quantum processor with Ising model. In this paper, we adopt three classical methods; i.e., the Williamson, Baumert-Hall, and Turyn, into quantum computing algorithms by formulating their corresponding Hamiltonians. Each of the quantum computing methods needs $O(M)$ logical qubits, which translates into ~$O(M^2)$ physical qubits when they are implemented on a QAM (Quantum Annealing Machine) type processor. In QGM (Quantum Gate Machine) processor, the number of required qubits will be proportional to the number of logical qubits in the Hamiltonians, i.e. $O(M)$. We can employ QAOA (Quantum Approximate Optimization Algorithm) to implement the proposed methods on the QGM processor.}
\end{figure*}

In general, existing quantum computers can be categorized into the universal quantum gate (QGM-Quantum Gate Machine) and quantum annealer (QAM-Quantum Annealing Machine). Regardless some issues related to noise and other non-ideal conditions, both of these types of quantum processors have been built and are accessible by public users through the Internet. The implementation scheme of the proposed methods for both of these kinds of quantum computers are illustrated in Fig.\ref{FIG_Method}. The \emph{direct method}, which work for QAM and has been described in our previous paper \cite{suksmono_minato_2019}, will be used as a reference. Three main proposed quantum computing methods are derived from non-quantum computing/classical H-matrix construction methods, which we will referred to as the Williamson, Baumert-Hall, and Turyn methods. For each of the method, we will described how to formulate its corresponding quantum Hamiltonian to be implemented on the quantum computers. 

The QAM processor, such as the D-Wave, only accepts problems in the form of a 2-body Hamiltonian, which generally can be expressed by
\begin{equation}
  \hat{H}_{pot} \left( \hat{\sigma} \right) \equiv -\sum_{i\neq j} J_{ij} \hat{\sigma}_i^z \hat{\sigma}_j^z -\sum_i h_i \hat{\sigma}_i^z
  \label{EQ_Hpot}
\end{equation}
which is a Hamiltonian of an Ising system, where $J_{ij}$ is a coupling constant or interaction strength between a spin at site $i$ with a spin at site $j$, $h_j$ is magnetic strength at site $j$, and $\{\hat{\sigma}_i^{\alpha}\}$ are Pauli's matrices of directions $\alpha=\{x,y,z\}$ at site-$i$. The processor performs quantum annealing by introducing a transverse field given by 
\begin{equation}
  \hat{H}_{kin}\left( \hat{\sigma} \right)\equiv -\Gamma \sum_i \hat{\sigma}_i^x    
  \label{EQ_Hkin}
\end{equation}
which is evolved over time according to the following equation
\begin{equation}
    \hat{H}_{QA}\left( \hat{\sigma}, t \right)=\left(1-\frac{t}{\tau} \right) \hat{H}_{kin} \left( \hat{\sigma} \right) + \frac{t}{\tau}\hat{H}_{pot}\left( \hat{\sigma} \right) 
  \label{EQ_HQA}
\end{equation}
where $t \in [0,\tau]$ denotes time \cite{kadowaki_nishimori_1988, boixo_2014}. The problem to solve should be encoded in $\hat{H}_{pot}$, which is represented by the Ising's coefficient $J_{ij}$ and $h_i$ for each of the problem. Some optimization problems have been solved by the quantum annealing methods; among others are: hand written digit recognition \cite{benedetti_2017}, computational biology \cite{li_2018}, and hydrologic inverse analysis \cite{omalley_2018}. 

On a QAM, the formulation of the H-SEARCH is started by calculation of its energy function $E(s)$ as a function of binary variables $s \in \{-1,+1\}$. For conciseness, we will represent the value of $s$ by its signs $\{-, +\}$. In general, $E(s)$ might contain high order $k$-body interaction terms so that we will denote it by $E_k(s)$, whereas the Ising model allows only up to 2-body terms in $E_2(s)$. To obtain the 2-body expression, and eventually a 2-body quantum Hamiltonian $\hat{H}_2\left(\hat{\sigma}\right)$, a sequence of transforms given by the following \emph{construction diagram} should be conducted \cite{suksmono_minato_2019},
\begin{equation}
  E_k(s) \rightarrow E_k(q) \rightarrow E_2(q) \rightarrow E_2(s)\rightarrow \hat{H}_2\left(\hat{\sigma}\right)
  \label{transdomain_diagram_MAIN}
\end{equation}
%
where $q\in \{0,1\}$ is a Boolean variable.

In the previous paper \cite{suksmono_minato_2019}, implementation of an $M$-order H-matrix on a QAM needs $M^2$ number of logical (binary) variables and additional $M^2\times (M-1)/2$ ancillary variables (\emph{ancillas}) so that the overall complexity is $O(M^3)$. In this paper, by adopting classical H-matrix construction methods, we can reduce the required number variables significantly into $O(M^2)$ which enables the search of higher order H-matrices than before. In the followings, we will address three quantum H-SEARCH methods, which are derived from the classical methods of Williamson, Baumert-Hall, and Turyn. In each of these method, we derive their corresponding Hamiltonians based on some criteria that are specifics for each of the cases. Low order cases can be calculated by hand, while higher order ones should symbolically be calculated by a computer due to the large number of  terms and variables. The complete lists of and expressions of the Hamiltonians will be provided in the \emph{Supporting Materials} section.  The proposed methods are illustrated in Fig.(\ref{FIG_Method}).

In the QGM quantum computing, we can employ QAOA (Quantum Approximate Optimization Algorithm), which is well-suited for solving an optimization problem on NISQ (Noisy Intermediate-Scale Quantum) processors. In principle, the general $k$-body Hamiltonian can directly be implemented on a QGM. Therefore, the required number of physical qubits will be about the same as the number of logical qubits. However, since the implementation needs direct connection to the actual machine, which is not available for us at this time, we will not address it in the current paper.

\section{Results}

\subsection{Williamson Based Quantum Computing Method}

The Sylvester construction method builds a larger H-matrix $H_{2^n}$ from smaller ones $H_{2^{n-1}}$ by iteratively applying the following tensor product, 
\begin{equation}
    H_{2^n}=H_2 \otimes H_{2^{n-1}} = 
    \begin{pmatrix}
      H_{2^{n-1}} & H_{2^{n-1}} \\
      H_{2^{n-1}} & -H_{2^{n-1}}
    \end{pmatrix}
    \nonumber
\end{equation}
where $H_2=\begin{pmatrix} + & + \\ + & - \end{pmatrix}$, i.e., it is a kind of plugging-in smaller H-matrices into a particular structure to obtain a larger H-matrix. Similarly, the Williamson method also builds a higher-order matrix from smaller ones, except that the smaller matrices are not necessarily an orthogonal one. In general, we can express the Williamson type H-matrices $W$ by \cite{williamson_1944, hedayat_1978, horadam_2007} 
\begin{equation}
    \label{EQ_MAIN_W}
    W=
    \begin{pmatrix}
      A  &  B &  C &  D\\
      -B &  A & -D &  C\\
      -C &  D &  A & -B\\
      -D & -C &  B & A
    \end{pmatrix}
\end{equation}
where $A,B,C, D$ are block matrices, whose any pair of them are commutative, i.e.,  $[A,B]=[A,C]=[A,D]=[B,C]=[B,D]=[C,D]=0$, with $[A,B]=A^TB-B^TA, \cdots,$ etc expressed the commutativity of a pair of matrices $A,B, \cdots$ etc. The orthogonality property of $W$ needs the following requirement to be satisfied, 
\begin{equation}
    \label{EQ_williamson}
    A^TA + B^TB + C^TC + D^TD = 4k I_k
\end{equation}
where $I_k$ is a $k\times k$ identity matrix. We will use the properties of the Williamson matrix; especially the one given by Eq.(\ref{EQ_williamson}), to formulate the Hamiltonian of Williamson-based quantum computing method. To further reduce the number of variables, we choose $A,B,C,D$ sub-matrices which are symmetric and circular.

For an illustration, consider $k=3$ which yields a $4k=12$-order H-matrix. The matrices can be expressed in terms of binary variables $s_i \in \{-1,+1\}$ by
\begin{equation}
    \label{EQ_MAIN_ABCD}
    A=
    \begin{pmatrix}
      s_0  &  s_1 &  s_1\\
      s_1  &  s_0 &  s_1\\
      s_1  &  s_1 &  s_0
    \end{pmatrix}
    ,
    B=
    \begin{pmatrix}
      s_2  &  s_3 &  s_3\\
      s_3  &  s_2 &  s_3\\
      s_3  &  s_3 &  s_2
    \end{pmatrix} ,
%
    C=
    \begin{pmatrix}
      s_4  &  s_5 &  s_5\\
      s_5  &  s_4 &  s_5\\
      s_5  &  s_5 &  s_4
    \end{pmatrix}
    ,
    D=
    \begin{pmatrix}
      s_6  &  s_7 &  s_7\\
      s_7  &  s_6 &  s_7\\
      s_7  &  s_7 &  s_6
    \end{pmatrix}
\end{equation}
Then, the requirement for Williamson matrix given by Eq.(\ref{EQ_williamson}) for $k=3$ becomes
\begin{equation}
    \label{EQ_MAIN_ABCD_SUM}
    A^TA + B^TB + C^TC + D^TD=
    \begin{pmatrix}
      12 &  v  &  v \\
       v &  12 &  v \\
       v &  v &  12
    \end{pmatrix}
    =12I_3
\end{equation}
where $ v=4+2\left(s_0s_1+s_2s_3+s_4s_5+s_6s_7\right)$. Suppose that $V \equiv [v_{i,j}]=A^TA+B^TB+C^TC+D^TD$. Naturally, we can define an $s$-dependent $k$-body energy function by
\begin{equation}
    \label{EQ_MAIN_ES_ABCD}
    E_k(s)=\sum_{i=0}^2 \sum_{i=0}^2 \left( v_{i,j}(s)-12\delta_{i,j}\right)^2
\end{equation}
where $\delta_{i,j}$ is a Kronecker delta function. The orthogonality requirement of $W$ will be satisfied when $E_k(s)=0$, which is the lowest value of the energy function in Eq.(\ref{EQ_MAIN_ES_ABCD}). For the $k=3$ case, the energy function  $E_k(s)$ can be expanded into
\begin{equation}
    \label{EQ_MAIN_Es01}
     E_k(s)=6\left( 4+2(s_0s_1+s_2s_3+s_4s_5+s_6s_7) \right)^2
\end{equation}

For implementing an energy function to a QAM processor; such as the D-Wave, the $k$-body energy function $E_k(s)$ should be transformed into a 2-body energy function $E_2(s)$ using the steps given by the construction diagram in Eq.(\ref{transdomain_diagram_MAIN}). In the process, we should choose a constant $\delta$ to translate the $k$-body into 2-body function, that should be larger than the maximum value $E_{max}$ of the energy function \cite{perdomo2008}. By taking $E_{max}=26,976$, which is the maximum value of $E_k(s)$ by assuming all of $s_i=+1$, then setting $\delta=2E_{max}$, we obtain the following result
\begin{equation}
  \begin{split}
     E_2(s)=13,728s_0 + 13,728s_1 + \cdots + 13,488s_0s_1 + \cdots+ 192s_{10}s_{11} + 162,720
  \end{split}
\end{equation}
This 2-body energy function gives the potential Hamiltonian $\hat{H}_{pot}(\hat{\sigma}) \equiv \hat{H}_2(\hat{\sigma}^z)$ as follows,
\begin{equation}
  \label{EQ_H2s_Williamson_3}
    \begin{split}
      \hat{H}_2(\hat{\sigma}^z)= 13,728\hat{\sigma}_0^z + 13,728\hat{\sigma}_1^z+ \cdots +13,488\hat{\sigma}_0^z\hat{\sigma}_1^z + \cdots +192\hat{\sigma}_{10}^z\hat{\sigma}_{11}^z  + 162,720
    \end{split}
\end{equation}
which can be encoded into a quantum annealing processor.

\begin{figure}
    \centering
	\includegraphics[width=0.45\columnwidth]{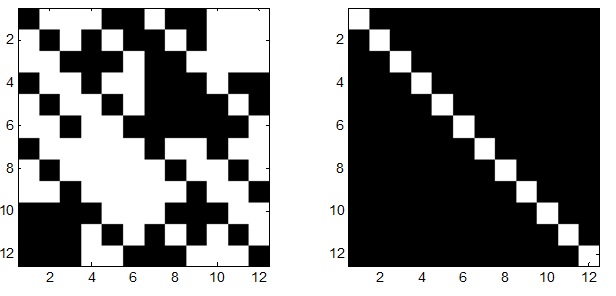}
	\includegraphics[width=0.45\columnwidth]{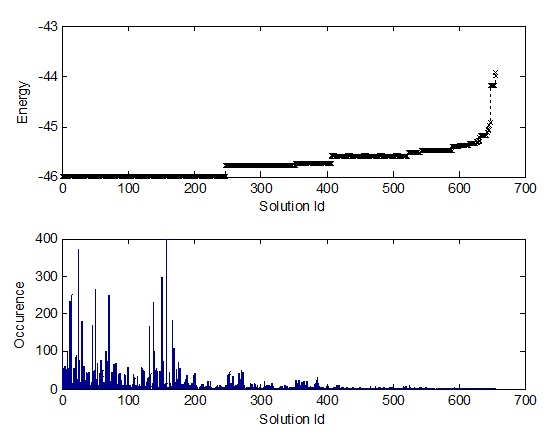}\\
    (a) \hspace{7cm} (b)\\
	\protect\caption{\label{FIG_12_Williamson} A Hamadard matrix of order 12 which is discovered by the Williamson's based quantum computing method. In (a), elements of H-matrix are represented by white color, which are $+1$, and black correspond to $-1$. In the indicator matrix, white corresponds to $M$, which is  $12$ in this case, and black to $0$. Statistics of the results are displayed in (b): energy distribution in top and occurrence of the solution at the bottom curve. The lowest energy solution at $E=-45.988$ corresponds to the $12$ order H-matrix. We see from the histogram at the bottom part of the figure that most of the solutions are concentrated around this energy value. 
	}
\end{figure}

In the experiment, we extract the Ising coefficients $\{J_{ij}, h_i\}$ then submit them to the D-Wave. We observe that the magnitude of the coefficients of the Hamiltonian are quite large, but they will be normalized by the system when they are entered into the D-Wave system. Additionally, the constant term, such as $162,720$ in $\hat{H}_2(\hat{\sigma}^z)$ of Eq.(\ref{EQ_H2s_Williamson_3}), will also be removed. Consequently, instead of zero, the minimum of the energy will be a negative value. We have set the number of to $10,000$ and obtain some solutions at minimum energy values. For $k=3$, which corresponds to H-matrix of order $12$, the required number of logical qubits was $8$ which translates into $36$ physical qubits, we have obtained the minimum energy at $-45.988$. The experimental results is displayed in Fig.(\ref{FIG_12_Williamson}), where (a) shows the found H-matrix $H$ and its indicator matrix $D\equiv H^TH$ and (b) is the energy distribution of all solutions.

Higher orders matrices, up to order $36$ that needs $49$ qubits to implement, has also been found successfully using the D-Wave. All of the complete expression of the Hamiltonians and the found matrices are listed in the \emph{Supplementary Information} section.



\subsection{Baumert-Hall Based Quantum Computing Method}
In principle, the Baumert-Hall quantum computing method works in a similar manner as the Williamson's by first finding the $A,B,C,D$ block matrices, except that the construction of the H-matrix is given by the following $12\times 12$ structure of block matrix \cite{baumert_hall_1965, hedayat_1978}:
%
\begin{equation}
    \label{EQ_MAIN_BH}
    H=
     \left( \begin{array}{rrrr rrrr rrrr}
      A &  A &  A &  B & -B &  C & -C & -D &  B &  C & -D & -D\\
      A & -A &  B & -A & -B & -D &  D & -C & -B & -D & -C & -C\\
      A & -B & -A &  A & -D &  D & -B &  B & -C & -D &  C & -C\\
      B &  A & -A & -A &  D &  D &  D &  C &  C & -B & -B & -C\\
      B & -D &  D &  D &  A &  A &  A &  C & -C &  B & -C &  B\\
      B &  C & -D &  D &  A & -A &  C & -A & -D &  C &  B & -B\\
      D & -C &  B & -B &  A & -C & -A &  A &  B &  C &  D & -D\\
     -C & -D & -C & -D &  C &  A & -A & -A & -D &  B & -B & -B\\
      D & -C & -B & -B & -B &  C &  C & -D &  A &  A &  A &  D\\
     -D & -B &  C &  C &  C &  B &  B & -D &  A & -A &  D & -A\\
      C & -B & -C &  C &  D & -B & -D & -B &  A & -D & -A &  A\\
     -C & -D & -D &  C & -C & -B &  B &  B &  D &  A & -A & -A
     \end{array}\right)
\end{equation}
%
Considering the usage efficiency of the variables, $A, B, C, D$ are also chosen to be symmetric circulant block matrices identical to the Williamsons's method described in the previous section. For a $k\times k$ size of the block matrices, Eq.(\ref{EQ_MAIN_BH}) yields a $12k \times 12k$ dimension of the H-matrix. The formulation of the energy function also follows the Williamsons method described previously.

\begin{figure}
    \centering
	\includegraphics[width=0.50\columnwidth]{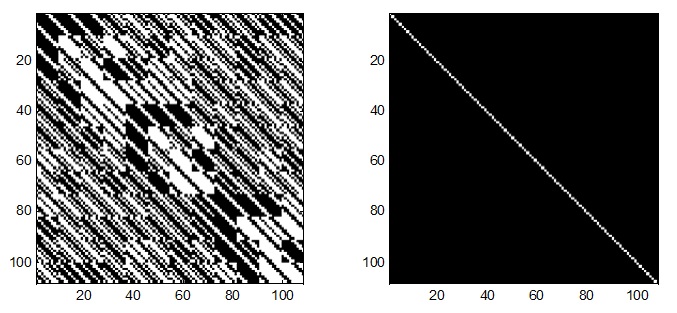}
	\includegraphics[width=0.45\columnwidth]{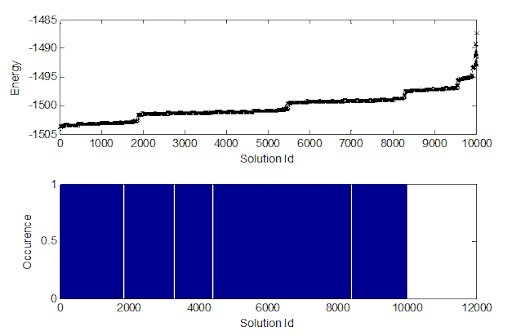}\\
    (a) \hspace{7cm} (b)\\
	\protect\caption{\label{FIG_H132_Baumert_Hall} Hamadard matrix of order $108$ obtained by the Baumert-Hall's based quantum computing method. Part (a) shows the H-matrix $H$ and its corresponding indicator matrix $D=H^TH$, black-white pixels corresponds to elements with $-1$ and $+1$, respectively for the H-matrix, and $108$ and $0$ values for the indicator matrix . The curve and histogram in (b)shows that the solution is evenly distributed across the energy. The minimum energy corresponding to the solution is achieved at $E= -1,499.9$.
	}
\end{figure}

Experiments on finding Baumert-Hall matrices using D-Wave quantum processor indicates that the capability of the method is limited by the available number of qubits and the capability of the embedding tool \cite{dwave_matlab_2018}. We have successfully find the Hadamard matrix up to order $108$ using this method. For the $108$-order case, initial energy function $E_k(s)$ to find this matrix is given by the following
\begin{equation}
  \label{EQ_Eks_williamson_108}
  \begin{split}
    E_k(s) = 432s_0s_2 + \cdots+ 720s_{18}s_{19} +\cdots + 432s_{16}s_{17}s_{18}s_{19} + 5,760
  \end{split}
\end{equation}
whose corresponding $k$-body Hamiltonian is given by
\begin{equation}
  \label{EQ_Hksigma_williamson_108}
  \begin{split}
   \hat{H}_k(\hat{\sigma}^z) = 432\hat{\sigma}^z_0\hat{\sigma}^z_2 +\cdots + 720\hat{\sigma}^z_{18}\hat{\sigma}^z_{19} + \cdots+ 432\hat{\sigma}^z_{16}\hat{\sigma}^z_{17}\hat{\sigma}^z_{18}\hat{\sigma}^z_{19} + 5,760
  \end{split}
\end{equation}
The 2-body Hamiltonian realized on the quantum annealing processor is given by,
\begin{equation}
  \label{EQ_H2sigma_williamson_108}
  \begin{split}
   \hat{H}_2(\hat{\sigma}^z) = 10,555,200\hat{\sigma}^z_{0} + \cdots+2,636,352\hat{\sigma}^z_{0}\hat{\sigma}^z_{1} +\cdots+ 1,728\hat{\sigma}^z_{54}\hat{\sigma}^z_{59} + 316,483,200
  \end{split}
\end{equation}
%
After extracting the Ising parameters and submitting to the D-Wave, we obtain the solutions containing correct values of $s_i$ for building the H-matrices. Figure \ref{FIG_H132_Baumert_Hall} shows a $108$ order H-matrix, which was found by the Baumert-Hall based method and its corresponding energy statistics as output of the quantum computer. All of the complete expression of the Hamiltonians and the found matrices by this method are listed in the \emph{Supplementary Information} section.

\subsection{The Turyn Based Quantum Computing Method}
In this method, first we find a set of 4-sequences $\{X,Y,Z,W\}$ that has particular properties, then use them to construct a H-matrix based on Goethals-Siedel method \cite{turyn_1974, kharaghani_rezaie_2004}. We translate the requirements into energy functions which then programmed into a quantum processor. In essence, the workflows of the Turyn based method are as follows
\begin{enumerate}
    \item Find an $(n, n, n, n-1)$ Turyn-Type (TT) sequence: TT-generating Algorithm $\rightarrow \{X,Y,Z,W\}$
    \item Construct base sequences: $\{X,Y,Z,W\}$ $\rightarrow$ $\{A,B,C,D\}$
    \item Construct T-sequences:$\{A,B,C,D\}$ $\rightarrow$ $\{ T_1,T_2,T_3,T_4\}$
    \item Construct seed sequences:$\{ T_1,T_2,T_3,T_4\}$ $\rightarrow$ $\{A_1,A_2,A_3,A_4\}$
    \item Construct block symmetric circular matrices: $\{A_1,A_2,A_3,A_4\}$ $\rightarrow$ $\{X_{A1},X_{A2},X_{A3},X_{A4}\}$
    \item Construct Hadamard matrix: $\{X_{A1},X_{A2},X_{A3},X_{A4}\}\rightarrow H$, which is given by
        \begin{equation}
            \label{EQ_MAIN_TGS}
            H=
            \begin{pmatrix}
               X_{A1}  &  X_{A2}R   &  X_{A3}R   &  X_{A4}R  \\
              -X_{A2}R &  X_{A1}    &  X_{A4}^TR & -X_{A3}^TR\\
              -X_{A3}R & -X_{A4}^TR &  X_{A1}    &  X_{A2}^TR\\
              -X_{A4}R &  X_{A3}^TR & -X_{A2}^TR &  X_{A1}
            \end{pmatrix}
        \end{equation}
    where $R$ is a back-diagonal identity matrix of size $k \times k$ as follows
        \begin{equation}
            \label{EQ_MAIN_R}
            R=
            \begin{pmatrix}
               0 & 0 &  \cdots   & 0 & 1  \\
               0 & 0 &  \cdots   & 1 & 0  \\
               \cdots & \cdots &\cdots & \cdots & \cdots \\ 
               0 & 1 &  \cdots   & 0 & 0 \\
               1 & 0 &  \cdots   & 0 & 0 \\
            \end{pmatrix}
        \end{equation}
\end{enumerate}
We derive the energy function from the requirement of a valid TT-sequences given by, 
\begin{equation}
  \label{EQ_TGS_MAIN_E0}
    N_X(r) +N_Y(r) +2N_Z(r) +2N_W(r)=0; r\geq 1
\end{equation}
where $N_X(r),N_Y(r),N_Z(r),N_W(r)$ are non-periodic auto-correlation functions of the sequences $\{X,Y,Z,W\}$ calculated at lag-$r$, respectively. The non-periodic auto-correlation function of a sequence $V=[v_0,v_1, \cdots, v_{n-1}]^T$ is given by,
\begin{equation}
    \label{EQ_MAIN_NAC}
    N_V(r)=\sum_{t=0}^{n-1-r} v_i v_{i+r}
\end{equation}
for $r = 0, 1,\cdots, n-1$ and $N_V(r)=0$ for $r\geq n$. Since the value given by the left-hand side of Eq.(\ref{EQ_TGS_MAIN_E0}) can be negative, whereas the annealing is performed to achieve a minimum value, we adopt a non-negative energy function which are sum of squared value of the auto-correlation function at each lag $r\geq 1$ as follows,
\begin{equation}
  \label{EQ_TGS_MAIN_E1}
    E\equiv \sum_{r\geq 1} \left(N_X(r)+N_Y(r)+2N_Z(r)+N_W(r)\right)^2
\end{equation}

To efficiently use available qubits in the quantum processor, it is important to reduce the number of variables encoded to the qubits as few as possible. We can achieve this by further employing the property of a TT-sequence. In this case, we can normalize the TT-sequence \cite{kharaghani_rezaie_2004} to obtain $X^T=(x_0, x_1, \cdots x_{n-1})$, $Y^T=(y_0, y_1, \cdots y_{n-1})^T$,  $Z^T=(z_0, z_1, \cdots z_{n-1})$, and $W^T=(w_0, w_1, \cdots w_{n-1})$, which have the following properties
\begin{itemize}
   \label{EQ_normalized_TT}
    \item $x_0=y_0=z_0=w_0=1$
    \item $x_{n-1}=y_{n-1}=-1, z_{n-1}=1$
    \item $x_1=x_{n-2}=1, y_1y_{n-2}=-1$
\end{itemize}

For clarity, in the followings we present an example of the Hamiltonian formulation for the lowest order of $k=4$ case. The first step as described previously is to find a $TT(4)$ -sequences $X,Y,Z,W$. By representing the elements of the sequences as binary (spin) variables $s_i\in \{-1,+1\}$, and applying the properties of a normalized sequence explained previously, a $TT(4)$ will be as follows,
\begin{equation}
\label{EQ_MAIN_XYZW}
  \begin{split}
   X=\left(1,1,1,-1\right)^T \\
   Y=\left(1,s_0,-s_0,-1\right)^T \\
   Z=\left(1,s_1,s_2,1\right)^T \\
   W=\left(1,s_3,s_4\right)^T
  \end{split}
\end{equation}
To determine the energy function, we have to calculate non-periodic auto-correlation functions $N_X,N_Y,N_Z,N_W$ given by Eq.(\ref{EQ_MAIN_NAC}). Since $s_i^2=1$, we get the following results after simplifications
\begin{equation}
\label{EQ_MAIN_R_XYZW}
  \begin{split}
   N_X=( 4, 1, 0, -1)^T \\
   N_Y= ( 4, 2s_0 - 1, -2s_0, -1)^T \\
   N_Z=\left( 4, s_1 + s_2 + s_1s_2, s_1 + s_2, 1\right)^T \\
   N_W=\left(3, s_3 + s_3s_4, s_4\right)^T
  \end{split}
\end{equation}
Therefore, the energy $E \equiv E_k(s)$ in Eq.(\ref{EQ_TGS_MAIN_E1}), whose terms may contain a product of $k$ variables, is now given by
\begin{equation}
  \label{EQ_Eks_MAIN}
  \begin{split}
   E_k(s) = 
   2s_1 + 2s_2 + 2s_4 + 4s_0s_3 + 4s_1s_2 - 4s_0s_4 + 2s_1s_3 + 2s_1s_4 + 2s_2s_3 + 2s_2s_4 \\
   + 4s_0s_1s_2+ 2s_1s_2s_3 + 4s_0s_3s_4 + 2s_1s_3s_4 + 2s_2s_3s_4 + 2s_1s_2s_3s_4 + 242 \\
  \end{split}
\end{equation}
%
\begin{figure}
    \centering
	\includegraphics[width=0.50\columnwidth]{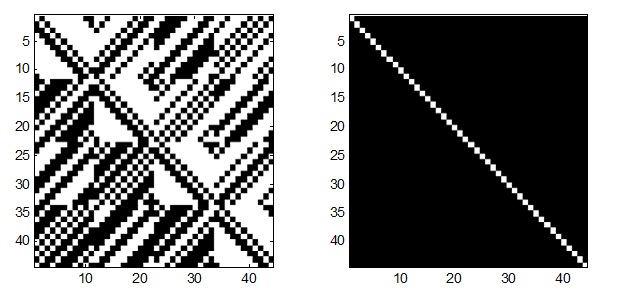}
	\includegraphics[width=0.45\columnwidth]{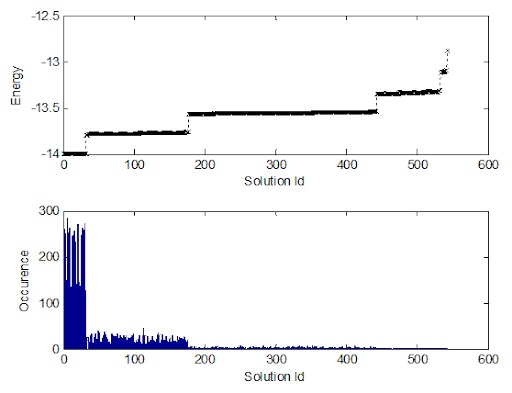}\\
    (a) \hspace{7cm} (b)\\
	\protect\caption{\label{FIG_H44_Turyn} Hamadard matrix of order $44$ obtained by the Turyn based quantum computing method. Part (a) shows the matrix and its corresponding orthogonality indicator, (b) display the statistics of the solutions obtained by D-Wave. Solutions are more concentrated in the low energy regime, where the matrix is found at $E= -14.9730$.
	}
\end{figure}

In the following steps, as described by the construction diagram in Eq.(\ref{transdomain_diagram_MAIN}), the energy function should be transformed into a 2-body interacting Ising Hamiltonian. Therefore, we have to change the $s$-dependent energy function into $q$-dependent $E_k(q)$. After simplification, this transform yields the following form,
\begin{equation}
  \label{EQ_MAIN_Ekq}
  \begin{split}
   E_k(q)=
    - 16q_0 - 40q_1 - 40q_2 - 40q_3 - 24q_4  + 16q_0q_1 + 16q_0q_2 + 32q_0q_3 + 48q_1q_2 \\
    + 32q_1q_3 + 24q_1q_4 + 32q_2q_3 + 24q_2q_4 + 40q_3q_4- 32q_0q_1q_2 - 32q_1q_2q_3 \\
    - 32q_0q_3q_4 - 16q_1q_2q_4 - 32q_1q_3q_4 - 32q_2q_3q_4 + 32q_1q_2q_3q_4 + 276
  \end{split}
\end{equation}
The conversion into 2-body energy function requires a constant $\delta_{ij}$ set to be larger than the maximum value of the energy function $E_{max}(k)$. Assuming it is at least an absolute sum of the  $E_k(q)$ coefficients as before, we have $E_{max} = 908$. By taking twice of this maximum value, we obtain $\delta_{ij}=1,816$, which transforms Eq.(\ref{EQ_MAIN_Ekq}) into
\begin{equation}
  \label{EQ_MAIN_E2q}
  \begin{split}
  E_2(q)=
  -16q_0 - 40q_1 - 40q_2 - 40q_3 - 24q_4 + 5,464q_5+ 5,480q_6 + 5,496q_7 + 5,480q_8 \\
  +5,480q_9 + 5,488q_{10} + 1,816q_0q_1 + 16q_0q_2 + 1,816q_0q_3 + 1,816q_1q_2+ 1,816q_1q_3\\
  -3,632q_0q_5 + 24q_1q_4 + 1,816q_2q_3- 3,632q_0q_6 - 3,632q_1q_5 + 24q_2q_4 - 32q_2q_5 \\
  + 1,816q_3q_4 - 3,632q_1q_7 - 3,632q_1q_8 - 3,632q_2q_7- 3,632q_3q_6 - 32q_3q_7-32q_4q_6 \\
  - 3,632q_2q_9- 3,632q_3q_8 - 16q_4q_7 - 3632q_3q_9 - 32q_4q_8 - 3,632q_3q_{10} - 32q_4q_9 \\
  - 3,632q_4q_{10} + 32q_7q_{10} + 276
  \end{split}
\end{equation}
Transforming back Eq.(\ref{EQ_MAIN_E2q}) to the $s$-domain yields the following expression,
\begin{equation}
  \label{EQ_E2s}
  \begin{split}
  E_2(s) =
  912s_{0} + 1376s_{1} + 926s_{2} + 1844s_{3} + 482s_{4} - 908s_{5} - 916s_{6} - 928s_{7}- 916s_{8} \\
  - 916s_{9} - 936s_{10} + 454s_{0}s_{1} + 4s_{0}s_{2} + 454s_{0}s_{3}+ 454s_{1}s_{2}+ 454s_{1}s_{3} - 908s_{0}s_{5} \\
  + 6s_{1}s_{4} + 454s_{2}s_{3} - 908s_{0}s_{6} - 908s_{1}s_{5} + 6s_{2}s_{4} - 8s_{2}s_{5} + 454s_{3}s_{4} - 908s_{1}s_{7} \\
  - 908s_{1}s_{8} - 908s_{2}s_{7}- 908s_{3}s_{6} - 8s_{3}s_{7} - 8s_{4}s_{6} - 908s_{2}s_{9} - 908s_{3}s_{8}  - 4s_{4}s_{7} \\
  - 908s_{3}s_{9} - 8s_{4}s_{8} - 908s_{3}s_{10} - 8s_{4}s_{9} - 908s_{4}s_{10} + 8s_{7}s_{10} + 8,448
  \end{split}
\end{equation}
which corresponds to the following 2-body Hamiltonian,
\begin{equation}
 \label{EQ_H2sigma}
 \begin{split}
   \hat{H}_2(\hat{\sigma}^z) =
    912\hat{\sigma}^z_{0} + 1376\hat{\sigma}^z_{1} + 926\hat{\sigma}^z_{2} + 1844\hat{\sigma}^z_{3} + 482\hat{\sigma}^z_{4} - 908\hat{\sigma}^z_{5}- 916\hat{\sigma}^z_{6} - 928\hat{\sigma}^z_{7} \\
    -916\hat{\sigma}^z_{8} - 916\hat{\sigma}^z_{9} - 936\hat{\sigma}^z_{10} + 454\hat{\sigma}^z_{0}\hat{\sigma}^z_{1} + 4\hat{\sigma}^z_{0}\hat{\sigma}^z_{2} + 454\hat{\sigma}^z_{0}\hat{\sigma}^z_{3} + 454\hat{\sigma}^z_{1}\hat{\sigma}^z_{2} + 454\hat{\sigma}^z_{1}\hat{\sigma}^z_{3} \\
    - 908\hat{\sigma}^z_{0}\hat{\sigma}^z_{5} + 6\hat{\sigma}^z_{1}\hat{\sigma}^z_{4}
    + 454\hat{\sigma}^z_{2}\hat{\sigma}^z_{3} - 908\hat{\sigma}^z_{0}\hat{\sigma}^z_{6} - 908\hat{\sigma}^z_{1}\hat{\sigma}^z_{5} + 6\hat{\sigma}^z_{2}\hat{\sigma}^z_{4} - 8\hat{\sigma}^z_{2}\hat{\sigma}^z_{5} + 454\hat{\sigma}^z_{3}\hat{\sigma}^z_{4}\\
    - 908\hat{\sigma}^z_{1}\hat{\sigma}^z_{7} - 908\hat{\sigma}^z_{1}\hat{\sigma}^z_{8} - 908\hat{\sigma}^z_{2}\hat{\sigma}^z_{7} - 908\hat{\sigma}^z_{3}\hat{\sigma}^z_{6} - 8\hat{\sigma}^z_{3}\hat{\sigma}^z_{7} - 8\hat{\sigma}^z_{4}\hat{\sigma}^z_{6} - 908\hat{\sigma}^z_{2}\hat{\sigma}^z_{9} - 908\hat{\sigma}^z_{3}\hat{\sigma}^z_{8} \\
    - 4\hat{\sigma}^z_{4}\hat{\sigma}^z_{7}- 908\hat{\sigma}^z_{3}\hat{\sigma}^z_{9}- 8\hat{\sigma}^z_{4}\hat{\sigma}^z_{8} - 908\hat{\sigma}^z_{3}\hat{\sigma}^z_{10} - 8\hat{\sigma}^z_{4}\hat{\sigma}^z_{9} - 908\hat{\sigma}^z_{4}\hat{\sigma}^z_{10} + 8\hat{\sigma}^z_{7}\hat{\sigma}^z_{10} + 8,448
  \end{split}
\end{equation}

In the experiment, we extract Ising coefficients from the 2-body Hamiltonian and submit them to the D-Wave system. We have successfully found the lowest order H-matrix by the Turyn-based method shown in Fig.(\ref{FIG_H44_Turyn}). Using this method, we successfully found an H-matrix of order $68$. Complete expression of the Hamiltonians and the found matrices are listed in the \emph{Supplementary Information} section.



\subsection{Balancing the Quantum and Classical Resources: Extension of The Turyn Based Quantum Computing Method}

Finding high order H-matrix through the Turyn's method can be achieved by checking all possible binary vector that satisfy the TT-sequences $\{X,Y,Z,W\}$ requirements. Exhaustive enumeration of all $(n, n, n, n-1)$ TT-sequence needs $2^{4n-1}$ steps, which is an exponentially increasing task. For finding even higher order H-matrices, we can explore the properties of the TT-sequence to reduce the number of binary sequence to enumerate \cite{kharaghani_rezaie_2004, best_2012}. In this method, instead of finding all $X,Y,Z,W$ at the same time, it will be more computationally realistic to start with filling some part of them, then subsequently imposing conditions and properties of the TT-sequence to limit the number of the sequences to check. Partially filled sequences $\{X^*, Y^*, Z^*, W^*\}$ with $m$-elements on the left part and another $m$-elements on the right one, are given as follows
\begin{equation}
    \label{EQ_XYZW_star}
     \begin{split}
        X^*=(x_0, x_1, ..., x_{m-1},*,*,\cdots,*,*, x_{n-m},\cdots,x_{n-1})^T\\
        Y^*=(y_0, y_1, ..., y_{m-1},*,*,\cdots,*,*, y_{n-m},\cdots, y_{n-1})^T\\
        Z^*=(z_0, z_1, ..., z_{m-1},*,*,\cdots,*,*, z_{n-m},\cdots, z_{n-1})^T\\
        W^*=(w_0, w_1, ..., w_{m-1},*,*,\cdots,*,*, w_{n-m},\cdots, w_{n-2})^T
     \end{split}
\end{equation}
The requirement of non-periodic auto-correlation sum for these sequences is now become
\begin{equation}
    \label{EQ_XYZW_star_autocorr}
    N_{X^*}(r) + N_{Y^*}(r) + 2N_{Z^*}(r) + 2N_{W^*}(r)=0; \  r\geq (n-m)
\end{equation}
We will refer all $\{X^*, Y^*, Z^*, W^*\}$ sequences satisfying condition given by Eq.(\ref{EQ_XYZW_star_autocorr}) as \emph{solution prototypes}.

Although increasing $m$ in Eq.(\ref{EQ_XYZW_star}) will reduce the number of sequence to check in the following steps, it also increases the number of the solution prototypes itself. There are about 2 millions prototypes for $2m=12$, which will increase into about $23$ millions for $2m=14$ \cite{kharaghani_rezaie_2004, best_2012, london_2013}. It has been reported that a few TT-sequence of up to $40$ can be found using classical computers, whereas higher order ones need more powerful computers which is impossible to be implemented at the moment. This is one of the main reasons that  H-matrix of order $668$ has not been found nor declared non-exists yet, assuming that such a matrix can be constructed by the Turyn's method.

On the other hand, we can use the solution prototypes to reduce the number of required qubits when a quantum computer is involved in the searching process. For clarity, in the followings we illustrate this method by a simple case which is implementable on a current quantum processor. We will consider a $(4,4,4,3)$ solution prototype to find a $(8,8,8,7)$ TT-sequence by using quantum computing; therefore, it is a kind of finding higher order sequence by extending the lower one. The extended TT-sequences can be expressed by 
\begin{equation}
  \begin{split}
    X=(x_0, x_1, s_0, s_1,s_2, s_3, x_2,x_3)^T \\
    Y=(y_0, y_1, s_4, s_5,s_6, s_7, y_2,y_3)^T \\
    Z=(z_0, z_1, s_8, s_9,s_{10}, s_{11}, z_2,z_3)^T \\
    W=(w_0, w_1, s_{12}, s_{13},s_{14}, s_{15}, w_2)^T \\
  \end{split}
\end{equation}
with known $x_0,\cdots,x_3, y_0,\cdots,y_3,\cdots,\cdots,w_0, w_2$ and unknown $s_0,s_1,\cdots,s_{15}$.

To find the unknown values represented by $s_i$, we calculate the energy of the Turyn's based method as before. Among all possible $\{X^*,Y^*,Z^*,W^*\}$ prototypes and the replacement of the unknowns with binary variables, we choose the following solution prototype as an example
\begin{equation}
 \begin{split}
  X = (1, 1, *, *, * , *, 1, -1)^T  \rightarrow (1,  1, s_0, s_1, s_2, s_3, 1, -1)^T\\
  Y = (1, -1, *, *, * , *, 1, -1)^T \rightarrow (1, -1, s_4, s_5, s_6, s_7, 1, -1 )^T \\
  Z = (1, -1,*, *, * , *,-1, 1)^T  \rightarrow (1, -1, s_8, s_9, s_{10}, s_{11}, -1, 1)^T\\
  W = (1, -1,*, *, * , *, 1)^T \rightarrow     (1, -1, s_{12}, s_{13}, s_{14}, s_{15}, 1)^T\\
 \end{split}
\end{equation}
Note that in the real case, we have to check all of the solution prototypes.

Symbolic computing of the energy functions yields the following $E_k(s)$, 
\begin{equation}
  \label{EQ_Eks_xturyn}
  \begin{split}
    E_k(s)= - 14s_0 + \cdots + 4s_0s_1s_2 + \cdots + 16s_{12}s_{13}s_{14}s_{15} + 264
  \end{split}
\end{equation}
Then, by using $\delta=65,936$, we obtained a 2-body energy function in $s$-domain as follows,  
\begin{equation}
  \begin{split}
    E_2(s)= 197,860s_0 + .. + 16,484s_0s_1+ ...+64s_{102}s_{107} + 4,551,232
  \end{split}
\end{equation}
and the Ising Hamiltonian $\hat{H}_2(\hat{\sigma}^z)$ is given by,
\begin{equation}
  \begin{split}
    \hat{H}_2(\hat{\sigma}^z)=197,860\hat{\sigma}^z_0 + .. + 16,484\hat{\sigma}^z_0\hat{\sigma}^z_1+ ...+64\hat{\sigma}^z_{102}\hat{\sigma}^z_{107} + 4,551,232
  \end{split}
\end{equation}
The detail expressions of all of them are listed in the \emph{Supplementary Information} section.

\begin{figure}[t]
    \centering
	\includegraphics[width=0.50\columnwidth]{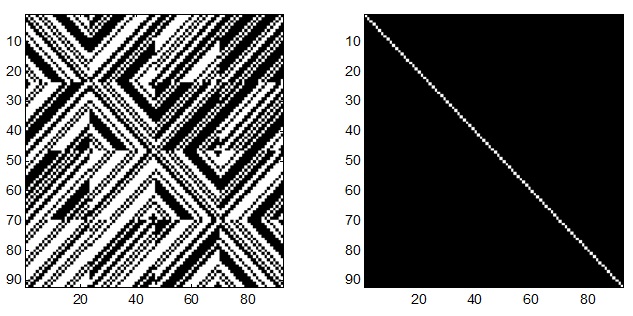}
	\includegraphics[width=0.45\columnwidth]{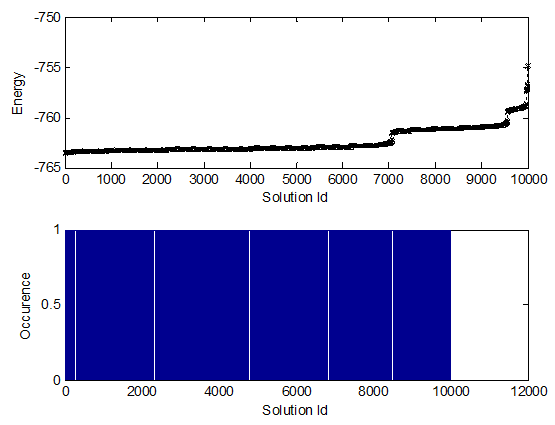}\\
    (a) \hspace{7cm} (b)\\
	\protect\caption{\label{FIG_H92_std} Hadamard matrix of order $92$ obtained by the extended Turyn based quantum computing method. Part (a) shows the H-matrix and its indicator, part (b) shows statistics of the quantum computer outputs.
	}
\end{figure}
%
After extracting Ising coefficients $\{h_i, J_ij\}$ from the Hamiltonian, we submitted them to the D-Wave. We need $108$ physical qubits to implement the problem, but embedding into the Chimera graph with the D-Wave provided embedding tools indicates that more qubits are required, which in this case is $860$. After quantum annealing, we get among others, the following solution

\begin{equation}
    \begin{split}
      X = (1, 1,-1, 1,-1, 1, 1,-1)^T \\
      Y = (1,-1, 1, 1, 1, 1, 1,-1^T \\
      Z = (1,-1,-1, 1, 1, 1,-1,1^T \\
      W = (1,-1,-1,-1,-1,-1,1)^T
    \end{split}
\end{equation}
Among the $10,000$ obtained results, we identified two correct solutions. One of the solution that has been constructed to a H-matrix, its corresponding indicator matrix, and solution statistics are displayed in Fig.(\ref{FIG_H92_std}). It is worth to note that the $92$ order H-matrix was a subject of interest when three NASA-JPL researchers tried to find, and eventually discover it, using computer search performed in a state-of-the art mainframe computer at that time \cite{baumert_golomb_hall_1962}.

\section{Discussions}

Difficulties in finding a H-matrix by classical computing methods, due to the exponential grows of the complexity, can be overcome by quantum-computing based search, such as by the \emph{direct method} \cite{suksmono_minato_2019} that represents each elements of the matrix directly into binary variables, which is then translated into qubits. However, the availability of quantum computing resource limits the implementation to only finding low order H-matrices. We have shown in this paper that classical construction/searching methods can be adopted to efficiently use the available resource to solve larger problems, i.e., finding higher order H-matrices.


\renewcommand{\arraystretch}{1.1}
\begin{table*}[t]
\centering 
\begin{tabular}{|c|c|c|c|c|c|c|c|c|}
	\hline
	\multirow{2}{*}{No}  & 
	\multirow{2}{*}{$k$} & 
	\multicolumn{2}{c|}{Order} & 
	\multicolumn{3}{c|}{Number of Qubits} & 
	\multirow{2}{*}{\shortstack{E/P\\Ratio}} & 
	\multirow{2}{*}{\shortstack{No. of Correct\\in 10,000}}\\
	\cline{3-7}
	 & & Williamson & Baumert-Hall & Logical & Physical & Embedded & &\\
	\hline
	\hline
	 1 &  3 & 12 &  36 &  8 &  36 &    51 & 1.4 & 322 \\
     2 &  5 & 20 &  60 & 12 &  78 &   207 & 2.7 & 493 \\
     3 &  7 & 28 &  84 & 16 & 316 &   688 & 5.1 & 282 \\
     4 &  9 & 36 & 108 & 20 & 210 & 1,492 & 7.1 &  18 \\
     5 & 11 & 44 & 132 & 24 & 300 &    NA &  NA &  NA \\
	\hline
	\end{tabular}
    \caption{Resource required in Willamson and Baumert-Hall based quantum computing methods. To find an $M$ order H-matrix, the required logical qubits grows by $O(M)$ and the physical qubits by $O(M^2)$. Embedding the connections implied by the Hamiltonian on existing Chimera graph further increases the required qubits, which ultimately limit the capability of the method. Decreasing percentage of correct solutions, knowing that only $10,000$ in a single run is allowed, indicates that repeated experiments will be needed to find higher order matrices.}
    \label{TB_WBH}
\end{table*}

The data displayed in Table.\ref{TB_WBH} show required resource and results in the Williamson and Baumert-Hall based methods for each order of the H-matrix. Since both of them share the same $A,B,C,D$ block matrices, we put them side-by-side on the same table. We observe in the table that the number of required \emph{logical qubits} grows linearly by $O(M)$ with respect to the order of the searched matrix, whereas the number of \emph{physical qubits} grows  quadratically as $O(M^2)$, which is caused by the ancillary qubits required to translate $k$-body into $2$-body Hamiltonians. In the implementation, the physical qubits and their connections should be mapped to the topology of qubits's connections in the quantum annealing processor; which is a Chimera graph in the \emph{DW-2000Q}. We have used (default) embedding tool provided by D-Wave \cite{dwave_matlab_2018} and the number of embedding qubits displayed in the table are taken from the output of the software. This mapping process, which is also called \emph{minor embedding}, further increases the number of required qubits. In the following discussions, the number of required embedded qubits will be labeled as the \emph{embedding qubits}. 

\begin{table*}[b]
    \centering
    \begin{tabular}{|c|c|c|c|c|c|c|c|}
    \hline
	\multirow{2}{*}{No}  & 
	\multirow{2}{*}{$k$} & 
	\multirow{2}{*}{Order} & 
	\multicolumn{3}{c|}{Number of Qubits} & 
	\multirow{2}{*}{\shortstack{E/P\\Ratio}} & 
	\multirow{2}{*}{\shortstack{No. of Correct\\in 10,000}}\\
	\cline{4-6}
    & & & Logical & Physical & Embedding & &\\
	\hline
    \hline
    1 & 4 & 44 &  5 &  11 & 25  & 2.3  & 17 \\
    2 & 6 & 68 & 13 &  36 & 397 & 11.0 & 35 \\
    3 & 8 & 92 & 31 & 199 & NA  & NA   & NA \\
    \hline
    \end{tabular}
    \caption{Resource needed in Turyn-based quantum computing method. Although the number of required physical qubits also grows by $O(M^2)$, the jump among the order is high so that the next one after $92$ cannot successfully be embedded in \emph{DW-2000Q}. We also cannot conclude the success rate for given $10,000$ number of reads due to lack of data, although we may suspect that it will also decrease as in the Williamsons and Baumert-Hall adopted methods.}
    \label{TB_Turyn}
\end{table*}

The Williamson and Baumert-Hall adopted methods can be implemented to all of matrix order as long as the embedding process is successful, which is $36$ for the Williamson and 108 for the Baumert-Hall. We observe from the output of embedding tool that the highest order needs $1,492$ qubits to implement, which is more than $6$ times of the required physical qubits. Observing that the trends of the embedding-to-physical qubits ratio increases with the H-matrix order, by taking a moderate estimate of $7$ times, the $300$ physical qubits for the order of $132$ matrix (in the Baumert-Hall based method) requires $2,100$ qubits to be implemented; which is more than currently available qubits in the DW-2000Q. We also observe from the experiment results in the quantum computer outputs that the number of correct solutions among $10,000$ number of reads tends to decrease with the increasing order of the matrix; i.e., it is about $ 4\%$ at the beginning then decreased to about $0.2 \%$ at order $108$ for the Baumert-Hall. Since the number of read is limited to $10,000$, repeated experiments should be done to find higher order H-matrices. 

\begin{figure}
    \centering
	\includegraphics[width=0.90\columnwidth]{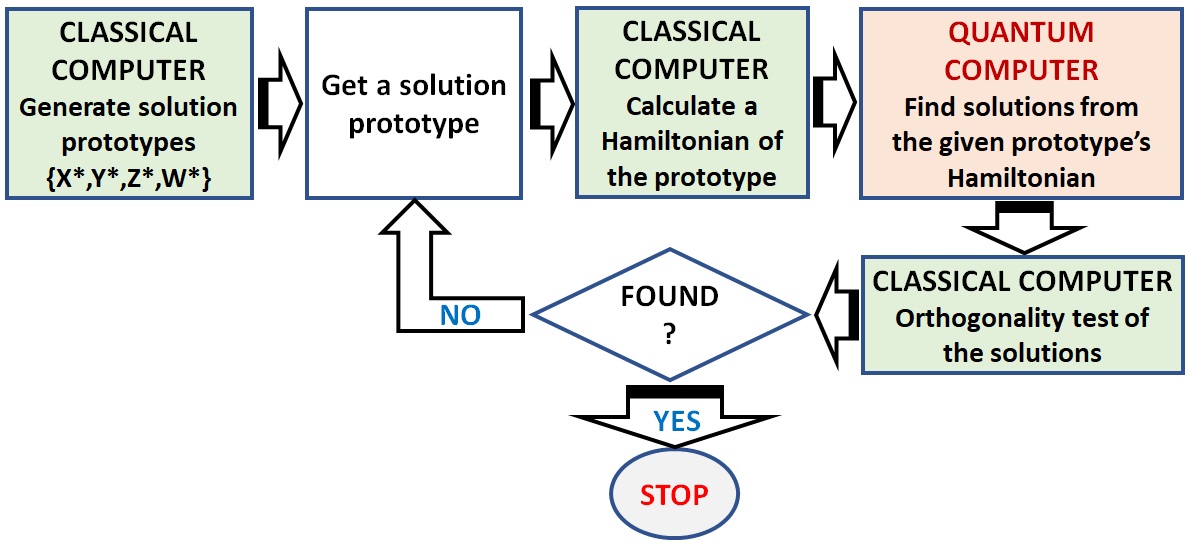}
	\protect\caption{\label{FIG_gen_qsolve_xyzw} Block diagram of classical and quantum processing in extended Turyn-based quantum computing method. The classical processing includes generation of the $\{X^*,Y^*,Z^*,W^*\}$ solution prototypes, construction of the Hamiltonian, and orthogonality test of the solutions. The quantum computing is solely employed to find the solution (ground state) of the problem defined in the Hamiltonian. In this scheme, the classical and quantum processing will be terminated after a valid solution is found.
	}
\end{figure}

Figure \ref{FIG_gen_qsolve_xyzw} shows a block diagram of extended Turyn-based quantum computing method, involving both of classical and quantum computing parts. The generation of $\{X^*,Y^*,Z^*,W^*\}$ solution prototypes and their corresponding Hamiltonians are done in a classical computer. They are fetch one-by-one and processed by a quantum computer which deliver solutions. In the next step, the classical computer then check the orthogonality of the matrices. Notes that the simplest way to check the orthogonality of an $M \times M$ matrix $H$ is by multiplication of $H^TH$ which consisting of $M$ times multiplications for each of all $M^2$ entries in the product followed by checking them whether the off diagonal are zeros and the diagonal entries are equal to $M^2$. Therefore, the orthogonality test can be done in $O(M^3)$.

Table \ref{TB_Turyn} shows required number of qubits and performance of the Turyn based quantum computing method. An H-matrix of order $44$ and $68$ have successfully been found, but higher orders matrices have not. We observe that the number of \emph{embedding qubits} compared to the number of physical qubits grows faster than the similar case in the Williamson and Baumert-Hall based methods; i.e, it is now about $11$ times at the order of $68$. Assuming this factor stay the same, higher order matrices of $92$, which needs $199$ physical qubits, might require about $2,189$ embedding qubits. This is more than the currently available number of qubits in the DW-2000Q quantum processor, and therefore the search of order $92$ H-matrix did not successful. We have proposed a solution for the limitation of quantum computer resource by the Extended Turyn based method described previously. 

\begin{table*}[b]
    \centering
    \begin{tabular}{|c|c|c|c|c|c|c|c|c|c|}
    \hline
	\multirow{2}{*}{No}  & 
	\multicolumn{2}{c|}{$k$} & 
	\multicolumn{2}{c|}{Order} & 
	\multicolumn{3}{c|}{Number of Qubits} & 
	\multirow{2}{*}{\shortstack{E/P\\Ratio}} & 
	\multirow{2}{*}{\shortstack{No. of Correct\\in 10,000}}\\
	\cline{2-8}
    & ORIG & EXTD & ORIG & EXTD & LOGI & PHYS & EMBD & &\\
	\hline
    \hline
     1 & 4 & 8 & 68 & 92 & 16 & 108 & 860 & 8 & 2 \\
    \hline
    \end{tabular}
    \caption{Resource required in Extended Turyn Method (ORIG: Original, EXTD: Extended, LOGI: Logical, PHYS: Physical, EMBD: Embedding, E/P: Embedding-to-Physical). The number of logical qubits is determined by the number of additional $k$ in the extension $\Delta k$, not by the final qubits. This table only show one of successful solution prototypes.}
    \label{TB_Extended_Turyn}
\end{table*}

Table \ref{TB_Extended_Turyn} shows the required resource and performance of extended Turyn based method. For extending $k=4$ into $k=8$, we need $108$ physical qubits; which is then increased into $860$ embedding qubits. An important feature of this method is that, as long as the number of additional/extension $\Delta k=4$ is kept the same, the required qubits to solve extended problem will also stay the same, regardless the targeted order. However, this advantage should be paid by increasing number of solution prototypes, implying that more classical computing resources is needed and the frequency usage of the quantum processor is also increased. We expect to have an optimal point where the combination of the classical and quantum resources delivers the best solution and achieves highest order of the searched H-matrix. 

At present, some of H-matrices of order under $1,000$ has not yet been found, such as $668, 716$, and $892$, due to huge computational resource required to perform the computation by existing classical methods. When using the Turyn-based quantum computing method, even after extension, H-SEARCH for such an order still cannot be implemented. As an illustration, with a $12$ pre-filled $\{X^*, Y^*, Z^*, W^*\}$, the required logical qubits for the $668$ case will be $4\cdot (56-12)= 176$ which becomes $176\cdot 175/2 = 15,400$ physical qubits. Assuming the similar embedding performance as before at a factor of $8$, the required qubits is $123,200$ which is beyond current capability of quantum annealing processors. 

\begin{figure}
    \centering
	\includegraphics[width=0.65\columnwidth]{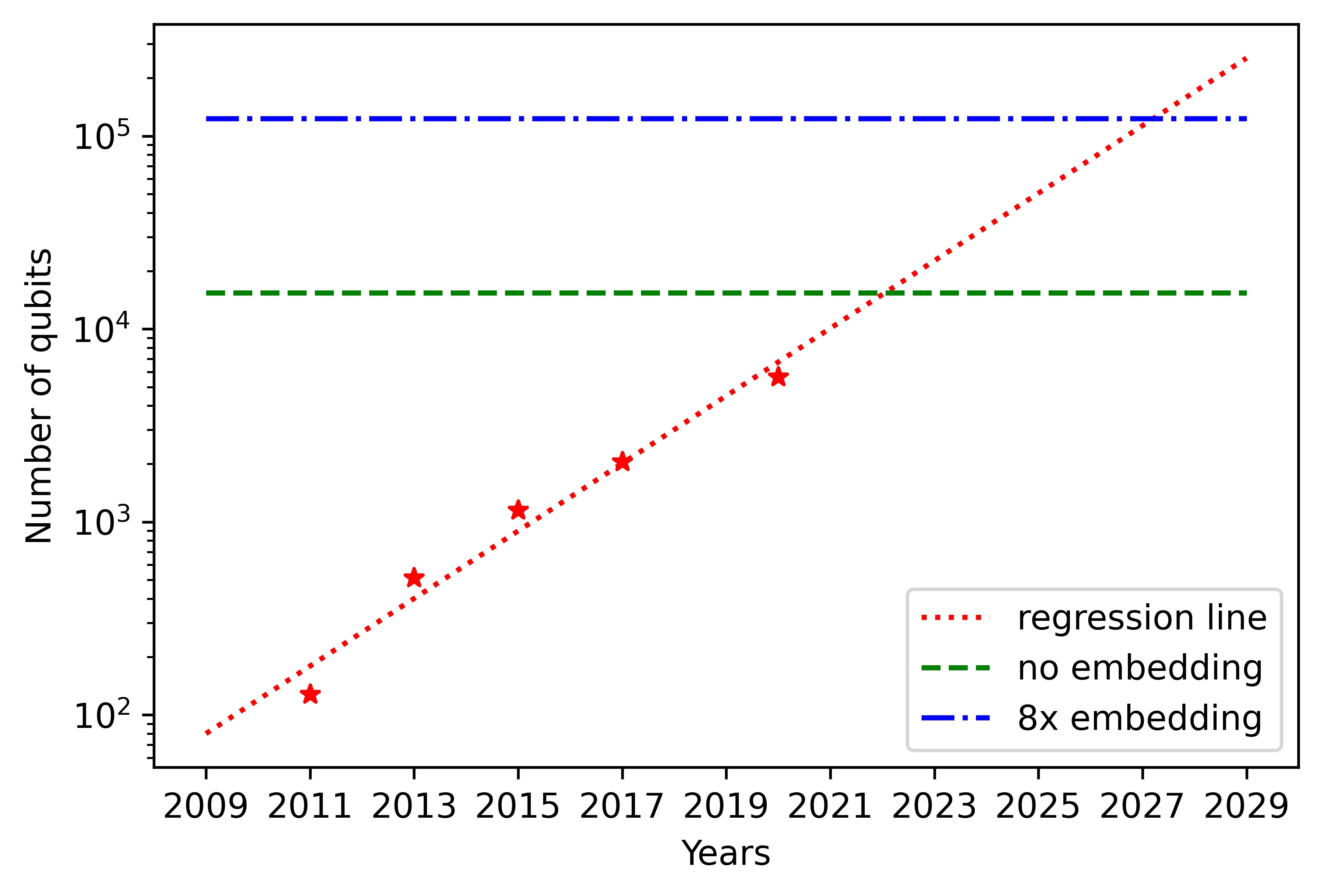}
	\protect\caption{\label{FIG_qubits_growth} Progress of required and available qubits for solving a H-SEARCH problem. The red-dotted curve is a regression line based on actual number of qubits (shown as red stars) produced by D-Wave \cite{dwave_2020}. The linear curve in semi-logarithmic plot indicates that the number of qubits is doubled every two years. The blue (dash-dot) line at the top shows required number of qubits in finding $668$ order H-matrix for each of a given $\{X^*,Y^*,Z^*,W^*\}$ by assuming an embedding factor of $8$, whereas the green (dashed) curve is the required number of qubits with no embedding factor, meaning an ideal complete graph connections among the qubits are available.
	}
\end{figure}

Figure \ref{FIG_qubits_growth} shows the progress of available qubits in D-Wave quantum annealers \cite{dwave_2020} and the decrease number of required qubits to implement H-matrix search of order $668$ by solving the $\{X^*, Y^*, Z^*, W^*\}$ problem. The points in the graph shows actual number of qubits achieved in every year since 2011. We can see that the number of qubits doubled every two years; therefore, using regression we get a linear line in a semi-logarithmic plot as shown by a dotted red curve. The middle dashed green horizontal line indicates the number of required qubits when no additional embedding qubits are required, which means that an ideal complete graph connection among the qubits is available. The top blue dashed dotted line indicates the number of required qubits with embedding factor of $8$. Assuming that the connections among qubits are also improved every year, we can expect the H-SEARCH of order 668 can be implemented between the year 2022 to 2029. Additionally, recent achievement of 64 qubits volume \cite{jurcevic_2020} and the 1000 qubits milestone \cite{cho_2020} of QGM processor, the H-SEARCH implementation through QAOA also very promising to explore.


\section*{Acknowledgments}
This work has been supported partially by the WCR (World Class Research) Program of Indonesian Ministry of Education (formerly Min. of Research and Higher Education), P3MI 2019 Program of ITB, and by the Blueqat Inc. (formerly MDR Inc.), Tokyo, Japan.


\clearpage
\section*{Supplementary Information I}
\bigskip
\subsection*{\bf Hadamard Matrices Found by Williamson Based Quantum Computing Method. }
\bigskip

\begin{figure}[h]
    \centering
	\includegraphics[width=0.65\columnwidth]{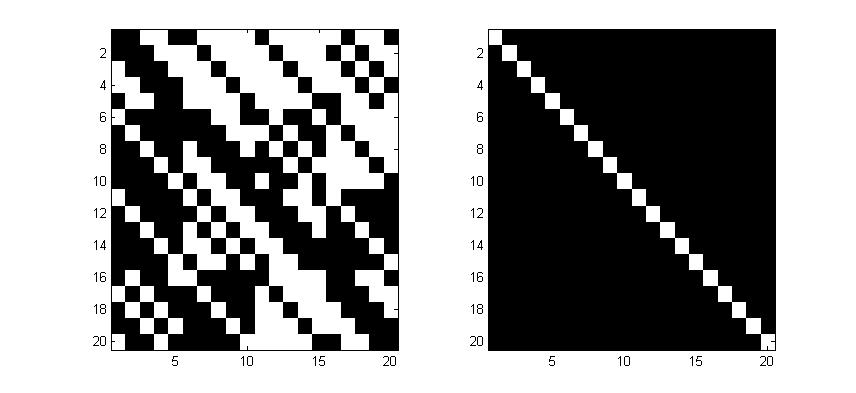}
	\protect\caption{\label{FIG_williamson_20} H-matrix of order $20$: left part is the H-matrix (white=$+1$, black=$-1$), right part is indicator matrix (black=$0$, white=$20$).
	}
\end{figure}
\begin{figure}[h]
    \centering
	\includegraphics[width=0.65\columnwidth]{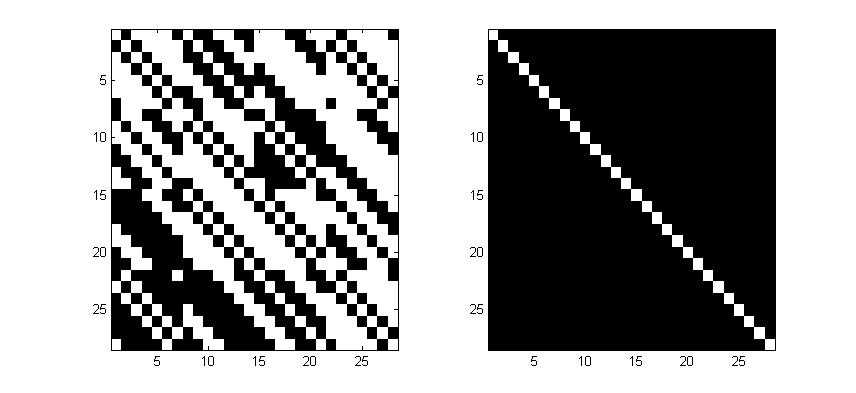}
	\protect\caption{\label{FIG_williamson_28} H-matrix of order 28: left part is the H-matrix (white= $+1$, black=-1), right part is indicator matrix (black=0, white=28).
	}
\end{figure}

\begin{figure}[h]
    \centering
	\includegraphics[width=0.65\columnwidth]{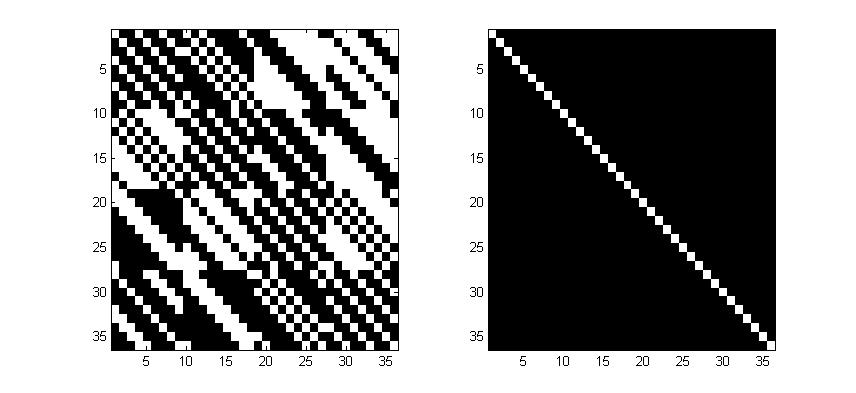}
	\protect\caption{\label{FIG_williamson_36} H-matrix of order 36: left part is the H-matrix (white=+1, black=-1), right part is indicator matrix (black=0, white=36).
	}
\end{figure}
\bigskip
\subsection*{\bf Hadamard Matrices Found by Baumert-Hall Based Quantum Computing Method. }
\bigskip

\begin{figure}[h]
    \centering
	\includegraphics[width=0.65\columnwidth]{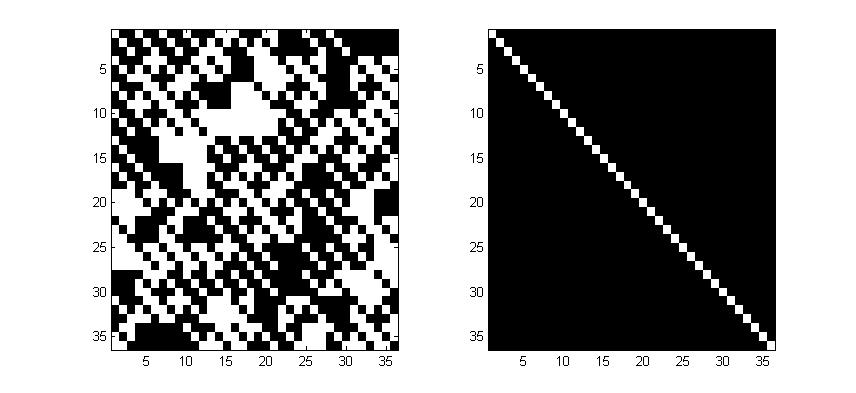}
	\protect\caption{\label{FIG_BauHall_36} H-matrix of order $36$: left part is the H-matrix (white=$+1$, black=$-1$), right part is indicator matrix (black=$0$, white=$36$).
	}
\end{figure}
%

\begin{figure}[h]
    \centering
	\includegraphics[width=0.65\columnwidth]{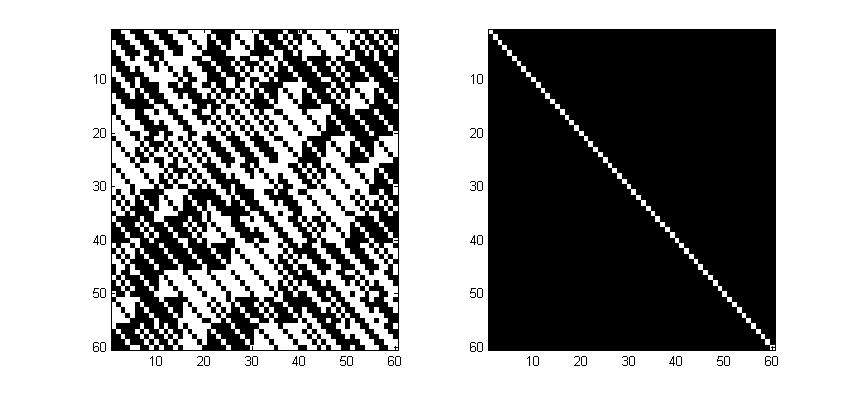}
	\protect\caption{\label{FIG_BauHall_60} H-matrix of order $60$: left part is the H-matrix (white=$+1$, black=$-1$), right part is indicator matrix (black=$0$, white=$60$).
	}
\end{figure}
%

\begin{figure}[h]
    \centering
	\includegraphics[width=0.65\columnwidth]{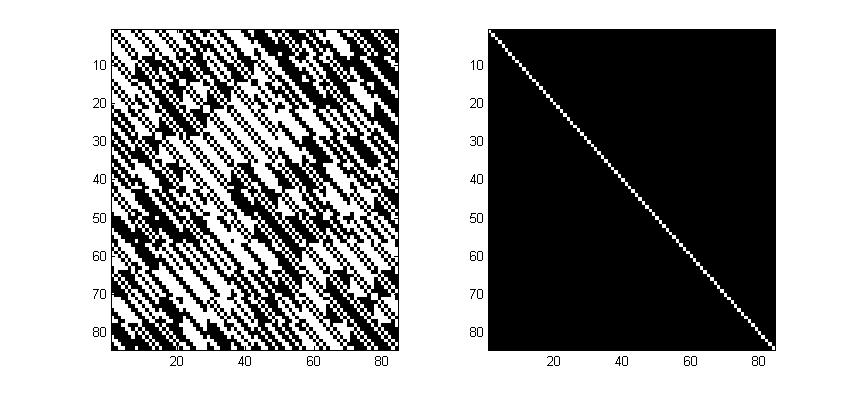}
	\protect\caption{\label{FIG_BauHall_84} H-matrix of order $84$: left part is the H-matrix (white=$+1$, black=$-1$), right part is indicator matrix (black=$0$, white=$84$).
	}
\end{figure}
\bigskip
\subsection*{\bf Hadamard Matrices Found by Turyn's Based Quantum Computing Method. }
\bigskip

\begin{figure}[h]
    \centering
	\includegraphics[width=0.65\columnwidth]{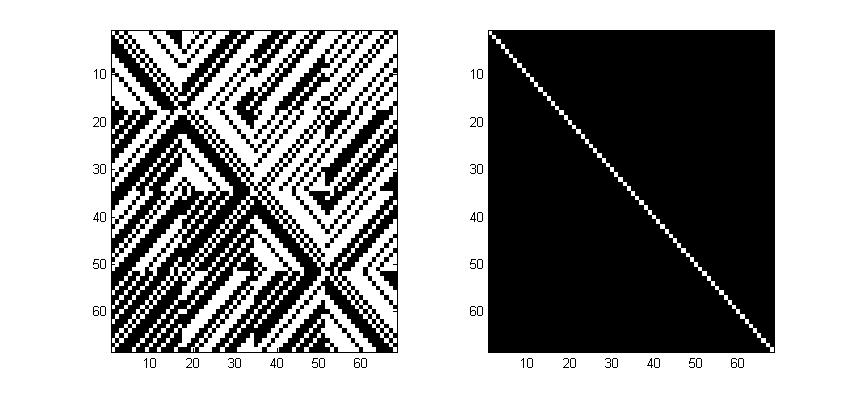}
	\protect\caption{\label{FIG_Turyn_44} H-matrix of order $68$: left part is the H-matrix (white=$+1$, black=$-1$), right part is indicator matrix (black=$0$, white=$68$).
	}
\end{figure}
%

%

\clearpage
\section*{Supplementary Information II}
\input{6_B_Appendix_Williamson}
\input{6_B_Appendix_Turyn}
\clearpage
\subsection*{ \bf Complete Expression in the Extended Turyn Based Method}
\bigskip

\subsection*{A complete expression of Eq.(\ref{EQ_Eks_xturyn})}
$E_k(s)=4s_2 - 14s_0 - 14s_3 + 18s_4 - 4s_5 + 4s_6 - 18s_7 + 24s_8 + 24s_{11} + 40s_{12}- 8s_{13} + 16s_{14} - 40s_{15} - 16s_0s_1 + 6s_0s_2 - 16s_1s_2 + 4s_0s_4 - 2s_1s_3 - 4s_0s_5 - 16s_2s_3 + 4s_0s_6 + 8s_1s_5 + 4s_2s_4 - 4s_0s_7 - 8s_1s_6 - 4s_2s_5 - 4s_3s_4 - 8s_0s_8 + 4s_2s_6 + 8s_1s_8 - 4s_2s_7 - 20s_4s_5 - 8s_1s_9 + 4s_3s_7 + 6s_4s_6 - 8s_0s_{11} - 8s_1s_{10} - 8s_3s_8 - 8s_4s_7 - 28s_5s_6 + 4s_0s_{12} + 8s_1s_{11} + 8s_3s_9 + 6s_5s_7 - 4s_0s_{13} + 8s_3s_{10} - 20s_6s_7 + 8s_0s_{14} + 12s_1s_{13} + 8s_2s_{12} - 8s_3s_{11} - 4s_0s_{15} - 8s_1s_{14} - 4s_2s_{13} - 4s_3s_{12} + 4s_1s_{15} + 8s_2s_{14} + 4s_3s_{13} + 12s_4s_{12} - 4s_2s_{15} - 4s_4s_{13} - 8s_5s_{12} - 48s_8s_9 + 4s_3s_{15} + 8s_4s_{14} + 20s_5s_{13} + 8s_6s_{12} - 8s_8s_{10} - 12s_4s_{15} - 16s_5s_{14} - 20s_6s_{13} - 12s_7s_{12} + 32s_8s_{11} - 16s_9s_{10} + 4s_5s_{15} + 16s_6s_{14} + 4s_7s_{13} + 8s_8s_{12} - 8s_9s_{11}- 4s_6s_{15} - 8s_7s_{14} - 8s_8s_{13} - 48s_{10}s_{11} + 12s_7s_{15} + 8s_9s_{13} + 8s_8s_{15}+ 8s_{10}s_{13} + 8s_{11}s_{12} - 8s_9s_{15} - 8s_{11}s_{13} - 8s_{10}s_{15} - 40s_{12}s_{13}+ 8s_{11}s_{15} + 24s_{12}s_{14} - 16s_{12}s_{15} - 56s_{13}s_{14} + 8s_{13}s_{15} - 40s_{14}s_{15} + 4s_0s_1s_2 + 8s_0s_1s_3 - 2s_0s_1s_4 + 2s_0s_2s_4 + 4s_1s_2s_3 - 2s_0s_2s_5 - 2s_1s_2s_4 + 2s_0s_1s_7+ 2s_0s_2s_6 + 4s_0s_3s_5 + 2s_1s_3s_4 - 4s_0s_1s_8 - 2s_0s_2s_7 - 4s_0s_3s_6 + 2s_0s_4s_5 - 2s_1s_3s_5- 2s_2s_3s_4 + 4s_0s_2s_8 + 2s_0s_4s_6 + 2s_1s_2s_7 + 2s_1s_3s_6 - 4s_0s_2s_9 + 2s_0s_5s_6 - 4s_1s_2s_8- 2s_1s_3s_7 + 2s_1s_4s_6 - 4s_0s_1s_{11} - 4s_0s_2s_{10} + 2s_0s_5s_7 + 4s_1s_3s_8 + 4s_1s_4s_7 + 2s_2s_3s_7 + 2s_2s_4s_6 + 2s_3s_4s_5 - 4s_0s_1s_{12} + 4s_0s_2s_{11} + 2s_0s_6s_7 - 4s_1s_3s_9 + 2s_1s_5s_7 - 4s_2s_3s_8 - 2s_3s_4s_6 + 4s_0s_2s_{12} - 4s_1s_2s_{11} - 4s_1s_3s_{10} + 2s_2s_5s_7 + 2s_3s_5s_6 - 4s_0s_2s_{13} - 4s_1s_2s_{12} + 4s_1s_3s_{11} - 2s_3s_5s_7 - 4s_4s_5s_6 + 4s_0s_1s_{15} + 4s_0s_2s_{14} + 8s_0s_3s_{13}+ 4s_1s_3s_{12} - 4s_2s_3s_{11} + 2s_3s_6s_7 + 8s_4s_5s_7 - 4s_0s_3s_{14} + 4s_0s_8s_9 - 4s_1s_3s_{13}- 4s_2s_3s_{12} - 4s_4s_5s_8 - 8s_4s_6s_7 + 4s_0s_8s_{10} + 4s_1s_2s_{15} + 4s_1s_3s_{14} + 4s_4s_6s_8 + 4s_5s_6s_7 + 4s_0s_9s_{10} + 4s_1s_8s_{10} - 4s_4s_6s_9 - 4s_5s_6s_8 + 4s_0s_9s_{11} + 8s_1s_8s_{11}+ 4s_2s_3s_{15} + 4s_2s_8s_{10} + 4s_3s_8s_9 - 4s_4s_5s_{11} - 4s_4s_6s_{10} + 4s_5s_7s_8 + 4s_0s_{10}s_{11}+ 4s_1s_9s_{11} - 4s_3s_8s_{10} - 4s_4s_5s_{12} + 4s_4s_6s_{11} - 4s_4s_8s_9 - 4s_5s_7s_9 - 4s_6s_7s_8 + 4s_2s_9s_{11} + 4s_3s_9s_{10} + 4s_4s_6s_{12} + 4s_4s_8s_{10} - 4s_5s_6s_{11} - 4s_5s_7s_{10} - 4s_3s_9s_{11}- 4s_4s_6s_{13} - 4s_4s_9s_{10} - 4s_5s_6s_{12} + 4s_5s_7s_{11} - 4s_5s_8s_{10} + 4s_3s_{10}s_{11} + 4s_4s_5s_{15}+ 4s_4s_6s_{14} + 8s_4s_7s_{13} + 4s_4s_9s_{11} + 4s_5s_7s_{12} + 8s_5s_8s_{11} - 4s_6s_7s_{11} + 4s_6s_8s_{10} + 4s_7s_8s_9 + 4s_0s_{12}s_{13} - 4s_4s_7s_{14} - 4s_4s_{10}s_{11} - 4s_5s_7s_{13} - 4s_5s_9s_{11} - 4s_6s_7s_{12} - 8s_6s_8s_{11} - 4s_7s_8s_{10} + 4s_0s_{12}s_{14} + 4s_5s_6s_{15} + 4s_5s_7s_{14} + 4s_6s_9s_{11} + 4s_7s_9s_{10} + 4s_0s_{13}s_{14} + 4s_1s_{12}s_{14} - 4s_7s_9s_{11} - 16s_8s_9s_{10} + 4s_0s_{13}s_{15} + 8s_1s_{12}s_{15} + 4s_2s_{12}s_{14} + 4s_3s_{12}s_{13} + 4s_6s_7s_{15} + 4s_7s_{10}s_{11} + 4s_0s_{14}s_{15} + 4s_1s_{13}s_{15} -4s_3s_{12}s_{14} - 4s_4s_{12}s_{13} - 8s_8s_9s_{12} + 4s_2s_{13}s_{15} + 4s_3s_{13}s_{14} + 4s_4s_{12}s_{14} + 8s_8s_{10}s_{12} - 16s_9s_{10}s_{11} - 4s_3s_{13}s_{15} - 4s_4s_{13}s_{14} - 4s_5s_{12}s_{14} - 8s_8s_{10}s_{13} - 8s_9s_{10}s_{12} + 4s_3s_{14}s_{15} + 4s_4s_{13}s_{15} + 8s_5s_{12}s_{15} + 4s_6s_{12}s_{14} + 4s_7s_{12}s_{13}+ 8s_8s_9s_{15} + 8s_8s_{10}s_{14} + 16s_8s_{11}s_{13} + 8s_9s_{11}s_{12} - 4s_4s_{14}s_{15} - 4s_5s_{13}s_{15} - 8s_6s_{12}s_{15} - 4s_7s_{12}s_{14} - 8s_8s_{11}s_{14} - 8s_8s_{12}s_{13} - 8s_9s_{11}s_{13} - 8s_{10}s_{11}s_{12} + 4s_6s_{13}s_{15} + 4s_7s_{13}s_{14} + 8s_8s_{12}s_{14} + 8s_9s_{10}s_{15} + 8s_9s_{11}s_{14} - 4s_7s_{13}s_{15} - 8s_8s_{13}s_{14} - 8s_9s_{12}s_{14} + 4s_7s_{14}s_{15} + 8s_8s_{13}s_{15} + 8s_{10}s_{11}s_{15} - 8s_{10}s_{12}s_{14} - 8s_{11}s_{12}s_{13} - 8s_8s_{14}s_{15} - 8s_9s_{13}s_{15} + 8s_{11}s_{12}s_{14} - 8s_{10}s_{13}s_{15} - 8s_{11}s_{13}s_{14}+ 8s_{11}s_{13}s_{15} - 16s_{12}s_{13}s_{14} - 8s_{11}s_{14}s_{15} + 32s_{12}s_{13}s_{15} - 16s_{12}s_{14}s_{15} + 16s_{13}s_{14}s_{15} + 4s_0s_1s_2s_3 + 2s_0s_1s_4s_5 + 2s_0s_1s_5s_6 + 2s_0s_2s_4s_6 + 2s_1s_2s_4s_5 + 2s_0s_1s_6s_7 + 2s_0s_2s_5s_7 + 2s_0s_3s_4s_7 + 2s_1s_2s_5s_6 + 2s_1s_3s_4s_6 + 2s_2s_3s_4s_5 + 2s_1s_2s_6s_7 + 2s_1s_3s_5s_7 + 2s_2s_3s_5s_6 + 4s_0s_1s_8s_9 + 2s_2s_3s_6s_7 + 4s_0s_1s_9s_{10} + 4s_0s_2s_8s_{10} + 4s_1s_2s_8s_9 + 4s_0s_1s_{10}s_{11} + 4s_0s_2s_9s_{11}+ 4s_0s_3s_8s_{11} + 4s_1s_2s_9s_{10} + 4s_1s_3s_8s_{10} + 4s_2s_3s_8s_9 + 4s_4s_5s_6s_7 + 4s_1s_2s_{10}s_{11} + 4s_1s_3s_9s_{11}+ 4s_2s_3s_9s_{10} + 4s_0s_1s_{12}s_{13} + 4s_2s_3s_{10}s_{11} + 4s_4s_5s_8s_9 + 4s_0s_1s_{13}s_{14} + 4s_0s_2s_{12}s_{14}+ 4s_1s_2s_{12}s_{13} + 4s_4s_5s_9s_{10} + 4s_4s_6s_8s_{10} + 4s_5s_6s_8s_9 + 4s_0s_1s_{14}s_{15} + 4s_0s_2s_{13}s_{15} + 4s_0s_3s_{12}s_{15} + 4s_1s_2s_{13}s_{14} + 4s_1s_3s_{12}s_{14} + 4s_2s_3s_{12}s_{13} + 4s_4s_5s_{10}s_{11}+ 4s_4s_6s_9s_{11} + 4s_4s_7s_8s_{11} + 4s_5s_6s_9s_{10} + 4s_5s_7s_8s_{10} + 4s_6s_7s_8s_9 + 4s_1s_2s_{14}s_{15}+ 4s_1s_3s_{13}s_{15} + 4s_2s_3s_{13}s_{14} + 4s_5s_6s_{10}s_{11} + 4s_5s_7s_9s_{11} + 4s_6s_7s_9s_{10} + 4s_2s_3s_{14}s_{15}+ 4s_4s_5s_{12}s_{13} + 4s_6s_7s_{10}s_{11} + 4s_4s_5s_{13}s_{14} + 4s_4s_6s_{12}s_{14} + 4s_5s_6s_{12}s_{13} + 4s_4s_5s_{14}s_{15} + 4s_4s_6s_{13}s_{15} + 4s_4s_7s_{12}s_{15} + 4s_5s_6s_{13}s_{14} + 4s_5s_7s_{12}s_{14} + 4s_6s_7s_{12}s_{13} + 16s_8s_9s_{10}s_{11} + 4s_5s_6s_{14}s_{15} + 4s_5s_7s_{13}s_{15} + 4s_6s_7s_{13}s_{14}+ 4s_6s_7s_{14}s_{15} + 8s_8s_9s_{12}s_{13} + 8s_8s_9s_{13}s_{14} + 8s_8s_{10}s_{12}s_{14} + 8s_9s_{10}s_{12}s_{13}+ 8s_8s_9s_{14}s_{15} + 8s_8s_{10}s_{13}s_{15} + 8s_8s_{11}s_{12}s_{15} + 8s_9s_{10}s_{13}s_{14} + 8s_9s_{11}s_{12}s_{14}+ 8s_{10}s_{11}s_{12}s_{13} + 8s_9s_{10}s_{14}s_{15} + 8s_9s_{11}s_{13}s_{15} + 8s_{10}s_{11}s_{13}s_{14} + 8s_{10}s_{11}s_{14}s_{15} + 16s_{12}s_{13}s_{14}s_{15} + 264$

\subsection*{A complete expression of $E_k(q)$ in 92-order H matrix using extended Turyn based method.}

$ E_k(q)=88q_{0}q_{1} - 204q_{1} - 172q_{2} - 116q_{3} - 132q_{4} - 92q_{5} - 76q_{6} - 84q_{7} - 240q_{8} - 80q_{9} - 80q_{10} - 240q_{11} - 272q_{12} - 160q_{13} - 176q_{14} - 224q_{15} - 180q_{0} + 152q_{0}q_{2} + 104q_{0}q_{3} + 72q_{1}q_{2} + 56q_{0}q_{4} + 152q_{1}q_{3} + 40q_{0}q_{5} + 40q_{1}q_{4} + 56q_{2}q_{3} + 56q_{0}q_{6} + 72q_{1}q_{5} + 40q_{2}q_{4} + 24q_{0}q_{7} + 24q_{1}q_{6} + 24q_{2}q_{5} + 8q_{3}q_{4} + 48q_{0}q_{8} + 56q_{1}q_{7} + 72q_{2}q_{6} + 40q_{3}q_{5} + 80q_{0}q_{9} + 112q_{1}q_{8} + 24q_{2}q_{7} + 24q_{3}q_{6} + 56q_{4}q_{5} + 80q_{0}q_{10} + 48q_{1}q_{9} + 48q_{2}q_{8} + 40q_{3}q_{7} + 104q_{4}q_{6} + 48q_{0}q_{11} + 48q_{1}q_{10} + 80q_{2}q_{9} + 16q_{3}q_{8} + 56q_{4}q_{7} + 8q_{5}q_{6} + 96q_{0}q_{12} + 112q_{1}q_{11} + 80q_{2}q_{10} + 80q_{3}q_{9} + 48q_{4}q_{8} + 200q_{5}q_{7} + 96q_{0}q_{13} + 80q_{1}q_{12} + 48q_{2}q_{11} + 80q_{3}q_{10} + 16q_{4}q_{9} + 48q_{5}q_{8} + 24q_{6}q_{7} + 128q_{0}q_{14} + 128q_{1}q_{13} + 80q_{2}q_{12} + 16q_{3}q_{11} + 16q_{4}q_{10} + 48q_{5}q_{9} + 16q_{6}q_{8} + 80q_{0}q_{15} + 80q_{1}q_{14} + 64q_{2}q_{13} + 32q_{3}q_{12} + 48q_{4}q_{11} + 48q_{5}q_{10} + 80q_{6}q_{9} + 48q_{7}q_{8} + 144q_{1}q_{15} + 144q_{2}q_{14} + 96q_{3}q_{13} + 96q_{4}q_{12} + 48q_{5}q_{11} + 80q_{6}q_{10} + 48q_{7}q_{9} + 80q_{2}q_{15} + 64q_{3}q_{14} + 32q_{4}q_{13} + 16q_{5}q_{12} + 16q_{6}q_{11} + 48q_{7}q_{10} + 32q_{8}q_{9} + 80q_{3}q_{15} + 64q_{4}q_{14} + 128q_{5}q_{13} + 48q_{6}q_{12} + 48q_{7}q_{11} + 160q_{8}q_{10} + 16q_{4}q_{15} + 16q_{5}q_{14} + 320q_{8}q_{11} + 96q_{9}q_{10} + 112q_{5}q_{15} + 176q_{6}q_{14} + 96q_{7}q_{13} + 128q_{8}q_{12} + 160q_{9}q_{11} + 48q_{6}q_{15} + 32q_{7}q_{14} + 64q_{8}q_{13} + 32q_{9}q_{12} + 32q_{10}q_{11} + 112q_{7}q_{15} + 64q_{8}q_{14} + 128q_{9}q_{13} + 32q_{10}q_{12} + 160q_{8}q_{15} + 160q_{9}q_{14} + 128q_{10}q_{13} + 128q_{11}q_{12} + 96q_{9}q_{15} + 160q_{10}q_{14} + 64q_{11}q_{13} + 96q_{10}q_{15} + 64q_{11}q_{14} + 128q_{12}q_{13} + 160q_{11}q_{15} + 192q_{12}q_{14} + 160q_{12}q_{15} + 448q_{13}q_{15} + 64q_{14}q_{15} - 64q_{0}q_{1}q_{2} - 96q_{0}q_{1}q_{3} - 32q_{0}q_{2}q_{3} - 32q_{0}q_{1}q_{5} - 32q_{0}q_{2}q_{4} - 64q_{1}q_{2}q_{3} - 32q_{0}q_{1}q_{6} - 16q_{0}q_{3}q_{4} - 32q_{0}q_{1}q_{7} - 32q_{0}q_{2}q_{6} - 32q_{0}q_{3}q_{5} - 32q_{1}q_{2}q_{5} - 32q_{1}q_{3}q_{4} + 32q_{0}q_{3}q_{6} - 32q_{0}q_{4}q_{5} - 32q_{1}q_{2}q_{6} - 64q_{0}q_{1}q_{9} - 64q_{0}q_{2}q_{8} - 16q_{0}q_{3}q_{7} - 32q_{0}q_{4}q_{6} - 32q_{1}q_{2}q_{7} - 32q_{1}q_{3}q_{6} - 32q_{1}q_{4}q_{5} - 32q_{2}q_{3}q_{5} - 64q_{0}q_{1}q_{10} - 32q_{0}q_{3}q_{8} - 16q_{0}q_{4}q_{7} - 32q_{0}q_{5}q_{6} - 32q_{1}q_{4}q_{6} - 32q_{2}q_{3}q_{6} - 32q_{2}q_{4}q_{5} - 32q_{0}q_{5}q_{7} - 64q_{1}q_{2}q_{9} - 64q_{1}q_{3}q_{8} - 32q_{1}q_{4}q_{7} - 32q_{1}q_{5}q_{6} - 32q_{2}q_{3}q_{7} - 32q_{2}q_{4}q_{6} - 32q_{3}q_{4}q_{5} - 64q_{0}q_{2}q_{11} - 32q_{0}q_{6}q_{7} - 64q_{1}q_{2}q_{10} - 32q_{1}q_{5}q_{7} - 32q_{2}q_{5}q_{6} - 64q_{0}q_{1}q_{13} - 64q_{0}q_{2}q_{12} - 32q_{0}q_{3}q_{11} - 32q_{1}q_{6}q_{7} - 64q_{2}q_{3}q_{9} - 32q_{2}q_{5}q_{7} - 16q_{3}q_{4}q_{7} - 32q_{3}q_{5}q_{6} - 64q_{0}q_{1}q_{14} - 32q_{0}q_{3}q_{12} - 64q_{1}q_{3}q_{11} - 64q_{2}q_{3}q_{10} - 32q_{2}q_{6}q_{7} - 64q_{0}q_{1}q_{15} - 64q_{0}q_{2}q_{14} - 64q_{0}q_{3}q_{13} - 64q_{1}q_{2}q_{13} - 64q_{1}q_{3}q_{12} - 32q_{3}q_{6}q_{7} - 96q_{4}q_{5}q_{7} - 32q_{0}q_{2}q_{15} + 32q_{0}q_{3}q_{14} - 64q_{0}q_{8}q_{9} - 64q_{1}q_{2}q_{14} + 32q_{4}q_{6}q_{7} - 32q_{0}q_{3}q_{15} - 64q_{0}q_{8}q_{10} - 64q_{1}q_{2}q_{15} - 64q_{1}q_{3}q_{14} - 64q_{1}q_{8}q_{9} - 64q_{2}q_{3}q_{13} - 64q_{4}q_{5}q_{9} - 64q_{4}q_{6}q_{8} - 64q_{5}q_{6}q_{7} - 32q_{0}q_{8}q_{11} - 64q_{0}q_{9}q_{10} - 32q_{1}q_{3}q_{15} - 64q_{1}q_{8}q_{10} - 64q_{2}q_{3}q_{14} - 64q_{2}q_{8}q_{9} - 64q_{4}q_{5}q_{10} - 32q_{4}q_{7}q_{8} - 64q_{0}q_{9}q_{11} - 64q_{1}q_{8}q_{11} - 64q_{1}q_{9}q_{10} - 64q_{2}q_{3}q_{15} - 64q_{2}q_{8}q_{10} - 64q_{3}q_{8}q_{9} - 64q_{5}q_{6}q_{9} - 64q_{5}q_{7}q_{8} - 64q_{0}q_{10}q_{11} - 64q_{1}q_{9}q_{11} - 64q_{2}q_{9}q_{10} - 64q_{4}q_{6}q_{11} - 64q_{5}q_{6}q_{10} - 64q_{1}q_{10}q_{11} - 64q_{2}q_{9}q_{11} - 32q_{3}q_{8}q_{11} - 64q_{3}q_{9}q_{10} - 64q_{4}q_{5}q_{13} - 64q_{4}q_{6}q_{12} - 32q_{4}q_{7}q_{11} - 64q_{4}q_{8}q_{10} - 64q_{5}q_{8}q_{9} - 64q_{6}q_{7}q_{9} - 64q_{2}q_{10}q_{11} - 64q_{4}q_{5}q_{14} - 32q_{4}q_{7}q_{12} - 32q_{4}q_{8}q_{11} - 64q_{5}q_{7}q_{11} - 64q_{6}q_{7}q_{10} - 64q_{6}q_{8}q_{9} - 64q_{3}q_{10}q_{11} - 64q_{4}q_{5}q_{15} - 64q_{4}q_{6}q_{14} - 64q_{4}q_{7}q_{13} - 64q_{4}q_{9}q_{11} - 64q_{5}q_{6}q_{13} - 64q_{5}q_{7}q_{12} - 64q_{5}q_{8}q_{11} - 64q_{5}q_{9}q_{10} - 64q_{6}q_{8}q_{10} - 64q_{7}q_{8}q_{9} - 64q_{0}q_{12}q_{13} - 32q_{4}q_{6}q_{15} + 32q_{4}q_{7}q_{14} - 64q_{5}q_{6}q_{14} + 64q_{6}q_{8}q_{11} - 64q_{6}q_{9}q_{10} - 64q_{0}q_{12}q_{14} - 64q_{1}q_{12}q_{13} - 32q_{4}q_{7}q_{15} - 64q_{5}q_{6}q_{15} - 64q_{5}q_{7}q_{14} - 64q_{5}q_{10}q_{11} - 64q_{6}q_{7}q_{13} - 64q_{6}q_{9}q_{11} - 32q_{7}q_{8}q_{11} - 64q_{7}q_{9}q_{10} - 32q_{0}q_{12}q_{15} - 64q_{0}q_{13}q_{14} - 64q_{1}q_{12}q_{14} - 64q_{2}q_{12}q_{13} - 32q_{5}q_{7}q_{15} - 64q_{6}q_{7}q_{14} - 64q_{6}q_{10}q_{11} - 64q_{0}q_{13}q_{15} - 64q_{1}q_{12}q_{15} - 64q_{1}q_{13}q_{14} - 64q_{2}q_{12}q_{14} - 64q_{3}q_{12}q_{13} - 64q_{6}q_{7}q_{15} - 64q_{7}q_{10}q_{11} - 128q_{8}q_{9}q_{11} - 64q_{0}q_{14}q_{15} - 64q_{1}q_{13}q_{15} - 64q_{2}q_{13}q_{14} - 128q_{8}q_{10}q_{11} - 64q_{1}q_{14}q_{15} - 64q_{2}q_{13}q_{15} - 32q_{3}q_{12}q_{15} - 64q_{3}q_{13}q_{14} - 64q_{4}q_{12}q_{14} - 64q_{5}q_{12}q_{13} - 128q_{8}q_{9}q_{13} - 128q_{8}q_{10}q_{12} - 64q_{2}q_{14}q_{15} - 32q_{4}q_{12}q_{15} - 64q_{6}q_{12}q_{13} - 128q_{8}q_{9}q_{14} - 64q_{8}q_{11}q_{12} - 64q_{3}q_{14}q_{15} - 64q_{4}q_{13}q_{15} - 64q_{5}q_{12}q_{15} - 64q_{5}q_{13}q_{14} - 64q_{6}q_{12}q_{14} - 64q_{7}q_{12}q_{13} - 128q_{8}q_{9}q_{15} - 128q_{8}q_{10}q_{14} - 128q_{8}q_{11}q_{13} - 128q_{9}q_{10}q_{13} - 128q_{9}q_{11}q_{12} + 64q_{6}q_{12}q_{15} - 64q_{6}q_{13}q_{14} - 64q_{8}q_{10}q_{15} + 64q_{8}q_{11}q_{14} - 128q_{9}q_{10}q_{14} - 64q_{5}q_{14}q_{15} - 64q_{6}q_{13}q_{15} - 32q_{7}q_{12}q_{15} - 64q_{7}q_{13}q_{14} - 64q_{8}q_{11}q_{15} - 128q_{8}q_{12}q_{14} - 128q_{9}q_{10}q_{15} - 128q_{9}q_{11}q_{14} - 128q_{9}q_{12}q_{13} - 128q_{10}q_{11}q_{13} - 64q_{6}q_{14}q_{15} - 64q_{8}q_{12}q_{15} - 64q_{9}q_{11}q_{15} - 128q_{10}q_{11}q_{14} - 128q_{10}q_{12}q_{13} - 64q_{7}q_{14}q_{15} - 128q_{8}q_{13}q_{15} - 128q_{9}q_{13}q_{14} - 128q_{10}q_{11}q_{15} - 128q_{10}q_{13}q_{14} - 128q_{11}q_{12}q_{14} - 128q_{9}q_{14}q_{15} - 64q_{11}q_{12}q_{15} - 128q_{10}q_{14}q_{15} - 128q_{11}q_{13}q_{15} - 384q_{12}q_{13}q_{15} - 256q_{13}q_{14}q_{15} + 64q_{0}q_{1}q_{2}q_{3} + 32q_{0}q_{1}q_{4}q_{5} + 32q_{0}q_{1}q_{5}q_{6} + 32q_{0}q_{2}q_{4}q_{6} + 32q_{1}q_{2}q_{4}q_{5} + 32q_{0}q_{1}q_{6}q_{7} + 32q_{0}q_{2}q_{5}q_{7} + 32q_{0}q_{3}q_{4}q_{7} + 32q_{1}q_{2}q_{5}q_{6} + 32q_{1}q_{3}q_{4}q_{6} + 32q_{2}q_{3}q_{4}q_{5} + 32q_{1}q_{2}q_{6}q_{7} + 32q_{1}q_{3}q_{5}q_{7} + 32q_{2}q_{3}q_{5}q_{6} + 64q_{0}q_{1}q_{8}q_{9} + 32q_{2}q_{3}q_{6}q_{7} + 64q_{0}q_{1}q_{9}q_{10} + 64q_{0}q_{2}q_{8}q_{10} + 64q_{1}q_{2}q_{8}q_{9} + 64q_{0}q_{1}q_{10}q_{11} + 64q_{0}q_{2}q_{9}q_{11} + 64q_{0}q_{3}q_{8}q_{11} + 64q_{1}q_{2}q_{9}q_{10} + 64q_{1}q_{3}q_{8}q_{10} + 64q_{2}q_{3}q_{8}q_{9} + 64q_{4}q_{5}q_{6}q_{7} + 64q_{1}q_{2}q_{10}q_{11} + 64q_{1}q_{3}q_{9}q_{11} + 64q_{2}q_{3}q_{9}q_{10} + 64q_{0}q_{1}q_{12}q_{13} + 64q_{2}q_{3}q_{10}q_{11} + 64q_{4}q_{5}q_{8}q_{9} + 64q_{0}q_{1}q_{13}q_{14} + 64q_{0}q_{2}q_{12}q_{14} + 64q_{1}q_{2}q_{12}q_{13} + 64q_{4}q_{5}q_{9}q_{10} + 64q_{4}q_{6}q_{8}q_{10} + 64q_{5}q_{6}q_{8}q_{9} + 64q_{0}q_{1}q_{14}q_{15} + 64q_{0}q_{2}q_{13}q_{15} + 64q_{0}q_{3}q_{12}q_{15} + 64q_{1}q_{2}q_{13}q_{14} + 64q_{1}q_{3}q_{12}q_{14} + 64q_{2}q_{3}q_{12}q_{13} + 64q_{4}q_{5}q_{10}q_{11} + 64q_{4}q_{6}q_{9}q_{11} + 64q_{4}q_{7}q_{8}q_{11} + 64q_{5}q_{6}q_{9}q_{10} + 64q_{5}q_{7}q_{8}q_{10} + 64q_{6}q_{7}q_{8}q_{9} + 64q_{1}q_{2}q_{14}q_{15} + 64q_{1}q_{3}q_{13}q_{15} + 64q_{2}q_{3}q_{13}q_{14} + 64q_{5}q_{6}q_{10}q_{11} + 64q_{5}q_{7}q_{9}q_{11} + 64q_{6}q_{7}q_{9}q_{10} + 64q_{2}q_{3}q_{14}q_{15} + 64q_{4}q_{5}q_{12}q_{13} + 64q_{6}q_{7}q_{10}q_{11} + 64q_{4}q_{5}q_{13}q_{14} + 64q_{4}q_{6}q_{12}q_{14} + 64q_{5}q_{6}q_{12}q_{13} + 64q_{4}q_{5}q_{14}q_{15} + 64q_{4}q_{6}q_{13}q_{15} + 64q_{4}q_{7}q_{12}q_{15} + 64q_{5}q_{6}q_{13}q_{14} + 64q_{5}q_{7}q_{12}q_{14} + 64q_{6}q_{7}q_{12}q_{13} + 256q_{8}q_{9}q_{10}q_{11} + 64q_{5}q_{6}q_{14}q_{15} + 64q_{5}q_{7}q_{13}q_{15} + 64q_{6}q_{7}q_{13}q_{14} + 64q_{6}q_{7}q_{14}q_{15} + 128q_{8}q_{9}q_{12}q_{13} + 128q_{8}q_{9}q_{13}q_{14} + 128q_{8}q_{10}q_{12}q_{14} + 128q_{9}q_{10}q_{12}q_{13} + 128q_{8}q_{9}q_{14}q_{15} + 128q_{8}q_{10}q_{13}q_{15} + 128q_{8}q_{11}q_{12}q_{15} + 128q_{9}q_{10}q_{13}q_{14} + 128q_{9}q_{11}q_{12}q_{14} + 128q_{10}q_{11}q_{12}q_{13} + 128q_{9}q_{10}q_{14}q_{15} + 128q_{9}q_{11}q_{13}q_{15} + 128q_{10}q_{11}q_{13}q_{14} + 128q_{10}q_{11}q_{14}q_{15} + 256q_{12}q_{13}q_{14}q_{15} + 472$

\subsection*{A complete expression of $E_2(q)$ in 92-order H matrix using extended Turyn based method, using $\delta=65,936$}

$E_2(q)=197896q_{16} - 204q_{1} - 172q_{2} - 116q_{3} - 132q_{4} - 92q_{5} - 76q_{6} - 84q_{7} - 240q_{8} - 80q_{9} - 80q_{10} - 240q_{11} - 272q_{12} - 160q_{13} - 176q_{14} - 224q_{15} - 180q_{0} + 197960q_{17} + 197912q_{18} + 197864q_{19} + 197848q_{20} + 197864q_{21} + 197856q_{22} + 197888q_{23} + 197888q_{24} + 197904q_{25} + 197904q_{26} + 197936q_{27} + 197880q_{28} + 197960q_{29} + 197848q_{30} + 197880q_{31} + 197832q_{32} + 197920q_{33} + 197856q_{34} + 197856q_{35} + 197888q_{36} + 197936q_{37} + 197888q_{38} + 197864q_{39} + 197848q_{40} + 197832q_{41} + 197880q_{42} + 197856q_{43} + 197888q_{44} + 197888q_{45} + 197888q_{46} + 197872q_{47} + 197952q_{48} + 197816q_{49} + 197848q_{50} + 197832q_{51} + 197824q_{52} + 197888q_{53} + 197888q_{54} + 197840q_{55} + 197904q_{56} + 197872q_{57} + 197864q_{58} + 197912q_{59} + 197864q_{60} + 197856q_{61} + 197824q_{62} + 197904q_{63} + 197840q_{64} + 197816q_{65} + 198008q_{66} + 197856q_{67} + 197856q_{68} + 197856q_{69} + 197824q_{70} + 197936q_{71} + 197824q_{72} + 197832q_{73} + 197824q_{74} + 197888q_{75} + 197888q_{76} + 197856q_{77} + 197808q_{78} + 197984q_{79} + 197856q_{80} + 197856q_{81} + 197856q_{82} + 197808q_{83} + 197904q_{84} + 197840q_{85} + 197840q_{86} + 197968q_{87} + 198128q_{88} + 197936q_{89} + 197872q_{90} + 197904q_{91} + 197968q_{8}3 + 197840q_{93} + 197936q_{94} + 197968q_{95} + 197840q_{96} + 197840q_{97} + 197936q_{98} + 197968q_{99} + 197936q_{100} + 197872q_{101} + 197936q_{102} + 198000q_{103} + 197968q_{104} + 197808q_{105} + 198256q_{106} + 197872q_{107} + 65936q_{0}q_{1} + 65936q_{0}q_{2} + 65936q_{0}q_{3} + 65936q_{1}q_{2} + 65936q_{0}q_{4} + 65936q_{1}q_{3} + 65936q_{0}q_{5} + 65936q_{1}q_{4} + 65936q_{2}q_{3} + 65936q_{0}q_{6} + 65936q_{1}q_{5} + 65936q_{2}q_{4} + 24q_{0}q_{7} + 65936q_{1}q_{6} + 65936q_{2}q_{5} + 65936q_{3}q_{4} + 65936q_{0}q_{8} + 56q_{1}q_{7} + 65936q_{2}q_{6} + 65936q_{3}q_{5} + 65936q_{0}q_{9} + 65936q_{1}q_{8} + 24q_{2}q_{7} + 65936q_{3}q_{6} + 65936q_{4}q_{5} + 65936q_{0}q_{10} + 65936q_{1}q_{9} + 65936q_{2}q_{8} + 40q_{3}q_{7} + 65936q_{4}q_{6} + 48q_{0}q_{11} + 65936q_{1}q_{10} + 65936q_{2}q_{9} + 65936q_{3}q_{8} + 65936q_{4}q_{7} + 65936q_{5}q_{6} + 65936q_{0}q_{12} + 112q_{1}q_{11} + 65936q_{2}q_{10} + 65936q_{3}q_{9} + 65936q_{4}q_{8} + 65936q_{5}q_{7} + 65936q_{0}q_{13} + 65936q_{1}q_{12} + 48q_{2}q_{11} + 65936q_{3}q_{10} + 65936q_{4}q_{9} + 65936q_{5}q_{8} + 65936q_{6}q_{7} + 65936q_{0}q_{14} + 65936q_{1}q_{13} + 65936q_{2}q_{12} + 16q_{3}q_{11} + 16q_{4}q_{10} + 65936q_{5}q_{9} + 65936q_{6}q_{8} + 80q_{0}q_{15} + 65936q_{1}q_{14} + 65936q_{2}q_{13} + 65936q_{3}q_{12} + 48q_{4}q_{11} + 65936q_{5}q_{10} + 65936q_{6}q_{9} + 65936q_{7}q_{8} - 131872q_{0}q_{16} + 144q_{1}q_{15} + 65936q_{2}q_{14} + 65936q_{3}q_{13} + 65936q_{4}q_{12} + 48q_{5}q_{11} + 65936q_{6}q_{10} + 65936q_{7}q_{9} - 131872q_{0}q_{17} - 131872q_{1}q_{16} + 80q_{2}q_{15} + 65936q_{3}q_{14} + 65936q_{4}q_{13} + 65936q_{5}q_{12} + 16q_{6}q_{11} + 65936q_{7}q_{10} + 65936q_{8}q_{9} - 131872q_{0}q_{18} - 64q_{2}q_{16} + 80q_{3}q_{15} + 64q_{4}q_{14} + 65936q_{5}q_{13} + 65936q_{6}q_{12} + 48q_{7}q_{11} + 65936q_{8}q_{10} - 131872q_{0}q_{19} - 131872q_{2}q_{17} - 96q_{3}q_{16} + 16q_{4}q_{15} + 65936q_{5}q_{14} + 65936q_{6}q_{13} + 65936q_{7}q_{12} + 65936q_{8}q_{11} + 65936q_{9}q_{10} - 131872q_{0}q_{20} - 32q_{3}q_{17} + 112q_{5}q_{15} + 65936q_{6}q_{14} + 65936q_{7}q_{13} + 65936q_{8}q_{12} + 65936q_{9}q_{11} - 131872q_{0}q_{21} - 131872q_{3}q_{18} - 32q_{4}q_{17} - 32q_{5}q_{16} + 48q_{6}q_{15} + 65936q_{7}q_{14} + 65936q_{8}q_{13} + 65936q_{9}q_{12} + 65936q_{10}q_{11} - 131872q_{0}q_{22} - 16q_{4}q_{18} - 32q_{6}q_{16} + 112q_{7}q_{15} + 64q_{8}q_{14} + 65936q_{9}q_{13} + 65936q_{10}q_{12} - 131872q_{0}q_{23} - 131872q_{4}q_{19} - 32q_{5}q_{18} - 32q_{6}q_{17} - 32q_{7}q_{16} + 160q_{8}q_{15} + 65936q_{9}q_{14} + 65936q_{10}q_{13} + 65936q_{11}q_{12} - 131872q_{0}q_{24} - 32q_{5}q_{19} + 32q_{6}q_{18} + 96q_{9}q_{15} + 65936q_{10}q_{14} + 65936q_{11}q_{13} - 131872q_{0}q_{25} - 131872q_{5}q_{20} - 32q_{6}q_{19} - 16q_{7}q_{18} - 64q_{8}q_{17} - 64q_{9}q_{16} + 96q_{10}q_{15} + 64q_{11}q_{14} + 65936q_{12}q_{13} - 131872q_{0}q_{26} - 32q_{6}q_{20} - 16q_{7}q_{19} - 32q_{8}q_{18} - 64q_{10}q_{16} + 160q_{11}q_{15} + 65936q_{12}q_{14} - 131872q_{0}q_{27} - 131872q_{6}q_{21} - 32q_{7}q_{20} + 65936q_{12}q_{15} + 65936q_{13}q_{14} - 32q_{7}q_{21} - 64q_{11}q_{17} + 65936q_{13}q_{15} - 131872q_{1}q_{28} - 32q_{11}q_{18} - 64q_{12}q_{17} - 64q_{13}q_{16} + 65936q_{14}q_{15} - 131872q_{1}q_{29} - 131872q_{2}q_{28} - 131872q_{8}q_{22} - 32q_{12}q_{18} - 64q_{14}q_{16} - 131872q_{1}q_{30} - 64q_{3}q_{28} - 64q_{9}q_{22} - 64q_{13}q_{18} - 64q_{14}q_{17} - 64q_{15}q_{16} - 131872q_{1}q_{31} - 131872q_{3}q_{29} - 131872q_{9}q_{23} - 64q_{10}q_{22} + 32q_{14}q_{18} - 32q_{15}q_{17} - 131872q_{1}q_{32} - 32q_{4}q_{29} - 32q_{5}q_{28} - 64q_{10}q_{23} - 32q_{11}q_{22} - 32q_{15}q_{18} - 131872q_{1}q_{33} - 131872q_{4}q_{30} - 32q_{6}q_{28} - 131872q_{10}q_{24} - 64q_{11}q_{23} - 131872q_{1}q_{34} - 32q_{5}q_{30} - 32q_{6}q_{29} - 32q_{7}q_{28} - 64q_{11}q_{24} - 131872q_{1}q_{35} - 131872q_{5}q_{31} - 32q_{6}q_{30} - 131872q_{1}q_{36} - 32q_{6}q_{31} - 32q_{7}q_{30} - 64q_{8}q_{29} - 64q_{9}q_{28} - 131872q_{12}q_{25} - 131872q_{1}q_{37} - 131872q_{6}q_{32} - 32q_{7}q_{31} - 64q_{10}q_{28} - 64q_{13}q_{25} - 131872q_{1}q_{38} - 32q_{7}q_{32} - 131872q_{13}q_{26} - 64q_{14}q_{25} - 64q_{11}q_{29} - 64q_{14}q_{26} - 32q_{15}q_{25} - 131872q_{2}q_{39} - 131872q_{8}q_{33} - 64q_{12}q_{29} - 64q_{13}q_{28} - 131872q_{14}q_{27} - 64q_{15}q_{26} - 131872q_{2}q_{40} - 131872q_{3}q_{39} - 64q_{9}q_{33} - 64q_{14}q_{28} - 64q_{15}q_{27} - 131872q_{2}q_{41} - 131872q_{9}q_{34} - 64q_{10}q_{33} - 64q_{14}q_{29} - 64q_{15}q_{28} - 131872q_{2}q_{42} - 131872q_{4}q_{40} - 32q_{5}q_{39} - 64q_{10}q_{34} - 64q_{11}q_{33} - 32q_{15}q_{29} - 131872q_{2}q_{43} - 32q_{5}q_{40} - 32q_{6}q_{39} - 131872q_{10}q_{35} - 64q_{11}q_{34} - 131872q_{2}q_{44} - 131872q_{5}q_{41} - 32q_{6}q_{40} - 32q_{7}q_{39} - 64q_{11}q_{35} - 131872q_{2}q_{45} - 32q_{6}q_{41} - 131872q_{2}q_{46} - 131872q_{6}q_{42} - 32q_{7}q_{41} - 64q_{9}q_{39} - 131872q_{12}q_{36} - 131872q_{2}q_{47} - 32q_{7}q_{42} - 64q_{10}q_{39} - 64q_{13}q_{36} - 131872q_{2}q_{48} - 131872q_{13}q_{37} - 64q_{14}q_{36} - 131872q_{8}q_{43} - 64q_{14}q_{37} - 64q_{15}q_{36} - 131872q_{3}q_{49} - 64q_{9}q_{43} - 64q_{13}q_{39} - 131872q_{14}q_{38} - 64q_{15}q_{37} - 131872q_{3}q_{50} - 131872q_{4}q_{49} - 131872q_{9}q_{44} - 64q_{10}q_{43} - 64q_{14}q_{39} - 64q_{15}q_{38} - 131872q_{3}q_{51} - 32q_{5}q_{49} - 64q_{10}q_{44} - 64q_{15}q_{39} - 131872q_{3}q_{52} - 131872q_{5}q_{50} - 131872q_{10}q_{45} - 64q_{11}q_{44} + 64q_{16}q_{39} - 131872q_{3}q_{53} - 32q_{6}q_{50} - 16q_{7}q_{49} - 64q_{11}q_{45} - 131872q_{3}q_{54} - 131872q_{6}q_{51} - 131872q_{3}q_{55} - 32q_{7}q_{51} - 131872q_{12}q_{46} - 131872q_{3}q_{56} - 64q_{13}q_{46} - 131872q_{3}q_{57} - 131872q_{8}q_{52} - 131872q_{13}q_{47} - 64q_{14}q_{46} - 64q_{9}q_{52} - 64q_{14}q_{47} - 131872q_{4}q_{58} - 131872q_{9}q_{53} - 131872q_{14}q_{48} - 64q_{15}q_{47} - 131872q_{4}q_{59} - 131872q_{5}q_{58} - 64q_{10}q_{53} - 32q_{11}q_{52} - 64q_{15}q_{48} - 131872q_{4}q_{60} - 131872q_{10}q_{54} - 131872q_{4}q_{61} - 131872q_{6}q_{59} - 96q_{7}q_{58} - 64q_{11}q_{54} - 131872q_{4}q_{62} + 32q_{7}q_{59} - 131872q_{4}q_{63} - 131872q_{7}q_{60} - 64q_{8}q_{59} - 64q_{9}q_{58} - 131872q_{12}q_{55} - 131872q_{4}q_{64} - 32q_{8}q_{60} - 64q_{10}q_{58} - 64q_{13}q_{55} - 131872q_{8}q_{61} - 131872q_{13}q_{56} - 131872q_{5}q_{65} - 64q_{11}q_{59} - 64q_{14}q_{56} - 32q_{15}q_{55} - 131872q_{5}q_{66} - 131872q_{6}q_{65} - 131872q_{9}q_{62} - 64q_{10}q_{61} - 32q_{11}q_{60} - 64q_{12}q_{59} - 64q_{13}q_{58} - 131872q_{14}q_{57} - 131872q_{5}q_{67} - 64q_{7}q_{65} - 32q_{11}q_{61} - 32q_{12}q_{60} - 64q_{14}q_{58} - 64q_{15}q_{57} - 131872q_{5}q_{68} - 131872q_{7}q_{66} - 64q_{11}q_{62} - 64q_{13}q_{60} - 64q_{14}q_{59} - 64q_{15}q_{58} - 131872q_{5}q_{69} - 64q_{8}q_{66} - 64q_{9}q_{65} + 32q_{14}q_{60} - 32q_{15}q_{59} + 32q_{16}q_{58} - 131872q_{5}q_{70} - 131872q_{8}q_{67} - 64q_{10}q_{65} - 131872q_{12}q_{63} - 32q_{15}q_{60} - 131872q_{5}q_{71} - 64q_{9}q_{67} + 32q_{17}q_{59} - 131872q_{5}q_{72} - 131872q_{9}q_{68} - 64q_{11}q_{66} - 131872q_{13}q_{64} - 64q_{14}q_{63} - 64q_{10}q_{68} - 64q_{11}q_{67} - 64q_{12}q_{66} - 64q_{13}q_{65} - 32q_{15}q_{63} + 32q_{18}q_{60} - 131872q_{6}q_{73} - 131872q_{10}q_{69} - 64q_{14}q_{65} - 64q_{15}q_{64} - 131872q_{6}q_{74} - 131872q_{7}q_{73} - 64q_{11}q_{69} - 64q_{14}q_{66} - 64q_{15}q_{65} - 131872q_{6}q_{75} - 32q_{15}q_{66} + 32q_{16}q_{65} - 131872q_{6}q_{76} - 131872q_{8}q_{74} - 64q_{9}q_{73} - 131872q_{12}q_{70} - 131872q_{6}q_{77} - 64q_{9}q_{74} - 64q_{10}q_{73} - 64q_{13}q_{70} + 32q_{17}q_{66} - 131872q_{6}q_{78} - 131872q_{9}q_{75} - 64q_{10}q_{74} - 131872q_{13}q_{71} - 131872q_{6}q_{79} - 64q_{10}q_{75} + 64q_{11}q_{74} - 64q_{14}q_{71} - 64q_{15}q_{70} - 131872q_{10}q_{76} - 64q_{11}q_{75} - 64q_{13}q_{73} - 131872q_{14}q_{72} + 32q_{28}q_{58} - 131872q_{7}q_{80} - 64q_{11}q_{76} - 64q_{14}q_{73} - 64q_{15}q_{72} - 131872q_{7}q_{81} - 131872q_{8}q_{80} - 64q_{15}q_{73} + 32q_{29}q_{59} - 131872q_{7}q_{82} - 64q_{9}q_{80} - 131872q_{12}q_{77} + 32q_{16}q_{73} - 131872q_{7}q_{83} - 131872q_{9}q_{81} - 64q_{13}q_{77} - 131872q_{7}q_{84} - 64q_{10}q_{81} - 32q_{11}q_{80} - 131872q_{13}q_{78} - 64q_{14}q_{77} - 131872q_{7}q_{85} - 131872q_{10}q_{82} - 64q_{14}q_{78} + 64q_{15}q_{77} - 64q_{11}q_{82} - 131872q_{14}q_{79} - 64q_{15}q_{78} + 32q_{28}q_{65} - 131872q_{8}q_{86} - 64q_{15}q_{79} - 131872q_{8}q_{87} - 131872q_{9}q_{86} - 131872q_{12}q_{83} + 32q_{29}q_{66} - 131872q_{8}q_{88} - 64q_{13}q_{83} - 131872q_{8}q_{89} - 131872q_{10}q_{87} - 128q_{11}q_{86} - 131872q_{13}q_{84} + 32q_{39}q_{58} - 131872q_{8}q_{90} - 128q_{11}q_{87} - 64q_{14}q_{84} - 32q_{15}q_{83} - 131872q_{11}q_{88} - 128q_{12}q_{87} - 128q_{13}q_{86} - 131872q_{14}q_{85} - 131872q_{9}q_{91} - 64q_{12}q_{88} - 128q_{14}q_{86} - 64q_{15}q_{85} - 131872q_{9}q_{8}3 - 131872q_{10}q_{91} - 131872q_{12}q_{89} - 128q_{13}q_{88} - 128q_{14}q_{87} - 128q_{15}q_{86} + 32q_{28}q_{73} - 131872q_{9}q_{93} + 64q_{14}q_{88} - 64q_{15}q_{87} + 64q_{16}q_{86} - 131872q_{9}q_{94} - 131872q_{11}q_{8}3 - 131872q_{13}q_{90} - 128q_{14}q_{89} - 64q_{15}q_{88} - 131872q_{9}q_{95} - 128q_{12}q_{8}3 - 128q_{13}q_{91} - 64q_{15}q_{89} + 64q_{17}q_{87} + 32q_{39}q_{65} - 131872q_{12}q_{93} - 128q_{14}q_{91} - 128q_{15}q_{90} - 131872q_{10}q_{96} - 128q_{13}q_{93} - 128q_{14}q_{8}3 - 128q_{15}q_{91} + 64q_{18}q_{88} - 131872q_{10}q_{97} - 131872q_{11}q_{96} - 131872q_{13}q_{94} - 64q_{15}q_{8}3 + 64q_{16}q_{91} - 131872q_{10}q_{98} - 128q_{14}q_{94} - 131872q_{10}q_{99} - 131872q_{12}q_{97} - 128q_{13}q_{96} - 131872q_{14}q_{95} + 64q_{17}q_{8}3 - 128q_{13}q_{97} - 128q_{14}q_{96} - 128q_{15}q_{95} - 131872q_{11}q_{100} - 131872q_{13}q_{98} - 128q_{15}q_{96} - 131872q_{11}q_{101} - 131872q_{12}q_{100} - 128q_{14}q_{98} + 64q_{16}q_{96} + 32q_{39}q_{73} - 131872q_{14}q_{99} - 131872q_{12}q_{102} - 131872q_{13}q_{101} - 128q_{14}q_{100} - 128q_{15}q_{99} + 64q_{28}q_{86} - 131872q_{12}q_{103} - 131872q_{13}q_{102} - 64q_{15}q_{100} - 131872q_{12}q_{104} - 128q_{15}q_{101} + 64q_{29}q_{87} - 131872q_{14}q_{103} - 384q_{15}q_{102} - 131872q_{13}q_{105} + 64q_{16}q_{102} - 131872q_{13}q_{106} - 131872q_{14}q_{105} - 131872q_{15}q_{104} + 64q_{28}q_{91} - 256q_{15}q_{105} + 64q_{17}q_{103} - 131872q_{14}q_{107} - 131872q_{15}q_{106} + 64q_{16}q_{105} + 64q_{29}q_{8}3 - 131872q_{15}q_{107} + 64q_{18}q_{104} + 64q_{16}q_{107} + 64q_{17}q_{106} + 64q_{28}q_{96} + 64q_{39}q_{86} + 64q_{28}q_{102} + 64q_{39}q_{91} + 64q_{58}q_{73} + 64q_{29}q_{103} + 64q_{28}q_{105} + 64q_{28}q_{107} + 64q_{29}q_{106} + 64q_{39}q_{96} + 64q_{39}q_{102} + 64q_{39}q_{105} + 64q_{58}q_{86} + 64q_{39}q_{107} + 64q_{59}q_{87} + 64q_{60}q_{88} + 64q_{58}q_{91} + 64q_{59}q_{8}3 + 64q_{65}q_{86} + 64q_{66}q_{87} + 64q_{58}q_{96} + 64q_{65}q_{91} + 64q_{66}q_{8}3 + 64q_{73}q_{86} + 64q_{58}q_{102} + 64q_{65}q_{96} + 64q_{59}q_{103} + 64q_{58}q_{105} + 64q_{60}q_{104} + 64q_{73}q_{91} + 64q_{58}q_{107} + 64q_{59}q_{106} + 64q_{65}q_{102} + 64q_{66}q_{103} + 64q_{73}q_{96} + 64q_{65}q_{105} + 64q_{65}q_{107} + 64q_{66}q_{106} + 64q_{73}q_{102} + 64q_{73}q_{105} + 64q_{73}q_{107} + 256q_{86}q_{96} + 128q_{86}q_{102} + 128q_{87}q_{103} + 128q_{86}q_{105} + 128q_{88}q_{104} + 128q_{86}q_{107} + 128q_{87}q_{106} + 128q_{91}q_{102} + 128q_{8}3q_{103} + 128q_{91}q_{105} + 128q_{91}q_{107} + 128q_{8}3q_{106} + 128q_{96}q_{102} + 128q_{96}q_{105} + 128q_{96}q_{107} + 256q_{102}q_{107} + 472$

\subsection{A complete expression of $E_2(s)$ in 92-order H matrix using extended-Turyn based method}

$E_2(s)=197860s_{0} + 197832s_{1} + 197872s_{2} + 197880s_{3} + 181374s_{4} + 214362s_{5} + 214402s_{6} + 148454s_{7} + 214436s_{8} + 231000s_{9} + 214592s_{10} + 82772s_{11} + 247556s_{12} + 247772s_{13} + 198456s_{14} + 50116s_{15} - 33004s_{16} - 33028s_{17} - 33012s_{18} - 32976s_{19} - 32972s_{20} - 32988s_{21} - 32952s_{22} - 32976s_{23} - 32992s_{24} - 32976s_{25} - 32984s_{26} - 33016s_{27} - 33004s_{28} - 33036s_{29} - 32964s_{30} - 32988s_{31} - 32972s_{32} - 32976s_{33} - 32960s_{34} - 32976s_{35} - 32960s_{36} - 33000s_{37} - 32992s_{38} - 33028s_{39} - 32972s_{40} - 32964s_{41} - 32996s_{42} - 32960s_{43} - 32976s_{44} - 32992s_{45} - 32976s_{46} - 32968s_{47} - 33024s_{48} - 32960s_{49} - 32980s_{50} - 32972s_{51} - 32952s_{52} - 32992s_{53} - 32992s_{54} - 32960s_{55} - 33000s_{56} - 32984s_{57} - 33028s_{58} - 33036s_{59} - 32996s_{60} - 32968s_{61} - 32960s_{62} - 32992s_{63} - 32968s_{64} - 32996s_{65} - 33076s_{66} - 32960s_{67} - 32976s_{68} - 32976s_{69} - 32944s_{70} - 33016s_{71} - 32960s_{72} - 33036s_{73} - 32960s_{74} - 32976s_{75} - 32992s_{76} - 32976s_{77} - 32936s_{78} - 33040s_{79} - 32968s_{80} - 32976s_{81} - 32976s_{82} - 32944s_{83} - 33000s_{84} - 32968s_{85} - 33112s_{86} - 33064s_{87} - 33144s_{88} - 32984s_{89} - 32968s_{90} - 33112s_{91} - 33096s_{92} - 32952s_{93} - 33000s_{94} - 33016s_{95} - 33144s_{96} - 32952s_{97} - 33000s_{98} - 33016s_{99} - 32984s_{100} - 32968s_{101} - 33192s_{102} - 33192s_{103} - 33112s_{104} - 33096s_{105} - 33320s_{106} - 33256s_{107} + 16484s_{0}s_{1} + 16484s_{0}s_{2} + 16484s_{0}s_{3} + 16484s_{1}s_{2} + 16484s_{0}s_{4} + 16484s_{1}s_{3} + 16484s_{0}s_{5} + 16484s_{1}s_{4} + 16484s_{2}s_{3} + 16484s_{0}s_{6} + 16484s_{1}s_{5} + 16484s_{2}s_{4} + 6s_{0}s_{7} + 16484s_{1}s_{6} + 16484s_{2}s_{5} + 16484s_{3}s_{4} + 16484s_{0}s_{8} + 14s_{1}s_{7} + 16484s_{2}s_{6} + 16484s_{3}s_{5} + 16484s_{0}s_{9} + 16484s_{1}s_{8} + 6s_{2}s_{7} + 16484s_{3}s_{6} + 16484s_{4}s_{5} + 16484s_{0}s_{10} + 16484s_{1}s_{9} + 16484s_{2}s_{8} + 10s_{3}s_{7} + 16484s_{4}s_{6} + 12s_{0}s_{11} + 16484s_{1}s_{10} + 16484s_{2}s_{9} + 16484s_{3}s_{8} + 16484s_{4}s_{7} + 16484s_{5}s_{6} + 16484s_{0}s_{12} + 28s_{1}s_{11} + 16484s_{2}s_{10} + 16484s_{3}s_{9} + 16484s_{4}s_{8} + 16484s_{5}s_{7} + 16484s_{0}s_{13} + 16484s_{1}s_{12} + 12s_{2}s_{11} + 16484s_{3}s_{10} + 16484s_{4}s_{9} + 16484s_{5}s_{8} + 16484s_{6}s_{7} + 16484s_{0}s_{14} + 16484s_{1}s_{13} + 16484s_{2}s_{12} + 4s_{3}s_{11} + 4s_{4}s_{10} + 16484s_{5}s_{9} + 16484s_{6}s_{8} + 20s_{0}s_{15} + 16484s_{1}s_{14} + 16484s_{2}s_{13} + 16484s_{3}s_{12} + 12s_{4}s_{11} + 16484s_{5}s_{10} + 16484s_{6}s_{9} + 16484s_{7}s_{8} - 32968s_{0}s_{16} + 36s_{1}s_{15} + 16484s_{2}s_{14} + 16484s_{3}s_{13} + 16484s_{4}s_{12} + 12s_{5}s_{11} + 16484s_{6}s_{10} + 16484s_{7}s_{9} - 32968s_{0}s_{17} - 32968s_{1}s_{16} + 20s_{2}s_{15} + 16484s_{3}s_{14} + 16484s_{4}s_{13} + 16484s_{5}s_{12} + 4s_{6}s_{11} + 16484s_{7}s_{10} + 16484s_{8}s_{9} - 32968s_{0}s_{18} - 16s_{2}s_{16} + 20s_{3}s_{15} + 16s_{4}s_{14} + 16484s_{5}s_{13} + 16484s_{6}s_{12} + 12s_{7}s_{11} + 16484s_{8}s_{10} - 32968s_{0}s_{19} - 32968s_{2}s_{17} - 24s_{3}s_{16} + 4s_{4}s_{15} + 16484s_{5}s_{14} + 16484s_{6}s_{13} + 16484s_{7}s_{12} + 16484s_{8}s_{11} + 16484s_{9}s_{10} - 32968s_{0}s_{20} - 8s_{3}s_{17} + 28s_{5}s_{15} + 16484s_{6}s_{14} + 16484s_{7}s_{13} + 16484s_{8}s_{12} + 16484s_{9}s_{11} - 32968s_{0}s_{21} - 32968s_{3}s_{18} - 8s_{4}s_{17} - 8s_{5}s_{16} + 12s_{6}s_{15} + 16484s_{7}s_{14} + 16484s_{8}s_{13} + 16484s_{9}s_{12} + 16484s_{10}s_{11} - 32968s_{0}s_{22} - 4s_{4}s_{18} - 8s_{6}s_{16} + 28s_{7}s_{15} + 16s_{8}s_{14} + 16484s_{9}s_{13} + 16484s_{10}s_{12} - 32968s_{0}s_{23} - 32968s_{4}s_{19} - 8s_{5}s_{18} - 8s_{6}s_{17} - 8s_{7}s_{16} + 40s_{8}s_{15} + 16484s_{9}s_{14} + 16484s_{10}s_{13} + 16484s_{11}s_{12} - 32968s_{0}s_{24} - 8s_{5}s_{19} + 8s_{6}s_{18} + 24s_{9}s_{15} + 16484s_{10}s_{14} + 16484s_{11}s_{13} - 32968s_{0}s_{25} - 32968s_{5}s_{20} - 8s_{6}s_{19} - 4s_{7}s_{18} - 16s_{8}s_{17} - 16s_{9}s_{16} + 24s_{10}s_{15} + 16s_{11}s_{14} + 16484s_{12}s_{13} - 32968s_{0}s_{26} - 8s_{6}s_{20} - 4s_{7}s_{19} - 8s_{8}s_{18} - 16s_{10}s_{16} + 40s_{11}s_{15} + 16484s_{12}s_{14} - 32968s_{0}s_{27} - 32968s_{6}s_{21} - 8s_{7}s_{20} + 16484s_{12}s_{15} + 16484s_{13}s_{14} - 8s_{7}s_{21} - 16s_{11}s_{17} + 16484s_{13}s_{15} - 32968s_{1}s_{28} - 8s_{11}s_{18} - 16s_{12}s_{17} - 16s_{13}s_{16} + 16484s_{14}s_{15} - 32968s_{1}s_{29} - 32968s_{2}s_{28} - 32968s_{8}s_{22} - 8s_{12}s_{18} - 16s_{14}s_{16} - 32968s_{1}s_{30} - 16s_{3}s_{28} - 16s_{9}s_{22} - 16s_{13}s_{18} - 16s_{14}s_{17} - 16s_{15}s_{16} - 32968s_{1}s_{31} - 32968s_{3}s_{29} - 32968s_{9}s_{23} - 16s_{10}s_{22} + 8s_{14}s_{18} - 8s_{15}s_{17} - 32968s_{1}s_{32} - 8s_{4}s_{29} - 8s_{5}s_{28} - 16s_{10}s_{23} - 8s_{11}s_{22} - 8s_{15}s_{18} - 32968s_{1}s_{33} - 32968s_{4}s_{30} - 8s_{6}s_{28} - 32968s_{10}s_{24} - 16s_{11}s_{23} - 32968s_{1}s_{34} - 8s_{5}s_{30} - 8s_{6}s_{29} - 8s_{7}s_{28} - 16s_{11}s_{24} - 32968s_{1}s_{35} - 32968s_{5}s_{31} - 8s_{6}s_{30} - 32968s_{1}s_{36} - 8s_{6}s_{31} - 8s_{7}s_{30} - 16s_{8}s_{29} - 16s_{9}s_{28} - 32968s_{12}s_{25} - 32968s_{1}s_{37} - 32968s_{6}s_{32} - 8s_{7}s_{31} - 16s_{10}s_{28} - 16s_{13}s_{25} - 32968s_{1}s_{38} - 8s_{7}s_{32} - 32968s_{13}s_{26} - 16s_{14}s_{25} - 16s_{11}s_{29} - 16s_{14}s_{26} - 8s_{15}s_{25} - 32968s_{2}s_{39} - 32968s_{8}s_{33} - 16s_{12}s_{29} - 16s_{13}s_{28} - 32968s_{14}s_{27} - 16s_{15}s_{26} - 32968s_{2}s_{40} - 32968s_{3}s_{39} - 16s_{9}s_{33} - 16s_{14}s_{28} - 16s_{15}s_{27} - 32968s_{2}s_{41} - 32968s_{9}s_{34} - 16s_{10}s_{33} - 16s_{14}s_{29} - 16s_{15}s_{28} - 32968s_{2}s_{42} - 32968s_{4}s_{40} - 8s_{5}s_{39} - 16s_{10}s_{34} - 16s_{11}s_{33} - 8s_{15}s_{29} - 32968s_{2}s_{43} - 8s_{5}s_{40} - 8s_{6}s_{39} - 32968s_{10}s_{35} - 16s_{11}s_{34} - 32968s_{2}s_{44} - 32968s_{5}s_{41} - 8s_{6}s_{40} - 8s_{7}s_{39} - 16s_{11}s_{35} - 32968s_{2}s_{45} - 8s_{6}s_{41} - 32968s_{2}s_{46} - 32968s_{6}s_{42} - 8s_{7}s_{41} - 16s_{9}s_{39} - 32968s_{12}s_{36} - 32968s_{2}s_{47} - 8s_{7}s_{42} - 16s_{10}s_{39} - 16s_{13}s_{36} - 32968s_{2}s_{48} - 32968s_{13}s_{37} - 16s_{14}s_{36} - 32968s_{8}s_{43} - 16s_{14}s_{37} - 16s_{15}s_{36} - 32968s_{3}s_{49} - 16s_{9}s_{43} - 16s_{13}s_{39} - 32968s_{14}s_{38} - 16s_{15}s_{37} - 32968s_{3}s_{50} - 32968s_{4}s_{49} - 32968s_{9}s_{44} - 16s_{10}s_{43} - 16s_{14}s_{39} - 16s_{15}s_{38} - 32968s_{3}s_{51} - 8s_{5}s_{49} - 16s_{10}s_{44} - 16s_{15}s_{39} - 32968s_{3}s_{52} - 32968s_{5}s_{50} - 32968s_{10}s_{45} - 16s_{11}s_{44} + 16s_{16}s_{39} - 32968s_{3}s_{53} - 8s_{6}s_{50} - 4s_{7}s_{49} - 16s_{11}s_{45} - 32968s_{3}s_{54} - 32968s_{6}s_{51} - 32968s_{3}s_{55} - 8s_{7}s_{51} - 32968s_{12}s_{46} - 32968s_{3}s_{56} - 16s_{13}s_{46} - 32968s_{3}s_{57} - 32968s_{8}s_{52} - 32968s_{13}s_{47} - 16s_{14}s_{46} - 16s_{9}s_{52} - 16s_{14}s_{47} - 32968s_{4}s_{58} - 32968s_{9}s_{53} - 32968s_{14}s_{48} - 16s_{15}s_{47} - 32968s_{4}s_{59} - 32968s_{5}s_{58} - 16s_{10}s_{53} - 8s_{11}s_{52} - 16s_{15}s_{48} - 32968s_{4}s_{60} - 32968s_{10}s_{54} - 32968s_{4}s_{61} - 32968s_{6}s_{59} - 24s_{7}s_{58} - 16s_{11}s_{54} - 32968s_{4}s_{62} + 8s_{7}s_{59} - 32968s_{4}s_{63} - 32968s_{7}s_{60} - 16s_{8}s_{59} - 16s_{9}s_{58} - 32968s_{12}s_{55} - 32968s_{4}s_{64} - 8s_{8}s_{60} - 16s_{10}s_{58} - 16s_{13}s_{55} - 32968s_{8}s_{61} - 32968s_{13}s_{56} - 32968s_{5}s_{65} - 16s_{11}s_{59} - 16s_{14}s_{56} - 8s_{15}s_{55} - 32968s_{5}s_{66} - 32968s_{6}s_{65} - 32968s_{9}s_{62} - 16s_{10}s_{61} - 8s_{11}s_{60} - 16s_{12}s_{59} - 16s_{13}s_{58} - 32968s_{14}s_{57} - 32968s_{5}s_{67} - 16s_{7}s_{65} - 8s_{11}s_{61} - 8s_{12}s_{60} - 16s_{14}s_{58} - 16s_{15}s_{57} - 32968s_{5}s_{68} - 32968s_{7}s_{66} - 16s_{11}s_{62} - 16s_{13}s_{60} - 16s_{14}s_{59} - 16s_{15}s_{58} - 32968s_{5}s_{69} - 16s_{8}s_{66} - 16s_{9}s_{65} + 8s_{14}s_{60} - 8s_{15}s_{59} + 8s_{16}s_{58} - 32968s_{5}s_{70} - 32968s_{8}s_{67} - 16s_{10}s_{65} - 32968s_{12}s_{63} - 8s_{15}s_{60} - 32968s_{5}s_{71} - 16s_{9}s_{67} + 8s_{17}s_{59} - 32968s_{5}s_{72} - 32968s_{9}s_{68} - 16s_{11}s_{66} - 32968s_{13}s_{64} - 16s_{14}s_{63} - 16s_{10}s_{68} - 16s_{11}s_{67} - 16s_{12}s_{66} - 16s_{13}s_{65} - 8s_{15}s_{63} + 8s_{18}s_{60} - 32968s_{6}s_{73} - 32968s_{10}s_{69} - 16s_{14}s_{65} - 16s_{15}s_{64} - 32968s_{6}s_{74} - 32968s_{7}s_{73} - 16s_{11}s_{69} - 16s_{14}s_{66} - 16s_{15}s_{65} - 32968s_{6}s_{75} - 8s_{15}s_{66} + 8s_{16}s_{65} - 32968s_{6}s_{76} - 32968s_{8}s_{74} - 16s_{9}s_{73} - 32968s_{12}s_{70} - 32968s_{6}s_{77} - 16s_{9}s_{74} - 16s_{10}s_{73} - 16s_{13}s_{70} + 8s_{17}s_{66} - 32968s_{6}s_{78} - 32968s_{9}s_{75} - 16s_{10}s_{74} - 32968s_{13}s_{71} - 32968s_{6}s_{79} - 16s_{10}s_{75} + 16s_{11}s_{74} - 16s_{14}s_{71} - 16s_{15}s_{70} - 32968s_{10}s_{76} - 16s_{11}s_{75} - 16s_{13}s_{73} - 32968s_{14}s_{72} + 8s_{28}s_{58} - 32968s_{7}s_{80} - 16s_{11}s_{76} - 16s_{14}s_{73} - 16s_{15}s_{72} - 32968s_{7}s_{81} - 32968s_{8}s_{80} - 16s_{15}s_{73} + 8s_{29}s_{59} - 32968s_{7}s_{82} - 16s_{9}s_{80} - 32968s_{12}s_{77} + 8s_{16}s_{73} - 32968s_{7}s_{83} - 32968s_{9}s_{81} - 16s_{13}s_{77} - 32968s_{7}s_{84} - 16s_{10}s_{81} - 8s_{11}s_{80} - 32968s_{13}s_{78} - 16s_{14}s_{77} - 32968s_{7}s_{85} - 32968s_{10}s_{82} - 16s_{14}s_{78} + 16s_{15}s_{77} - 16s_{11}s_{82} - 32968s_{14}s_{79} - 16s_{15}s_{78} + 8s_{28}s_{65} - 32968s_{8}s_{86} - 16s_{15}s_{79} - 32968s_{8}s_{87} - 32968s_{9}s_{86} - 32968s_{12}s_{83} + 8s_{29}s_{66} - 32968s_{8}s_{88} - 16s_{13}s_{83} - 32968s_{8}s_{89} - 32968s_{10}s_{87} - 32s_{11}s_{86} - 32968s_{13}s_{84} + 8s_{39}s_{58} - 32968s_{8}s_{90} - 32s_{11}s_{87} - 16s_{14}s_{84} - 8s_{15}s_{83} - 32968s_{11}s_{88} - 32s_{12}s_{87} - 32s_{13}s_{86} - 32968s_{14}s_{85} - 32968s_{9}s_{91} - 16s_{12}s_{88} - 32s_{14}s_{86} - 16s_{15}s_{85} - 32968s_{9}s_{92} - 32968s_{10}s_{91} - 32968s_{12}s_{89} - 32s_{13}s_{88} - 32s_{14}s_{87} - 32s_{15}s_{86} + 8s_{28}s_{73} - 32968s_{9}s_{93} + 16s_{14}s_{88} - 16s_{15}s_{87} + 16s_{16}s_{86} - 32968s_{9}s_{94} - 32968s_{11}s_{92} - 32968s_{13}s_{90} - 32s_{14}s_{89} - 16s_{15}s_{88} - 32968s_{9}s_{95} - 32s_{12}s_{92} - 32s_{13}s_{91} - 16s_{15}s_{89} + 16s_{17}s_{87} + 8s_{39}s_{65} - 32968s_{12}s_{93} - 32s_{14}s_{91} - 32s_{15}s_{90} - 32968s_{10}s_{96} - 32s_{13}s_{93} - 32s_{14}s_{92} - 32s_{15}s_{91} + 16s_{18}s_{88} - 32968s_{10}s_{97} - 32968s_{11}s_{96} - 32968s_{13}s_{94} - 16s_{15}s_{92} + 16s_{16}s_{91} - 32968s_{10}s_{98} - 32s_{14}s_{94} - 32968s_{10}s_{99} - 32968s_{12}s_{97} - 32s_{13}s_{96} - 32968s_{14}s_{95} + 16s_{17}s_{92} - 32s_{13}s_{97} - 32s_{14}s_{96} - 32s_{15}s_{95} - 32968s_{11}s_{100} - 32968s_{13}s_{98} - 32s_{15}s_{96} - 32968s_{11}s_{101} - 32968s_{12}s_{100} - 32s_{14}s_{98} + 16s_{16}s_{96} + 8s_{39}s_{73} - 32968s_{14}s_{99} - 32968s_{12}s_{102} - 32968s_{13}s_{101} - 32s_{14}s_{100} - 32s_{15}s_{99} + 16s_{28}s_{86} - 32968s_{12}s_{103} - 32968s_{13}s_{102} - 16s_{15}s_{100} - 32968s_{12}s_{104} - 32s_{15}s_{101} + 16s_{29}s_{87} - 32968s_{14}s_{103} - 96s_{15}s_{102} - 32968s_{13}s_{105} + 16s_{16}s_{102} - 32968s_{13}s_{106} - 32968s_{14}s_{105} - 32968s_{15}s_{104} + 16s_{28}s_{91} - 64s_{15}s_{105} + 16s_{17}s_{103} - 32968s_{14}s_{107} - 32968s_{15}s_{106} + 16s_{16}s_{105} + 16s_{29}s_{92} - 32968s_{15}s_{107} + 16s_{18}s_{104} + 16s_{16}s_{107} + 16s_{17}s_{106} + 16s_{28}s_{96} + 16s_{39}s_{86} + 16s_{28}s_{102} + 16s_{39}s_{91} + 16s_{58}s_{73} + 16s_{29}s_{103} + 16s_{28}s_{105} + 16s_{28}s_{107} + 16s_{29}s_{106} + 16s_{39}s_{96} + 16s_{39}s_{102} + 16s_{39}s_{105} + 16s_{58}s_{86} + 16s_{39}s_{107} + 16s_{59}s_{87} + 16s_{60}s_{88} + 16s_{58}s_{91} + 16s_{59}s_{92} + 16s_{65}s_{86} + 16s_{66}s_{87} + 16s_{58}s_{96} + 16s_{65}s_{91} + 16s_{66}s_{92} + 16s_{73}s_{86} + 16s_{58}s_{102} + 16s_{65}s_{96} + 16s_{59}s_{103} + 16s_{58}s_{105} + 16s_{60}s_{104} + 16s_{73}s_{91} + 16s_{58}s_{107} + 16s_{59}s_{106} + 16s_{65}s_{102} + 16s_{66}s_{103} + 16s_{73}s_{96} + 16s_{65}s_{105} + 16s_{65}s_{107} + 16s_{66}s_{106} + 16s_{73}s_{102} + 16s_{73}s_{105} + 16s_{73}s_{107} + 64s_{86}s_{96} + 32s_{86}s_{102} + 32s_{87}s_{103} + 32s_{86}s_{105} + 32s_{88}s_{104} + 32s_{86}s_{107} + 32s_{87}s_{106} + 32s_{91}s_{102} + 32s_{92}s_{103} + 32s_{91}s_{105} + 32s_{91}s_{107} + 32s_{92}s_{106} + 32s_{96}s_{102} + 32s_{96}s_{105} + 32s_{96}s_{107} + 64s_{102}s_{107} + 4551232$

\subsection{A complete expression of $\hat{H}_2\left(\hat{\sigma}^z\right)$ in 92-order H matrix using extended-Turyn based method}

$\hat{H}_2\left(\hat{\sigma}^z\right)=197860\hat{\sigma}^z_{0} + 197832\hat{\sigma}^z_{1} + 197872\hat{\sigma}^z_{2} + 197880\hat{\sigma}^z_{3} + 181374\hat{\sigma}^z_{4} + 214362\hat{\sigma}^z_{5} + 214402\hat{\sigma}^z_{6} + 148454\hat{\sigma}^z_{7} + 214436\hat{\sigma}^z_{8} + 231000\hat{\sigma}^z_{9} + 214592\hat{\sigma}^z_{10} + 82772\hat{\sigma}^z_{11} + 247556\hat{\sigma}^z_{12} + 247772\hat{\sigma}^z_{13} + 198456\hat{\sigma}^z_{14} + 50116\hat{\sigma}^z_{15} - 33004\hat{\sigma}^z_{16} - 33028\hat{\sigma}^z_{17} - 33012\hat{\sigma}^z_{18} - 32976\hat{\sigma}^z_{19} - 32972\hat{\sigma}^z_{20} - 32988\hat{\sigma}^z_{21} - 32952\hat{\sigma}^z_{22} - 32976\hat{\sigma}^z_{23} - 32992\hat{\sigma}^z_{24} - 32976\hat{\sigma}^z_{25} - 32984\hat{\sigma}^z_{26} - 33016\hat{\sigma}^z_{27} - 33004\hat{\sigma}^z_{28} - 33036\hat{\sigma}^z_{29} - 32964\hat{\sigma}^z_{30} - 32988\hat{\sigma}^z_{31} - 32972\hat{\sigma}^z_{32} - 32976\hat{\sigma}^z_{33} - 32960\hat{\sigma}^z_{34} - 32976\hat{\sigma}^z_{35} - 32960\hat{\sigma}^z_{36} - 33000\hat{\sigma}^z_{37} - 32992\hat{\sigma}^z_{38} - 33028\hat{\sigma}^z_{39} - 32972\hat{\sigma}^z_{40} - 32964\hat{\sigma}^z_{41} - 32996\hat{\sigma}^z_{42} - 32960\hat{\sigma}^z_{43} - 32976\hat{\sigma}^z_{44} - 32992\hat{\sigma}^z_{45} - 32976\hat{\sigma}^z_{46} - 32968\hat{\sigma}^z_{47} - 33024\hat{\sigma}^z_{48} - 32960\hat{\sigma}^z_{49} - 32980\hat{\sigma}^z_{50} - 32972\hat{\sigma}^z_{51} - 32952\hat{\sigma}^z_{52} - 32992\hat{\sigma}^z_{53} - 32992\hat{\sigma}^z_{54} - 32960\hat{\sigma}^z_{55} - 33000\hat{\sigma}^z_{56} - 32984\hat{\sigma}^z_{57} - 33028\hat{\sigma}^z_{58} - 33036\hat{\sigma}^z_{59} - 32996\hat{\sigma}^z_{60} - 32968\hat{\sigma}^z_{61} - 32960\hat{\sigma}^z_{62} - 32992\hat{\sigma}^z_{63} - 32968\hat{\sigma}^z_{64} - 32996\hat{\sigma}^z_{65} - 33076\hat{\sigma}^z_{66} - 32960\hat{\sigma}^z_{67} - 32976\hat{\sigma}^z_{68} - 32976\hat{\sigma}^z_{69} - 32944\hat{\sigma}^z_{70} - 33016\hat{\sigma}^z_{71} - 32960\hat{\sigma}^z_{72} - 33036\hat{\sigma}^z_{73} - 32960\hat{\sigma}^z_{74} - 32976\hat{\sigma}^z_{75} - 32992\hat{\sigma}^z_{76} - 32976\hat{\sigma}^z_{77} - 32936\hat{\sigma}^z_{78} - 33040\hat{\sigma}^z_{79} - 32968\hat{\sigma}^z_{80} - 32976\hat{\sigma}^z_{81} - 32976\hat{\sigma}^z_{82} - 32944\hat{\sigma}^z_{83} - 33000\hat{\sigma}^z_{84} - 32968\hat{\sigma}^z_{85} - 33112\hat{\sigma}^z_{86} - 33064\hat{\sigma}^z_{87} - 33144\hat{\sigma}^z_{88} - 32984\hat{\sigma}^z_{89} - 32968\hat{\sigma}^z_{90} - 33112\hat{\sigma}^z_{91} - 33096\hat{\sigma}^z_{92} - 32952\hat{\sigma}^z_{93} - 33000\hat{\sigma}^z_{94} - 33016\hat{\sigma}^z_{95} - 33144\hat{\sigma}^z_{96} - 32952\hat{\sigma}^z_{97} - 33000\hat{\sigma}^z_{98} - 33016\hat{\sigma}^z_{99} - 32984\hat{\sigma}^z_{100} - 32968\hat{\sigma}^z_{101} - 33192\hat{\sigma}^z_{102} - 33192\hat{\sigma}^z_{103} - 33112\hat{\sigma}^z_{104} - 33096\hat{\sigma}^z_{105} - 33320\hat{\sigma}^z_{106} - 33256\hat{\sigma}^z_{107} + 16484\hat{\sigma}^z_{0}\hat{\sigma}^z_{1} + 16484\hat{\sigma}^z_{0}\hat{\sigma}^z_{2} + 16484\hat{\sigma}^z_{0}\hat{\sigma}^z_{3} + 16484\hat{\sigma}^z_{1}\hat{\sigma}^z_{2} + 16484\hat{\sigma}^z_{0}\hat{\sigma}^z_{4} + 16484\hat{\sigma}^z_{1}\hat{\sigma}^z_{3} + 16484\hat{\sigma}^z_{0}\hat{\sigma}^z_{5} + 16484\hat{\sigma}^z_{1}\hat{\sigma}^z_{4} + 16484\hat{\sigma}^z_{2}\hat{\sigma}^z_{3} + 16484\hat{\sigma}^z_{0}\hat{\sigma}^z_{6} + 16484\hat{\sigma}^z_{1}\hat{\sigma}^z_{5} + 16484\hat{\sigma}^z_{2}\hat{\sigma}^z_{4} + 6\hat{\sigma}^z_{0}\hat{\sigma}^z_{7} + 16484\hat{\sigma}^z_{1}\hat{\sigma}^z_{6} + 16484\hat{\sigma}^z_{2}\hat{\sigma}^z_{5} + 16484\hat{\sigma}^z_{3}\hat{\sigma}^z_{4} + 16484\hat{\sigma}^z_{0}\hat{\sigma}^z_{8} + 14\hat{\sigma}^z_{1}\hat{\sigma}^z_{7} + 16484\hat{\sigma}^z_{2}\hat{\sigma}^z_{6} + 16484\hat{\sigma}^z_{3}\hat{\sigma}^z_{5} + 16484\hat{\sigma}^z_{0}\hat{\sigma}^z_{9} + 16484\hat{\sigma}^z_{1}\hat{\sigma}^z_{8} + 6\hat{\sigma}^z_{2}\hat{\sigma}^z_{7} + 16484\hat{\sigma}^z_{3}\hat{\sigma}^z_{6} + 16484\hat{\sigma}^z_{4}\hat{\sigma}^z_{5} + 16484\hat{\sigma}^z_{0}\hat{\sigma}^z_{10} + 16484\hat{\sigma}^z_{1}\hat{\sigma}^z_{9} + 16484\hat{\sigma}^z_{2}\hat{\sigma}^z_{8} + 10\hat{\sigma}^z_{3}\hat{\sigma}^z_{7} + 16484\hat{\sigma}^z_{4}\hat{\sigma}^z_{6} + 12\hat{\sigma}^z_{0}\hat{\sigma}^z_{11} + 16484\hat{\sigma}^z_{1}\hat{\sigma}^z_{10} + 16484\hat{\sigma}^z_{2}\hat{\sigma}^z_{9} + 16484\hat{\sigma}^z_{3}\hat{\sigma}^z_{8} + 16484\hat{\sigma}^z_{4}\hat{\sigma}^z_{7} + 16484\hat{\sigma}^z_{5}\hat{\sigma}^z_{6} + 16484\hat{\sigma}^z_{0}\hat{\sigma}^z_{12} + 28\hat{\sigma}^z_{1}\hat{\sigma}^z_{11} + 16484\hat{\sigma}^z_{2}\hat{\sigma}^z_{10} + 16484\hat{\sigma}^z_{3}\hat{\sigma}^z_{9} + 16484\hat{\sigma}^z_{4}\hat{\sigma}^z_{8} + 16484\hat{\sigma}^z_{5}\hat{\sigma}^z_{7} + 16484\hat{\sigma}^z_{0}\hat{\sigma}^z_{13} + 16484\hat{\sigma}^z_{1}\hat{\sigma}^z_{12} + 12\hat{\sigma}^z_{2}\hat{\sigma}^z_{11} + 16484\hat{\sigma}^z_{3}\hat{\sigma}^z_{10} + 16484\hat{\sigma}^z_{4}\hat{\sigma}^z_{9} + 16484\hat{\sigma}^z_{5}\hat{\sigma}^z_{8} + 16484\hat{\sigma}^z_{6}\hat{\sigma}^z_{7} + 16484\hat{\sigma}^z_{0}\hat{\sigma}^z_{14} + 16484\hat{\sigma}^z_{1}\hat{\sigma}^z_{13} + 16484\hat{\sigma}^z_{2}\hat{\sigma}^z_{12} + 4\hat{\sigma}^z_{3}\hat{\sigma}^z_{11} + 4\hat{\sigma}^z_{4}\hat{\sigma}^z_{10} + 16484\hat{\sigma}^z_{5}\hat{\sigma}^z_{9} + 16484\hat{\sigma}^z_{6}\hat{\sigma}^z_{8} + 20\hat{\sigma}^z_{0}\hat{\sigma}^z_{15} + 16484\hat{\sigma}^z_{1}\hat{\sigma}^z_{14} + 16484\hat{\sigma}^z_{2}\hat{\sigma}^z_{13} + 16484\hat{\sigma}^z_{3}\hat{\sigma}^z_{12} + 12\hat{\sigma}^z_{4}\hat{\sigma}^z_{11} + 16484\hat{\sigma}^z_{5}\hat{\sigma}^z_{10} + 16484\hat{\sigma}^z_{6}\hat{\sigma}^z_{9} + 16484\hat{\sigma}^z_{7}\hat{\sigma}^z_{8} - 32968\hat{\sigma}^z_{0}\hat{\sigma}^z_{16} + 36\hat{\sigma}^z_{1}\hat{\sigma}^z_{15} + 16484\hat{\sigma}^z_{2}\hat{\sigma}^z_{14} + 16484\hat{\sigma}^z_{3}\hat{\sigma}^z_{13} + 16484\hat{\sigma}^z_{4}\hat{\sigma}^z_{12} + 12\hat{\sigma}^z_{5}\hat{\sigma}^z_{11} + 16484\hat{\sigma}^z_{6}\hat{\sigma}^z_{10} + 16484\hat{\sigma}^z_{7}\hat{\sigma}^z_{9} - 32968\hat{\sigma}^z_{0}\hat{\sigma}^z_{17} - 32968\hat{\sigma}^z_{1}\hat{\sigma}^z_{16} + 20\hat{\sigma}^z_{2}\hat{\sigma}^z_{15} + 16484\hat{\sigma}^z_{3}\hat{\sigma}^z_{14} + 16484\hat{\sigma}^z_{4}\hat{\sigma}^z_{13} + 16484\hat{\sigma}^z_{5}\hat{\sigma}^z_{12} + 4\hat{\sigma}^z_{6}\hat{\sigma}^z_{11} + 16484\hat{\sigma}^z_{7}\hat{\sigma}^z_{10} + 16484\hat{\sigma}^z_{8}\hat{\sigma}^z_{9} - 32968\hat{\sigma}^z_{0}\hat{\sigma}^z_{18} - 16\hat{\sigma}^z_{2}\hat{\sigma}^z_{16} + 20\hat{\sigma}^z_{3}\hat{\sigma}^z_{15} + 16\hat{\sigma}^z_{4}\hat{\sigma}^z_{14} + 16484\hat{\sigma}^z_{5}\hat{\sigma}^z_{13} + 16484\hat{\sigma}^z_{6}\hat{\sigma}^z_{12} + 12\hat{\sigma}^z_{7}\hat{\sigma}^z_{11} + 16484\hat{\sigma}^z_{8}\hat{\sigma}^z_{10} - 32968\hat{\sigma}^z_{0}\hat{\sigma}^z_{19} - 32968\hat{\sigma}^z_{2}\hat{\sigma}^z_{17} - 24\hat{\sigma}^z_{3}\hat{\sigma}^z_{16} + 4\hat{\sigma}^z_{4}\hat{\sigma}^z_{15} + 16484\hat{\sigma}^z_{5}\hat{\sigma}^z_{14} + 16484\hat{\sigma}^z_{6}\hat{\sigma}^z_{13} + 16484\hat{\sigma}^z_{7}\hat{\sigma}^z_{12} + 16484\hat{\sigma}^z_{8}\hat{\sigma}^z_{11} + 16484\hat{\sigma}^z_{9}\hat{\sigma}^z_{10} - 32968\hat{\sigma}^z_{0}\hat{\sigma}^z_{20} - 8\hat{\sigma}^z_{3}\hat{\sigma}^z_{17} + 28\hat{\sigma}^z_{5}\hat{\sigma}^z_{15} + 16484\hat{\sigma}^z_{6}\hat{\sigma}^z_{14} + 16484\hat{\sigma}^z_{7}\hat{\sigma}^z_{13} + 16484\hat{\sigma}^z_{8}\hat{\sigma}^z_{12} + 16484\hat{\sigma}^z_{9}\hat{\sigma}^z_{11} - 32968\hat{\sigma}^z_{0}\hat{\sigma}^z_{21} - 32968\hat{\sigma}^z_{3}\hat{\sigma}^z_{18} - 8\hat{\sigma}^z_{4}\hat{\sigma}^z_{17} - 8\hat{\sigma}^z_{5}\hat{\sigma}^z_{16} + 12\hat{\sigma}^z_{6}\hat{\sigma}^z_{15} + 16484\hat{\sigma}^z_{7}\hat{\sigma}^z_{14} + 16484\hat{\sigma}^z_{8}\hat{\sigma}^z_{13} + 16484\hat{\sigma}^z_{9}\hat{\sigma}^z_{12} + 16484\hat{\sigma}^z_{10}\hat{\sigma}^z_{11} - 32968\hat{\sigma}^z_{0}\hat{\sigma}^z_{22} - 4\hat{\sigma}^z_{4}\hat{\sigma}^z_{18} - 8\hat{\sigma}^z_{6}\hat{\sigma}^z_{16} + 28\hat{\sigma}^z_{7}\hat{\sigma}^z_{15} + 16\hat{\sigma}^z_{8}\hat{\sigma}^z_{14} + 16484\hat{\sigma}^z_{9}\hat{\sigma}^z_{13} + 16484\hat{\sigma}^z_{10}\hat{\sigma}^z_{12} - 32968\hat{\sigma}^z_{0}\hat{\sigma}^z_{23} - 32968\hat{\sigma}^z_{4}\hat{\sigma}^z_{19} - 8\hat{\sigma}^z_{5}\hat{\sigma}^z_{18} - 8\hat{\sigma}^z_{6}\hat{\sigma}^z_{17} - 8\hat{\sigma}^z_{7}\hat{\sigma}^z_{16} + 40\hat{\sigma}^z_{8}\hat{\sigma}^z_{15} + 16484\hat{\sigma}^z_{9}\hat{\sigma}^z_{14} + 16484\hat{\sigma}^z_{10}\hat{\sigma}^z_{13} + 16484\hat{\sigma}^z_{11}\hat{\sigma}^z_{12} - 32968\hat{\sigma}^z_{0}\hat{\sigma}^z_{24} - 8\hat{\sigma}^z_{5}\hat{\sigma}^z_{19} + 8\hat{\sigma}^z_{6}\hat{\sigma}^z_{18} + 24\hat{\sigma}^z_{9}\hat{\sigma}^z_{15} + 16484\hat{\sigma}^z_{10}\hat{\sigma}^z_{14} + 16484\hat{\sigma}^z_{11}\hat{\sigma}^z_{13} - 32968\hat{\sigma}^z_{0}\hat{\sigma}^z_{25} - 32968\hat{\sigma}^z_{5}\hat{\sigma}^z_{20} - 8\hat{\sigma}^z_{6}\hat{\sigma}^z_{19} - 4\hat{\sigma}^z_{7}\hat{\sigma}^z_{18} - 16\hat{\sigma}^z_{8}\hat{\sigma}^z_{17} - 16\hat{\sigma}^z_{9}\hat{\sigma}^z_{16} + 24\hat{\sigma}^z_{10}\hat{\sigma}^z_{15} + 16\hat{\sigma}^z_{11}\hat{\sigma}^z_{14} + 16484\hat{\sigma}^z_{12}\hat{\sigma}^z_{13} - 32968\hat{\sigma}^z_{0}\hat{\sigma}^z_{26} - 8\hat{\sigma}^z_{6}\hat{\sigma}^z_{20} - 4\hat{\sigma}^z_{7}\hat{\sigma}^z_{19} - 8\hat{\sigma}^z_{8}\hat{\sigma}^z_{18} - 16\hat{\sigma}^z_{10}\hat{\sigma}^z_{16} + 40\hat{\sigma}^z_{11}\hat{\sigma}^z_{15} + 16484\hat{\sigma}^z_{12}\hat{\sigma}^z_{14} - 32968\hat{\sigma}^z_{0}\hat{\sigma}^z_{27} - 32968\hat{\sigma}^z_{6}\hat{\sigma}^z_{21} - 8\hat{\sigma}^z_{7}\hat{\sigma}^z_{20} + 16484\hat{\sigma}^z_{12}\hat{\sigma}^z_{15} + 16484\hat{\sigma}^z_{13}\hat{\sigma}^z_{14} - 8\hat{\sigma}^z_{7}\hat{\sigma}^z_{21} - 16\hat{\sigma}^z_{11}\hat{\sigma}^z_{17} + 16484\hat{\sigma}^z_{13}\hat{\sigma}^z_{15} - 32968\hat{\sigma}^z_{1}\hat{\sigma}^z_{28} - 8\hat{\sigma}^z_{11}\hat{\sigma}^z_{18} - 16\hat{\sigma}^z_{12}\hat{\sigma}^z_{17} - 16\hat{\sigma}^z_{13}\hat{\sigma}^z_{16} + 16484\hat{\sigma}^z_{14}\hat{\sigma}^z_{15} - 32968\hat{\sigma}^z_{1}\hat{\sigma}^z_{29} - 32968\hat{\sigma}^z_{2}\hat{\sigma}^z_{28} - 32968\hat{\sigma}^z_{8}\hat{\sigma}^z_{22} - 8\hat{\sigma}^z_{12}\hat{\sigma}^z_{18} - 16\hat{\sigma}^z_{14}\hat{\sigma}^z_{16} - 32968\hat{\sigma}^z_{1}\hat{\sigma}^z_{30} - 16\hat{\sigma}^z_{3}\hat{\sigma}^z_{28} - 16\hat{\sigma}^z_{9}\hat{\sigma}^z_{22} - 16\hat{\sigma}^z_{13}\hat{\sigma}^z_{18} - 16\hat{\sigma}^z_{14}\hat{\sigma}^z_{17} - 16\hat{\sigma}^z_{15}\hat{\sigma}^z_{16} - 32968\hat{\sigma}^z_{1}\hat{\sigma}^z_{31} - 32968\hat{\sigma}^z_{3}\hat{\sigma}^z_{29} - 32968\hat{\sigma}^z_{9}\hat{\sigma}^z_{23} - 16\hat{\sigma}^z_{10}\hat{\sigma}^z_{22} + 8\hat{\sigma}^z_{14}\hat{\sigma}^z_{18} - 8\hat{\sigma}^z_{15}\hat{\sigma}^z_{17} - 32968\hat{\sigma}^z_{1}\hat{\sigma}^z_{32} - 8\hat{\sigma}^z_{4}\hat{\sigma}^z_{29} - 8\hat{\sigma}^z_{5}\hat{\sigma}^z_{28} - 16\hat{\sigma}^z_{10}\hat{\sigma}^z_{23} - 8\hat{\sigma}^z_{11}\hat{\sigma}^z_{22} - 8\hat{\sigma}^z_{15}\hat{\sigma}^z_{18} - 32968\hat{\sigma}^z_{1}\hat{\sigma}^z_{33} - 32968\hat{\sigma}^z_{4}\hat{\sigma}^z_{30} - 8\hat{\sigma}^z_{6}\hat{\sigma}^z_{28} - 32968\hat{\sigma}^z_{10}\hat{\sigma}^z_{24} - 16\hat{\sigma}^z_{11}\hat{\sigma}^z_{23} - 32968\hat{\sigma}^z_{1}\hat{\sigma}^z_{34} - 8\hat{\sigma}^z_{5}\hat{\sigma}^z_{30} - 8\hat{\sigma}^z_{6}\hat{\sigma}^z_{29} - 8\hat{\sigma}^z_{7}\hat{\sigma}^z_{28} - 16\hat{\sigma}^z_{11}\hat{\sigma}^z_{24} - 32968\hat{\sigma}^z_{1}\hat{\sigma}^z_{35} - 32968\hat{\sigma}^z_{5}\hat{\sigma}^z_{31} - 8\hat{\sigma}^z_{6}\hat{\sigma}^z_{30} - 32968\hat{\sigma}^z_{1}\hat{\sigma}^z_{36} - 8\hat{\sigma}^z_{6}\hat{\sigma}^z_{31} - 8\hat{\sigma}^z_{7}\hat{\sigma}^z_{30} - 16\hat{\sigma}^z_{8}\hat{\sigma}^z_{29} - 16\hat{\sigma}^z_{9}\hat{\sigma}^z_{28} - 32968\hat{\sigma}^z_{12}\hat{\sigma}^z_{25} - 32968\hat{\sigma}^z_{1}\hat{\sigma}^z_{37} - 32968\hat{\sigma}^z_{6}\hat{\sigma}^z_{32} - 8\hat{\sigma}^z_{7}\hat{\sigma}^z_{31} - 16\hat{\sigma}^z_{10}\hat{\sigma}^z_{28} - 16\hat{\sigma}^z_{13}\hat{\sigma}^z_{25} - 32968\hat{\sigma}^z_{1}\hat{\sigma}^z_{38} - 8\hat{\sigma}^z_{7}\hat{\sigma}^z_{32} - 32968\hat{\sigma}^z_{13}\hat{\sigma}^z_{26} - 16\hat{\sigma}^z_{14}\hat{\sigma}^z_{25} - 16\hat{\sigma}^z_{11}\hat{\sigma}^z_{29} - 16\hat{\sigma}^z_{14}\hat{\sigma}^z_{26} - 8\hat{\sigma}^z_{15}\hat{\sigma}^z_{25} - 32968\hat{\sigma}^z_{2}\hat{\sigma}^z_{39} - 32968\hat{\sigma}^z_{8}\hat{\sigma}^z_{33} - 16\hat{\sigma}^z_{12}\hat{\sigma}^z_{29} - 16\hat{\sigma}^z_{13}\hat{\sigma}^z_{28} - 32968\hat{\sigma}^z_{14}\hat{\sigma}^z_{27} - 16\hat{\sigma}^z_{15}\hat{\sigma}^z_{26} - 32968\hat{\sigma}^z_{2}\hat{\sigma}^z_{40} - 32968\hat{\sigma}^z_{3}\hat{\sigma}^z_{39} - 16\hat{\sigma}^z_{9}\hat{\sigma}^z_{33} - 16\hat{\sigma}^z_{14}\hat{\sigma}^z_{28} - 16\hat{\sigma}^z_{15}\hat{\sigma}^z_{27} - 32968\hat{\sigma}^z_{2}\hat{\sigma}^z_{41} - 32968\hat{\sigma}^z_{9}\hat{\sigma}^z_{34} - 16\hat{\sigma}^z_{10}\hat{\sigma}^z_{33} - 16\hat{\sigma}^z_{14}\hat{\sigma}^z_{29} - 16\hat{\sigma}^z_{15}\hat{\sigma}^z_{28} - 32968\hat{\sigma}^z_{2}\hat{\sigma}^z_{42} - 32968\hat{\sigma}^z_{4}\hat{\sigma}^z_{40} - 8\hat{\sigma}^z_{5}\hat{\sigma}^z_{39} - 16\hat{\sigma}^z_{10}\hat{\sigma}^z_{34} - 16\hat{\sigma}^z_{11}\hat{\sigma}^z_{33} - 8\hat{\sigma}^z_{15}\hat{\sigma}^z_{29} - 32968\hat{\sigma}^z_{2}\hat{\sigma}^z_{43} - 8\hat{\sigma}^z_{5}\hat{\sigma}^z_{40} - 8\hat{\sigma}^z_{6}\hat{\sigma}^z_{39} - 32968\hat{\sigma}^z_{10}\hat{\sigma}^z_{35} - 16\hat{\sigma}^z_{11}\hat{\sigma}^z_{34} - 32968\hat{\sigma}^z_{2}\hat{\sigma}^z_{44} - 32968\hat{\sigma}^z_{5}\hat{\sigma}^z_{41} - 8\hat{\sigma}^z_{6}\hat{\sigma}^z_{40} - 8\hat{\sigma}^z_{7}\hat{\sigma}^z_{39} - 16\hat{\sigma}^z_{11}\hat{\sigma}^z_{35} - 32968\hat{\sigma}^z_{2}\hat{\sigma}^z_{45} - 8\hat{\sigma}^z_{6}\hat{\sigma}^z_{41} - 32968\hat{\sigma}^z_{2}\hat{\sigma}^z_{46} - 32968\hat{\sigma}^z_{6}\hat{\sigma}^z_{42} - 8\hat{\sigma}^z_{7}\hat{\sigma}^z_{41} - 16\hat{\sigma}^z_{9}\hat{\sigma}^z_{39} - 32968\hat{\sigma}^z_{12}\hat{\sigma}^z_{36} - 32968\hat{\sigma}^z_{2}\hat{\sigma}^z_{47} - 8\hat{\sigma}^z_{7}\hat{\sigma}^z_{42} - 16\hat{\sigma}^z_{10}\hat{\sigma}^z_{39} - 16\hat{\sigma}^z_{13}\hat{\sigma}^z_{36} - 32968\hat{\sigma}^z_{2}\hat{\sigma}^z_{48} - 32968\hat{\sigma}^z_{13}\hat{\sigma}^z_{37} - 16\hat{\sigma}^z_{14}\hat{\sigma}^z_{36} - 32968\hat{\sigma}^z_{8}\hat{\sigma}^z_{43} - 16\hat{\sigma}^z_{14}\hat{\sigma}^z_{37} - 16\hat{\sigma}^z_{15}\hat{\sigma}^z_{36} - 32968\hat{\sigma}^z_{3}\hat{\sigma}^z_{49} - 16\hat{\sigma}^z_{9}\hat{\sigma}^z_{43} - 16\hat{\sigma}^z_{13}\hat{\sigma}^z_{39} - 32968\hat{\sigma}^z_{14}\hat{\sigma}^z_{38} - 16\hat{\sigma}^z_{15}\hat{\sigma}^z_{37} - 32968\hat{\sigma}^z_{3}\hat{\sigma}^z_{50} - 32968\hat{\sigma}^z_{4}\hat{\sigma}^z_{49} - 32968\hat{\sigma}^z_{9}\hat{\sigma}^z_{44} - 16\hat{\sigma}^z_{10}\hat{\sigma}^z_{43} - 16\hat{\sigma}^z_{14}\hat{\sigma}^z_{39} - 16\hat{\sigma}^z_{15}\hat{\sigma}^z_{38} - 32968\hat{\sigma}^z_{3}\hat{\sigma}^z_{51} - 8\hat{\sigma}^z_{5}\hat{\sigma}^z_{49} - 16\hat{\sigma}^z_{10}\hat{\sigma}^z_{44} - 16\hat{\sigma}^z_{15}\hat{\sigma}^z_{39} - 32968\hat{\sigma}^z_{3}\hat{\sigma}^z_{52} - 32968\hat{\sigma}^z_{5}\hat{\sigma}^z_{50} - 32968\hat{\sigma}^z_{10}\hat{\sigma}^z_{45} - 16\hat{\sigma}^z_{11}\hat{\sigma}^z_{44} + 16\hat{\sigma}^z_{16}\hat{\sigma}^z_{39} - 32968\hat{\sigma}^z_{3}\hat{\sigma}^z_{53} - 8\hat{\sigma}^z_{6}\hat{\sigma}^z_{50} - 4\hat{\sigma}^z_{7}\hat{\sigma}^z_{49} - 16\hat{\sigma}^z_{11}\hat{\sigma}^z_{45} - 32968\hat{\sigma}^z_{3}\hat{\sigma}^z_{54} - 32968\hat{\sigma}^z_{6}\hat{\sigma}^z_{51} - 32968\hat{\sigma}^z_{3}\hat{\sigma}^z_{55} - 8\hat{\sigma}^z_{7}\hat{\sigma}^z_{51} - 32968\hat{\sigma}^z_{12}\hat{\sigma}^z_{46} - 32968\hat{\sigma}^z_{3}\hat{\sigma}^z_{56} - 16\hat{\sigma}^z_{13}\hat{\sigma}^z_{46} - 32968\hat{\sigma}^z_{3}\hat{\sigma}^z_{57} - 32968\hat{\sigma}^z_{8}\hat{\sigma}^z_{52} - 32968\hat{\sigma}^z_{13}\hat{\sigma}^z_{47} - 16\hat{\sigma}^z_{14}\hat{\sigma}^z_{46} - 16\hat{\sigma}^z_{9}\hat{\sigma}^z_{52} - 16\hat{\sigma}^z_{14}\hat{\sigma}^z_{47} - 32968\hat{\sigma}^z_{4}\hat{\sigma}^z_{58} - 32968\hat{\sigma}^z_{9}\hat{\sigma}^z_{53} - 32968\hat{\sigma}^z_{14}\hat{\sigma}^z_{48} - 16\hat{\sigma}^z_{15}\hat{\sigma}^z_{47} - 32968\hat{\sigma}^z_{4}\hat{\sigma}^z_{59} - 32968\hat{\sigma}^z_{5}\hat{\sigma}^z_{58} - 16\hat{\sigma}^z_{10}\hat{\sigma}^z_{53} - 8\hat{\sigma}^z_{11}\hat{\sigma}^z_{52} - 16\hat{\sigma}^z_{15}\hat{\sigma}^z_{48} - 32968\hat{\sigma}^z_{4}\hat{\sigma}^z_{60} - 32968\hat{\sigma}^z_{10}\hat{\sigma}^z_{54} - 32968\hat{\sigma}^z_{4}\hat{\sigma}^z_{61} - 32968\hat{\sigma}^z_{6}\hat{\sigma}^z_{59} - 24\hat{\sigma}^z_{7}\hat{\sigma}^z_{58} - 16\hat{\sigma}^z_{11}\hat{\sigma}^z_{54} - 32968\hat{\sigma}^z_{4}\hat{\sigma}^z_{62} + 8\hat{\sigma}^z_{7}\hat{\sigma}^z_{59} - 32968\hat{\sigma}^z_{4}\hat{\sigma}^z_{63} - 32968\hat{\sigma}^z_{7}\hat{\sigma}^z_{60} - 16\hat{\sigma}^z_{8}\hat{\sigma}^z_{59} - 16\hat{\sigma}^z_{9}\hat{\sigma}^z_{58} - 32968\hat{\sigma}^z_{12}\hat{\sigma}^z_{55} - 32968\hat{\sigma}^z_{4}\hat{\sigma}^z_{64} - 8\hat{\sigma}^z_{8}\hat{\sigma}^z_{60} - 16\hat{\sigma}^z_{10}\hat{\sigma}^z_{58} - 16\hat{\sigma}^z_{13}\hat{\sigma}^z_{55} - 32968\hat{\sigma}^z_{8}\hat{\sigma}^z_{61} - 32968\hat{\sigma}^z_{13}\hat{\sigma}^z_{56} - 32968\hat{\sigma}^z_{5}\hat{\sigma}^z_{65} - 16\hat{\sigma}^z_{11}\hat{\sigma}^z_{59} - 16\hat{\sigma}^z_{14}\hat{\sigma}^z_{56} - 8\hat{\sigma}^z_{15}\hat{\sigma}^z_{55} - 32968\hat{\sigma}^z_{5}\hat{\sigma}^z_{66} - 32968\hat{\sigma}^z_{6}\hat{\sigma}^z_{65} - 32968\hat{\sigma}^z_{9}\hat{\sigma}^z_{62} - 16\hat{\sigma}^z_{10}\hat{\sigma}^z_{61} - 8\hat{\sigma}^z_{11}\hat{\sigma}^z_{60} - 16\hat{\sigma}^z_{12}\hat{\sigma}^z_{59} - 16\hat{\sigma}^z_{13}\hat{\sigma}^z_{58} - 32968\hat{\sigma}^z_{14}\hat{\sigma}^z_{57} - 32968\hat{\sigma}^z_{5}\hat{\sigma}^z_{67} - 16\hat{\sigma}^z_{7}\hat{\sigma}^z_{65} - 8\hat{\sigma}^z_{11}\hat{\sigma}^z_{61} - 8\hat{\sigma}^z_{12}\hat{\sigma}^z_{60} - 16\hat{\sigma}^z_{14}\hat{\sigma}^z_{58} - 16\hat{\sigma}^z_{15}\hat{\sigma}^z_{57} - 32968\hat{\sigma}^z_{5}\hat{\sigma}^z_{68} - 32968\hat{\sigma}^z_{7}\hat{\sigma}^z_{66} - 16\hat{\sigma}^z_{11}\hat{\sigma}^z_{62} - 16\hat{\sigma}^z_{13}\hat{\sigma}^z_{60} - 16\hat{\sigma}^z_{14}\hat{\sigma}^z_{59} - 16\hat{\sigma}^z_{15}\hat{\sigma}^z_{58} - 32968\hat{\sigma}^z_{5}\hat{\sigma}^z_{69} - 16\hat{\sigma}^z_{8}\hat{\sigma}^z_{66} - 16\hat{\sigma}^z_{9}\hat{\sigma}^z_{65} + 8\hat{\sigma}^z_{14}\hat{\sigma}^z_{60} - 8\hat{\sigma}^z_{15}\hat{\sigma}^z_{59} + 8\hat{\sigma}^z_{16}\hat{\sigma}^z_{58} - 32968\hat{\sigma}^z_{5}\hat{\sigma}^z_{70} - 32968\hat{\sigma}^z_{8}\hat{\sigma}^z_{67} - 16\hat{\sigma}^z_{10}\hat{\sigma}^z_{65} - 32968\hat{\sigma}^z_{12}\hat{\sigma}^z_{63} - 8\hat{\sigma}^z_{15}\hat{\sigma}^z_{60} - 32968\hat{\sigma}^z_{5}\hat{\sigma}^z_{71} - 16\hat{\sigma}^z_{9}\hat{\sigma}^z_{67} + 8\hat{\sigma}^z_{17}\hat{\sigma}^z_{59} - 32968\hat{\sigma}^z_{5}\hat{\sigma}^z_{72} - 32968\hat{\sigma}^z_{9}\hat{\sigma}^z_{68} - 16\hat{\sigma}^z_{11}\hat{\sigma}^z_{66} - 32968\hat{\sigma}^z_{13}\hat{\sigma}^z_{64} - 16\hat{\sigma}^z_{14}\hat{\sigma}^z_{63} - 16\hat{\sigma}^z_{10}\hat{\sigma}^z_{68} - 16\hat{\sigma}^z_{11}\hat{\sigma}^z_{67} - 16\hat{\sigma}^z_{12}\hat{\sigma}^z_{66} - 16\hat{\sigma}^z_{13}\hat{\sigma}^z_{65} - 8\hat{\sigma}^z_{15}\hat{\sigma}^z_{63} + 8\hat{\sigma}^z_{18}\hat{\sigma}^z_{60} - 32968\hat{\sigma}^z_{6}\hat{\sigma}^z_{73} - 32968\hat{\sigma}^z_{10}\hat{\sigma}^z_{69} - 16\hat{\sigma}^z_{14}\hat{\sigma}^z_{65} - 16\hat{\sigma}^z_{15}\hat{\sigma}^z_{64} - 32968\hat{\sigma}^z_{6}\hat{\sigma}^z_{74} - 32968\hat{\sigma}^z_{7}\hat{\sigma}^z_{73} - 16\hat{\sigma}^z_{11}\hat{\sigma}^z_{69} - 16\hat{\sigma}^z_{14}\hat{\sigma}^z_{66} - 16\hat{\sigma}^z_{15}\hat{\sigma}^z_{65} - 32968\hat{\sigma}^z_{6}\hat{\sigma}^z_{75} - 8\hat{\sigma}^z_{15}\hat{\sigma}^z_{66} + 8\hat{\sigma}^z_{16}\hat{\sigma}^z_{65} - 32968\hat{\sigma}^z_{6}\hat{\sigma}^z_{76} - 32968\hat{\sigma}^z_{8}\hat{\sigma}^z_{74} - 16\hat{\sigma}^z_{9}\hat{\sigma}^z_{73} - 32968\hat{\sigma}^z_{12}\hat{\sigma}^z_{70} - 32968\hat{\sigma}^z_{6}\hat{\sigma}^z_{77} - 16\hat{\sigma}^z_{9}\hat{\sigma}^z_{74} - 16\hat{\sigma}^z_{10}\hat{\sigma}^z_{73} - 16\hat{\sigma}^z_{13}\hat{\sigma}^z_{70} + 8\hat{\sigma}^z_{17}\hat{\sigma}^z_{66} - 32968\hat{\sigma}^z_{6}\hat{\sigma}^z_{78} - 32968\hat{\sigma}^z_{9}\hat{\sigma}^z_{75} - 16\hat{\sigma}^z_{10}\hat{\sigma}^z_{74} - 32968\hat{\sigma}^z_{13}\hat{\sigma}^z_{71} - 32968\hat{\sigma}^z_{6}\hat{\sigma}^z_{79} - 16\hat{\sigma}^z_{10}\hat{\sigma}^z_{75} + 16\hat{\sigma}^z_{11}\hat{\sigma}^z_{74} - 16\hat{\sigma}^z_{14}\hat{\sigma}^z_{71} - 16\hat{\sigma}^z_{15}\hat{\sigma}^z_{70} - 32968\hat{\sigma}^z_{10}\hat{\sigma}^z_{76} - 16\hat{\sigma}^z_{11}\hat{\sigma}^z_{75} - 16\hat{\sigma}^z_{13}\hat{\sigma}^z_{73} - 32968\hat{\sigma}^z_{14}\hat{\sigma}^z_{72} + 8\hat{\sigma}^z_{28}\hat{\sigma}^z_{58} - 32968\hat{\sigma}^z_{7}\hat{\sigma}^z_{80} - 16\hat{\sigma}^z_{11}\hat{\sigma}^z_{76} - 16\hat{\sigma}^z_{14}\hat{\sigma}^z_{73} - 16\hat{\sigma}^z_{15}\hat{\sigma}^z_{72} - 32968\hat{\sigma}^z_{7}\hat{\sigma}^z_{81} - 32968\hat{\sigma}^z_{8}\hat{\sigma}^z_{80} - 16\hat{\sigma}^z_{15}\hat{\sigma}^z_{73} + 8\hat{\sigma}^z_{29}\hat{\sigma}^z_{59} - 32968\hat{\sigma}^z_{7}\hat{\sigma}^z_{82} - 16\hat{\sigma}^z_{9}\hat{\sigma}^z_{80} - 32968\hat{\sigma}^z_{12}\hat{\sigma}^z_{77} + 8\hat{\sigma}^z_{16}\hat{\sigma}^z_{73} - 32968\hat{\sigma}^z_{7}\hat{\sigma}^z_{83} - 32968\hat{\sigma}^z_{9}\hat{\sigma}^z_{81} - 16\hat{\sigma}^z_{13}\hat{\sigma}^z_{77} - 32968\hat{\sigma}^z_{7}\hat{\sigma}^z_{84} - 16\hat{\sigma}^z_{10}\hat{\sigma}^z_{81} - 8\hat{\sigma}^z_{11}\hat{\sigma}^z_{80} - 32968\hat{\sigma}^z_{13}\hat{\sigma}^z_{78} - 16\hat{\sigma}^z_{14}\hat{\sigma}^z_{77} - 32968\hat{\sigma}^z_{7}\hat{\sigma}^z_{85} - 32968\hat{\sigma}^z_{10}\hat{\sigma}^z_{82} - 16\hat{\sigma}^z_{14}\hat{\sigma}^z_{78} + 16\hat{\sigma}^z_{15}\hat{\sigma}^z_{77} - 16\hat{\sigma}^z_{11}\hat{\sigma}^z_{82} - 32968\hat{\sigma}^z_{14}\hat{\sigma}^z_{79} - 16\hat{\sigma}^z_{15}\hat{\sigma}^z_{78} + 8\hat{\sigma}^z_{28}\hat{\sigma}^z_{65} - 32968\hat{\sigma}^z_{8}\hat{\sigma}^z_{86} - 16\hat{\sigma}^z_{15}\hat{\sigma}^z_{79} - 32968\hat{\sigma}^z_{8}\hat{\sigma}^z_{87} - 32968\hat{\sigma}^z_{9}\hat{\sigma}^z_{86} - 32968\hat{\sigma}^z_{12}\hat{\sigma}^z_{83} + 8\hat{\sigma}^z_{29}\hat{\sigma}^z_{66} - 32968\hat{\sigma}^z_{8}\hat{\sigma}^z_{88} - 16\hat{\sigma}^z_{13}\hat{\sigma}^z_{83} - 32968\hat{\sigma}^z_{8}\hat{\sigma}^z_{89} - 32968\hat{\sigma}^z_{10}\hat{\sigma}^z_{87} - 32\hat{\sigma}^z_{11}\hat{\sigma}^z_{86} - 32968\hat{\sigma}^z_{13}\hat{\sigma}^z_{84} + 8\hat{\sigma}^z_{39}\hat{\sigma}^z_{58} - 32968\hat{\sigma}^z_{8}\hat{\sigma}^z_{90} - 32\hat{\sigma}^z_{11}\hat{\sigma}^z_{87} - 16\hat{\sigma}^z_{14}\hat{\sigma}^z_{84} - 8\hat{\sigma}^z_{15}\hat{\sigma}^z_{83} - 32968\hat{\sigma}^z_{11}\hat{\sigma}^z_{88} - 32\hat{\sigma}^z_{12}\hat{\sigma}^z_{87} - 32\hat{\sigma}^z_{13}\hat{\sigma}^z_{86} - 32968\hat{\sigma}^z_{14}\hat{\sigma}^z_{85} - 32968\hat{\sigma}^z_{9}\hat{\sigma}^z_{91} - 16\hat{\sigma}^z_{12}\hat{\sigma}^z_{88} - 32\hat{\sigma}^z_{14}\hat{\sigma}^z_{86} - 16\hat{\sigma}^z_{15}\hat{\sigma}^z_{85} - 32968\hat{\sigma}^z_{9}\hat{\sigma}^z_{92} - 32968\hat{\sigma}^z_{10}\hat{\sigma}^z_{91} - 32968\hat{\sigma}^z_{12}\hat{\sigma}^z_{89} - 32\hat{\sigma}^z_{13}\hat{\sigma}^z_{88} - 32\hat{\sigma}^z_{14}\hat{\sigma}^z_{87} - 32\hat{\sigma}^z_{15}\hat{\sigma}^z_{86} + 8\hat{\sigma}^z_{28}\hat{\sigma}^z_{73} - 32968\hat{\sigma}^z_{9}\hat{\sigma}^z_{93} + 16\hat{\sigma}^z_{14}\hat{\sigma}^z_{88} - 16\hat{\sigma}^z_{15}\hat{\sigma}^z_{87} + 16\hat{\sigma}^z_{16}\hat{\sigma}^z_{86} - 32968\hat{\sigma}^z_{9}\hat{\sigma}^z_{94} - 32968\hat{\sigma}^z_{11}\hat{\sigma}^z_{92} - 32968\hat{\sigma}^z_{13}\hat{\sigma}^z_{90} - 32\hat{\sigma}^z_{14}\hat{\sigma}^z_{89} - 16\hat{\sigma}^z_{15}\hat{\sigma}^z_{88} - 32968\hat{\sigma}^z_{9}\hat{\sigma}^z_{95} - 32\hat{\sigma}^z_{12}\hat{\sigma}^z_{92} - 32\hat{\sigma}^z_{13}\hat{\sigma}^z_{91} - 16\hat{\sigma}^z_{15}\hat{\sigma}^z_{89} + 16\hat{\sigma}^z_{17}\hat{\sigma}^z_{87} + 8\hat{\sigma}^z_{39}\hat{\sigma}^z_{65} - 32968\hat{\sigma}^z_{12}\hat{\sigma}^z_{93} - 32\hat{\sigma}^z_{14}\hat{\sigma}^z_{91} - 32\hat{\sigma}^z_{15}\hat{\sigma}^z_{90} - 32968\hat{\sigma}^z_{10}\hat{\sigma}^z_{96} - 32\hat{\sigma}^z_{13}\hat{\sigma}^z_{93} - 32\hat{\sigma}^z_{14}\hat{\sigma}^z_{92} - 32\hat{\sigma}^z_{15}\hat{\sigma}^z_{91} + 16\hat{\sigma}^z_{18}\hat{\sigma}^z_{88} - 32968\hat{\sigma}^z_{10}\hat{\sigma}^z_{97} - 32968\hat{\sigma}^z_{11}\hat{\sigma}^z_{96} - 32968\hat{\sigma}^z_{13}\hat{\sigma}^z_{94} - 16\hat{\sigma}^z_{15}\hat{\sigma}^z_{92} + 16\hat{\sigma}^z_{16}\hat{\sigma}^z_{91} - 32968\hat{\sigma}^z_{10}\hat{\sigma}^z_{98} - 32\hat{\sigma}^z_{14}\hat{\sigma}^z_{94} - 32968\hat{\sigma}^z_{10}\hat{\sigma}^z_{99} - 32968\hat{\sigma}^z_{12}\hat{\sigma}^z_{97} - 32\hat{\sigma}^z_{13}\hat{\sigma}^z_{96} - 32968\hat{\sigma}^z_{14}\hat{\sigma}^z_{95} + 16\hat{\sigma}^z_{17}\hat{\sigma}^z_{92} - 32\hat{\sigma}^z_{13}\hat{\sigma}^z_{97} - 32\hat{\sigma}^z_{14}\hat{\sigma}^z_{96} - 32\hat{\sigma}^z_{15}\hat{\sigma}^z_{95} - 32968\hat{\sigma}^z_{11}\hat{\sigma}^z_{100} - 32968\hat{\sigma}^z_{13}\hat{\sigma}^z_{98} - 32\hat{\sigma}^z_{15}\hat{\sigma}^z_{96} - 32968\hat{\sigma}^z_{11}\hat{\sigma}^z_{101} - 32968\hat{\sigma}^z_{12}\hat{\sigma}^z_{100} - 32\hat{\sigma}^z_{14}\hat{\sigma}^z_{98} + 16\hat{\sigma}^z_{16}\hat{\sigma}^z_{96} + 8\hat{\sigma}^z_{39}\hat{\sigma}^z_{73} - 32968\hat{\sigma}^z_{14}\hat{\sigma}^z_{99} - 32968\hat{\sigma}^z_{12}\hat{\sigma}^z_{102} - 32968\hat{\sigma}^z_{13}\hat{\sigma}^z_{101} - 32\hat{\sigma}^z_{14}\hat{\sigma}^z_{100} - 32\hat{\sigma}^z_{15}\hat{\sigma}^z_{99} + 16\hat{\sigma}^z_{28}\hat{\sigma}^z_{86} - 32968\hat{\sigma}^z_{12}\hat{\sigma}^z_{103} - 32968\hat{\sigma}^z_{13}\hat{\sigma}^z_{102} - 16\hat{\sigma}^z_{15}\hat{\sigma}^z_{100} - 32968\hat{\sigma}^z_{12}\hat{\sigma}^z_{104} - 32\hat{\sigma}^z_{15}\hat{\sigma}^z_{101} + 16\hat{\sigma}^z_{29}\hat{\sigma}^z_{87} - 32968\hat{\sigma}^z_{14}\hat{\sigma}^z_{103} - 96\hat{\sigma}^z_{15}\hat{\sigma}^z_{102} - 32968\hat{\sigma}^z_{13}\hat{\sigma}^z_{105} + 16\hat{\sigma}^z_{16}\hat{\sigma}^z_{102} - 32968\hat{\sigma}^z_{13}\hat{\sigma}^z_{106} - 32968\hat{\sigma}^z_{14}\hat{\sigma}^z_{105} - 32968\hat{\sigma}^z_{15}\hat{\sigma}^z_{104} + 16\hat{\sigma}^z_{28}\hat{\sigma}^z_{91} - 64\hat{\sigma}^z_{15}\hat{\sigma}^z_{105} + 16\hat{\sigma}^z_{17}\hat{\sigma}^z_{103} - 32968\hat{\sigma}^z_{14}\hat{\sigma}^z_{107} - 32968\hat{\sigma}^z_{15}\hat{\sigma}^z_{106} + 16\hat{\sigma}^z_{16}\hat{\sigma}^z_{105} + 16\hat{\sigma}^z_{29}\hat{\sigma}^z_{92} - 32968\hat{\sigma}^z_{15}\hat{\sigma}^z_{107} + 16\hat{\sigma}^z_{18}\hat{\sigma}^z_{104} + 16\hat{\sigma}^z_{16}\hat{\sigma}^z_{107} + 16\hat{\sigma}^z_{17}\hat{\sigma}^z_{106} + 16\hat{\sigma}^z_{28}\hat{\sigma}^z_{96} + 16\hat{\sigma}^z_{39}\hat{\sigma}^z_{86} + 16\hat{\sigma}^z_{28}\hat{\sigma}^z_{102} + 16\hat{\sigma}^z_{39}\hat{\sigma}^z_{91} + 16\hat{\sigma}^z_{58}\hat{\sigma}^z_{73} + 16\hat{\sigma}^z_{29}\hat{\sigma}^z_{103} + 16\hat{\sigma}^z_{28}\hat{\sigma}^z_{105} + 16\hat{\sigma}^z_{28}\hat{\sigma}^z_{107} + 16\hat{\sigma}^z_{29}\hat{\sigma}^z_{106} + 16\hat{\sigma}^z_{39}\hat{\sigma}^z_{96} + 16\hat{\sigma}^z_{39}\hat{\sigma}^z_{102} + 16\hat{\sigma}^z_{39}\hat{\sigma}^z_{105} + 16\hat{\sigma}^z_{58}\hat{\sigma}^z_{86} + 16\hat{\sigma}^z_{39}\hat{\sigma}^z_{107} + 16\hat{\sigma}^z_{59}\hat{\sigma}^z_{87} + 16\hat{\sigma}^z_{60}\hat{\sigma}^z_{88} + 16\hat{\sigma}^z_{58}\hat{\sigma}^z_{91} + 16\hat{\sigma}^z_{59}\hat{\sigma}^z_{92} + 16\hat{\sigma}^z_{65}\hat{\sigma}^z_{86} + 16\hat{\sigma}^z_{66}\hat{\sigma}^z_{87} + 16\hat{\sigma}^z_{58}\hat{\sigma}^z_{96} + 16\hat{\sigma}^z_{65}\hat{\sigma}^z_{91} + 16\hat{\sigma}^z_{66}\hat{\sigma}^z_{92} + 16\hat{\sigma}^z_{73}\hat{\sigma}^z_{86} + 16\hat{\sigma}^z_{58}\hat{\sigma}^z_{102} + 16\hat{\sigma}^z_{65}\hat{\sigma}^z_{96} + 16\hat{\sigma}^z_{59}\hat{\sigma}^z_{103} + 16\hat{\sigma}^z_{58}\hat{\sigma}^z_{105} + 16\hat{\sigma}^z_{60}\hat{\sigma}^z_{104} + 16\hat{\sigma}^z_{73}\hat{\sigma}^z_{91} + 16\hat{\sigma}^z_{58}\hat{\sigma}^z_{107} + 16\hat{\sigma}^z_{59}\hat{\sigma}^z_{106} + 16\hat{\sigma}^z_{65}\hat{\sigma}^z_{102} + 16\hat{\sigma}^z_{66}\hat{\sigma}^z_{103} + 16\hat{\sigma}^z_{73}\hat{\sigma}^z_{96} + 16\hat{\sigma}^z_{65}\hat{\sigma}^z_{105} + 16\hat{\sigma}^z_{65}\hat{\sigma}^z_{107} + 16\hat{\sigma}^z_{66}\hat{\sigma}^z_{106} + 16\hat{\sigma}^z_{73}\hat{\sigma}^z_{102} + 16\hat{\sigma}^z_{73}\hat{\sigma}^z_{105} + 16\hat{\sigma}^z_{73}\hat{\sigma}^z_{107} + 64\hat{\sigma}^z_{86}\hat{\sigma}^z_{96} + 32\hat{\sigma}^z_{86}\hat{\sigma}^z_{102} + 32\hat{\sigma}^z_{87}\hat{\sigma}^z_{103} + 32\hat{\sigma}^z_{86}\hat{\sigma}^z_{105} + 32\hat{\sigma}^z_{88}\hat{\sigma}^z_{104} + 32\hat{\sigma}^z_{86}\hat{\sigma}^z_{107} + 32\hat{\sigma}^z_{87}\hat{\sigma}^z_{106} + 32\hat{\sigma}^z_{91}\hat{\sigma}^z_{102} + 32\hat{\sigma}^z_{92}\hat{\sigma}^z_{103} + 32\hat{\sigma}^z_{91}\hat{\sigma}^z_{105} + 32\hat{\sigma}^z_{91}\hat{\sigma}^z_{107} + 32\hat{\sigma}^z_{92}\hat{\sigma}^z_{106} + 32\hat{\sigma}^z_{96}\hat{\sigma}^z_{102} + 32\hat{\sigma}^z_{96}\hat{\sigma}^z_{105} + 32\hat{\sigma}^z_{96}\hat{\sigma}^z_{107} + 64\hat{\sigma}^z_{102}\hat{\sigma}^z_{107} + 4551232$


\section*{References}
\begin{thebibliography}{9}
%
\bibitem{sylvester1867}
Sylvester, J.J.
Thoughts on inverse orthogonal matrices, simultaneous sign successions, and tessellated pavements in two or more colours, with applications to Newton's Rule, ornamental tile-work, and the theory of numbers.
{\em Phil. Mag.}, {\bf 34}, 461-475 (1867).

\bibitem{hadamard_1893}
Hadamard, J.
Resolution d'une question relative aux determinants.
{\em Bulletin des Sciences Mathematiques}, {\bf 17}, 240–246 (1893).

\bibitem{paley_1933}
Paley, R.E.A.C.
On Orthogonal Matrices.
{\em Journal of Mathematics and Physics}, {\bf 12}, 311–320(1933).

\bibitem{williamson_1944}
Williamson, J.
Hadamard's determinant theorem and the sum of four squares.
{\em Duke Math. J.}, {\bf 11}, 65-81(1944).

\bibitem{baumert_golomb_hall_1962}
Baumert, L., Golomb, S.W., and Hall, M.
Discovery of an Hadamard matrix of order 92.
{\emph Bull. Amer. Math. Soc.},
{\bf 68}, No. 3, 237-238 (1962).

\bibitem{baumert_hall_1965}
Baumert, L.D., and Hall Jr, M.
A new construction for Hadamard matrices.
{\em Bull. Amer. Math. Soc.}, {\bf 71}, pp. 169–170 (1965).

\bibitem{turyn_1974}
Turyn, R.J. 
Hadamard Matrices, Baumert-Hall Units, Four-Symbol Sequences, Pulse Compression, and Surface Wave Encodings.
{\em Journal of Combinatorial Theory (A)}, {\bf 16}, 313-333 (1974).

\bibitem{hedayat_1978}
Hedayat, A. and Wallis, W.D. 
Hadamard Matrices and Their Applications.
{\em The Ann. Of Statistics}, Princeton University Press, {\bf 6}, 1184-1238 (1978).

\bibitem{kharaghani_rezaie_2004}
Kharaghani,H. and Tayfeh‐Rezaie, B.
A Hadamard matrix of order 428.
{\em Journal of Combinatorial Designs}, {\bf 13}, 6 (2004).

\bibitem{horadam_2007}
Horadam, K.J.  
{\em Hadamard Matrices and Their Applications},
Princeton University Press (2007), ISBN: 9781400842902.

\bibitem{best_2012}
Best, D., Dokovic, D.Z., Kharaghani, H. and  Ramp, H. 
Turyn‐Type Sequences: Classification, Enumeration, and Construction.
{\em Journal of Combinatorial Designs}, {\bf 21}, 1 (16 May 2012).

\bibitem{london_2013}
London, S.
Constructing New Turyn Type Sequences, T-Sequences and Hadamard Matrices.
{\em PhD. Thesis, Grad. College}, University of Illinois at Chicago (2013).

\bibitem{harrow_2017}
Harrow, A.W. and Montanaro, A.
Quantum computational supremacy.
{\em Nature}, {\bf 549}, 7671, September 2017.

\bibitem{arute_2019}
Arute, F. et al. 
Quantum supremacy using a programmable superconducting processor.
{\em Nature}, {\bf 574}, 7779: 505–510, 23 October 2019. 

\bibitem{santoro_2002}
Santoro, G.E., Martonak, R., Tosatti, E., and Car, E.
Theory of quantum annealing of an Ising spin glass.
{\em Science}, {\bf 295}, 2427-30(2002).

\bibitem{suksmono_minato_2019}
Suksmono, A.B. and Minato, Y.  
Finding Hadamard Matrices by a Quantum Annealing Machine.
{\em Scientific Reports}, {\bf 9}, Article number: 14380 (2019). 
%
\bibitem{kadowaki_nishimori_1988}
Kadowaki, T. and Nishimori, H. 
Quantum annealing in the transverse Ising model.
{\em Phys. Rev.E}, {\bf 58}, 10.1103/PhysRevE.58.5355 (1988).

\bibitem{boixo_2014}
Boixo, S. et al.
Evidence for quantum annealing with more than one hundred qubits.
{\em Nature Physics}, {\bf 10}, 218–224(2014).

\bibitem{benedetti_2017}
Benedetti, M., Realpe-Gómez,  J., Biswas,  R.,  and Perdomo-Ortiz, A.
Quantum-assisted learning of hardware-embedded probabilistic graphical models.
{\em Phys. Rev.X}, {\bf 7}, 041052 (2017).

\bibitem{li_2018}
Li, R., Felice, R., Rohs, R., and Lidar, D. 
Quantum annealing versus classical machine learning applied to a simplified computational biology problem.
{\em Npj Quantum Information}, {\bf 4}, (2018).

\bibitem{omalley_2018}
O’Malley, D.
An approach to quantum-computational hydrologic inverse analysis.
{\em Scientific Reports}, {\bf 8}, (2018).

\bibitem{perdomo2008}
Perdomo, A., Truncik, C., Tubert-Brohman, I., Rose, G. and Aspuru-Guzik, A.
Construction of model hamiltonians for adiabatic quantum computation and its application to finding low-energy conformations of lattice protein models. 
{\em Phys. Rev. A}, {\bf 78}, 012320–15 (2008).

\bibitem{dwave_matlab_2018}
D-Wave System, Inc., 
\emph{Developer Guide for MATLAB: User Manual}
D-Wave System Inc. (2018).

\bibitem{dwave_2020}
\emph{D-Wave Systems}, 16 September 2020.
Retrieved from \url{https://en.wikipedia.org/wiki/D-Wave_Systems}

\bibitem{jurcevic_2020}
P. Jurcevic, et al.
Demonstration of quantum volume 64 on a superconducting quantum computing system.
\emph{arXiv:2008.08571 [quant-ph]}

\bibitem{cho_2020}
A.Cho,
IBM promises 1000-qubit quantum computer—a milestone—by 2023.
{\em Science}, Sep. 15, 2020.
\end{thebibliography}
\end{document}